\documentclass[sn-mathphys-ay]{sn-jnl}
\usepackage{graphicx}
\usepackage{multirow}
\usepackage{amsmath,amssymb,amsfonts}
\usepackage{amsthm}
\usepackage{mathrsfs}
\usepackage[title]{appendix}
\usepackage{xcolor}
\usepackage{textcomp}
\usepackage{manyfoot}
\usepackage{booktabs}
\usepackage{algorithm}
\usepackage{algorithmicx}
\usepackage{algpseudocode}
\usepackage{listings}
\begin{document}

\title[Article Title]{P53 Orchestrates Cancer Metabolism: Unveiling Strategies to Reverse the Warburg Effect}
\author*[1,2]{\fnm{Roba} \sur{Abukwaik}}\email{rabukwaik@kau.edu.sa}

\author[2,3]{\fnm{Elias} \sur{Vera-Siguenza}}\email{e.vera-siguenza@bham.ac.uk}

\author[3]{\fnm{Daniel} \sur{Tennant}}\email{d.tennant@bham.ac.uk}

\author*[2]{\fnm{Fabian} \sur{Spill}}\email{f.spill@bham.ac.uk}
\affil*[1]{\orgdiv{Mathematics Department}, \orgname{King Abdulaziz University}, \orgaddress{\city{Rabigh}, \country{Saudi Arabia}}}

\affil*[2]{\orgdiv{School of Mathematics}, \orgname{University of Birmingham}, \orgaddress{\city{Birmingham}, \postcode{B15 2TS}, \country{United Kingdom}}}

\affil[3]{\orgdiv{Institute of Metabolism and Systems Research}, \orgname{University of Birmingham}, \orgaddress{\city{Birmingham}, \postcode{B15 2TT},  \country{United Kingdom}}}
\abstract{Cancer cells exhibit significant alterations in their metabolism, characterised by a reduction in oxidative phosphorylation (OXPHOS) and an increased reliance on glycolysis, even in the presence of oxygen. This metabolic shift, known as the Warburg effect, is pivotal in fuelling cancer's uncontrolled growth, invasion, and therapeutic resistance. While dysregulation of many genes contributes to this metabolic shift, the tumour suppressor gene p53 emerges as a master player. Yet, the molecular mechanisms remain elusive. This study introduces a comprehensive mathematical model, integrating essential p53 targets, offering insights into how p53 orchestrates its targets to redirect cancer metabolism towards an OXPHOS-dominant state. Simulation outcomes align closely with experimental data comparing glucose metabolism in colon cancer cells with wild-type and mutated p53. Additionally, our findings reveal the dynamic capability of elevated p53 activation to fully reverse the Warburg effect, highlighting the significance of its activity levels not just in triggering apoptosis (programmed cell death) post-chemotherapy but also in modifying the metabolic pathways implicated in treatment resistance. In scenarios of p53 mutations, our analysis suggests targeting glycolysis-instigating signalling pathways as an alternative strategy, whereas targeting solely synthesis of cytochrome c oxidase 2 (SCO2) does support mitochondrial respiration but may not effectively suppress the glycolysis pathway, potentially boosting the energy production and cancer cell viability.}
\keywords{Cancer Metabolism, p53, Warburg effect, Glycolysis, Hypoxia, Mathematical Biology}
\maketitle
\section{Introduction}
Cancer cells undergo profound metabolic alterations facilitating their proliferation, invasion, metastasis, and even drug resistance \citep{han2013does,rahman2015cancer}. Unlike normal cells, cancer cells derive a substantial amount of their energy from glycolysis, converting a majority of incoming glucose into lactate in the cytoplasm rather than metabolising it in the mitochondria through oxidative phosphorylation (OXPHOS) \citep{cairns2011regulation,simabuco2018p53}. This metabolic adaptation, recognised as the Warburg effect or aerobic glycolysis, leads to decreased oxygen consumption required by mitochondrial respiration while generating an increased amount of lactate \citep{simabuco2018p53}.\\

By favouring glycolysis over OXPHOS, cancer cells ensure the availability of essential building blocks for biomass synthesis and meet the energy demands necessary for their rapid growth \citep{hanahan2011hallmarks,simabuco2018p53}. While glycolysis can produce adenosine triphosphate (ATP), the major cellular energy unit, more rapidly than oxidative phosphorylation, it is significantly less efficient in terms of ATP generated per unit of glucose consumed \citep{cairns2011regulation,simabuco2018p53}. Consequently, tumour cells increase their glucose uptake at an exceptionally high rate to adequately satisfy their elevated energy and biosynthesis needs \citep{cairns2011regulation,simabuco2018p53}.\\

The glycolytic phenotype of cancer cells is influenced by various molecular mechanisms extending beyond hypoxic conditions. Disruptions in signalling pathways downstream of growth factor receptors have been observed to affect glucose metabolism in cancer cells \citep{zhong2000modulation,laughner2001her2}. Specifically, the PI3K/AKT/mTOR pathway, which is activated in the vast majority of human cancers \citep{hennessy2005exploiting,danielsen2015portrait,vara2004pi3k,malinowsky2014activation,wang2013april}, and seen to instigate the glycolytic activity of cancer cells by upregulating the hypoxia-inducible factor 1 (HIF1) and its downstream targets \citep{cairns2011regulation,valvona2016regulation,laughner2001her2,zhong2000modulation}.\\

Another crucial event that can impact cancer metabolism and is commonly observed in cancer is the inactivation of the tumour suppressor gene p53. Depending on the cellular conditions, p53 suppresses tumorigenesis by multiple mechanisms, including cell cycle regulation, initiation of DNA repair, and induction of apoptosis  (programmed cell death) \citep{wanka2012synthesis,simabuco2018p53}. Moreover, p53 has recently emerged as a significant metabolic regulator in cancer cells, whether by inhibiting the PI3K/AKT/mTOR pathway \citep{feng2010regulation}, thereby disrupting the glycolytic phenotype or by supporting mitochondrial respiration activity  \citep{vousden2009p53,zhang2010role,lago2011p53,wanka2012synthesis,liang2013regulation,floter2017regulation,simabuco2018p53,liu2019tumor}. This idea was investigated by Matoba et al. in 2006, where they examined the impact of p53 alterations on the cellular metabolism of human colon cancer cells \citep{matoba2006p53}. Experimental results revealed that p53-deficient cells produced nearly the same amount of ATP but with substantially higher levels of lactate and lower oxygen consumption, highlighting the influence of p53 mutations in changing the energy production mode to one favouring glycolysis \citep{matoba2006p53}.\\

The metabolic response controlled by p53 is mediated through the AMP-activated protein kinase (AMPK), a sensor attuned to cellular metabolic stress conditions \citep{jones2005amp}. When activated by AMPK, typically in response to metabolic adversity such as those experienced by cancer cells, p53 restrains cell growth and division, conserves energy, and shifts the cell towards oxidative phosphorylation for more efficient energy production \citep{feng2010regulation}. This may explain why cancer cells with p53 mutations tend to rely more on glycolysis and have a higher ability to grow and survive even under stress conditions.\\

Previous mathematical modelling sheds light on different aspects of cancer metabolism —ranging from the effect of reactive oxygen species (ROS) on HIF1 stabilisation in ischemic conditions \citep{qutub2008reactive} to the identification of metabolic targets to hinder cancer migration \citep{yizhak2014computational}. Despite these insights, the genetic complexities underpinning the Warburg effect remain elusive. Recently, Linglin et al. made a notable contribution by discussing the genetic regulation of the interplay between glycolysis and oxidative phosphorylation \citep{yu2017modeling}. Nonetheless, this work did not account for the crucial influence of p53 or demonstrate its impact on metabolic pathways as observed experimentally in cancer cells.\\

Undoubtedly, the tumour-suppressive role of p53 has been studied through several computational modelling approaches. Ma et al. modelled the oscillatory dynamics of p53, emphasising the correlation between the DNA damage severity and the average number of p53 pulses \citep{ma2005plausible}. Further studies by Zhang et al. have elucidated how the frequency of p53 pulses can influence cellular decisions between repair and apoptosis \citep{zhang2007exploring} and described the transition from cell repair to apoptosis with an inverse relationship between the apoptosis time and damage strength \citep{zhang2009computational}. Another study has introduced a two-phase dynamic of p53, distinguishing between partial and complete activation of p53 \citep{zhang2011two}. Our recent study expanded on these findings by demonstrating that p53 can switch among three dynamic modes in a DNA damage strength-dependent manner following chemotherapy \citep{abukwaik2023interplay}. Despite these advances, earlier works have mainly concentrated on p53 dynamic behaviour in response to DNA damage stimuli and the subsequent cell fate, leaving a substantial gap in understanding its mechanisms regarding cellular metabolism.\\

To address this gap, this study develops a mathematical model that unveils the intricate machinery behind the regulation of cancer glucose metabolism by p53. While numerous p53 targets involved in cellular metabolism have been identified, their complex molecular interactions across different scenarios remain largely unexplored. By transforming existing experimental data into a mathematical framework, we uncovered hard-to-detect mechanisms and quantitatively analysed the activities of glycolysis and OXPHOS pathways under different cellular states. This methodology provides valuable insights for developing targeted therapeutic approaches aimed at disrupting cancer metabolism and combating the aggressive behaviour of cancer. Although our primary focus is on colon cancer cells, the model's applicability extends to many cancer types experiencing similar conditions.
\section{Methods}
\begin{figure}
\centering
\includegraphics[width=1\textwidth]{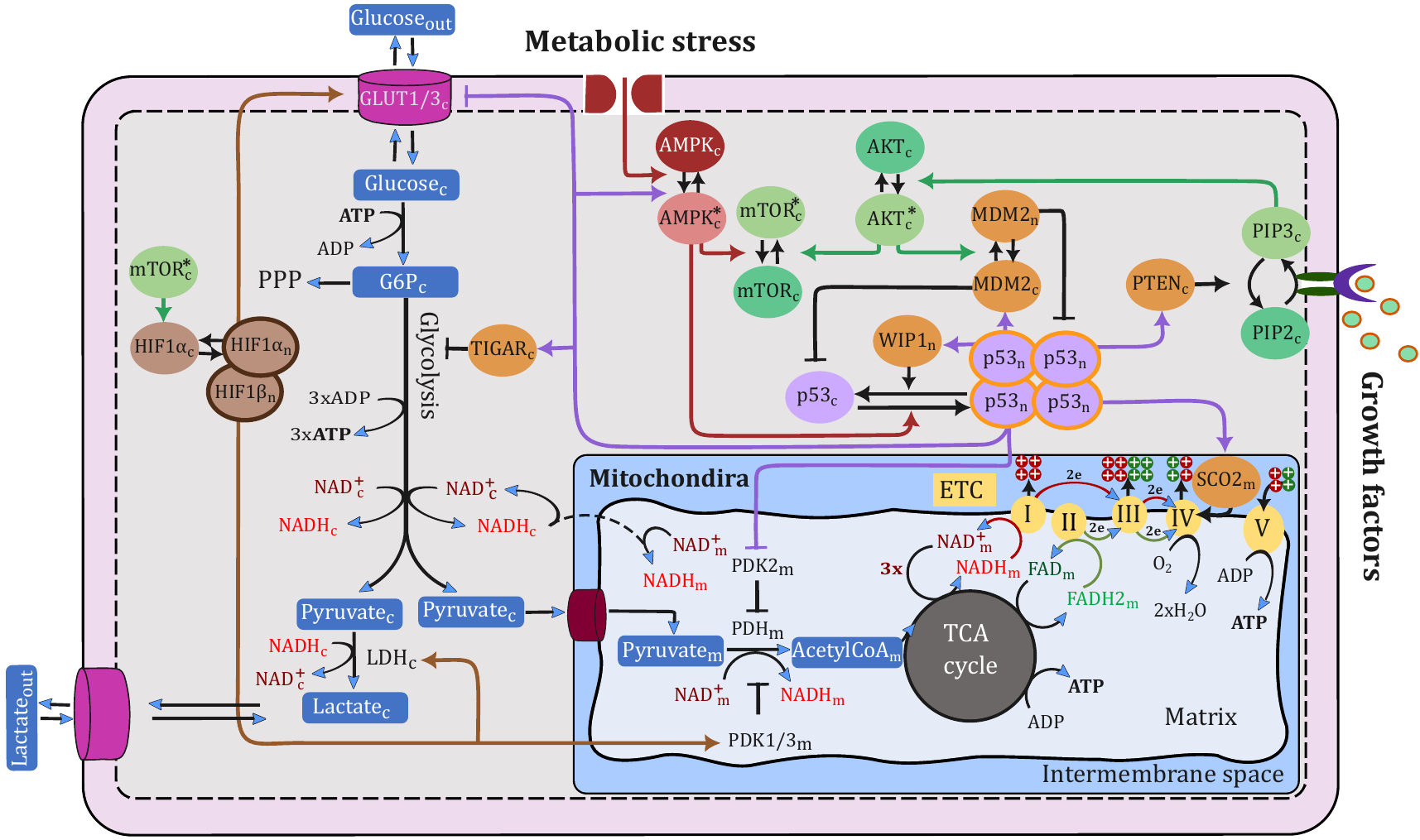}
\caption{Schematic diagram depicting the signalling pathway of key molecules involved in glucose oxidation, spanning glycolysis, the tricarboxylic acid (TCA) cycle, and the electron transport chain (ETC). In our notation, cytoplasmic, nuclear, and mitochondrial molecular species are indicated by subscripts `c', `n', and `m', respectively, while the `$*$' superscript symbol is used to denote active species in the case of species that exist in two states (active and inactive)}
\label{Fig1}
\end{figure}

We constructed a comprehensive theoretical framework aiming to delineate the role of p53 on cellular metabolism, particularly its involvement in the Warburg effect. While cancer cells engage in diverse metabolic pathways, the Warburg effect is closely associated with alterations in glucose metabolism. Consequently, our primary attention was devoted to glucose metabolism, investigating its main pathways: glycolysis and oxidative phosphorylation. Integrating information from literature, our model incorporated all well-established p53 targets that markedly manipulate these pathways alongside the signalling pathways commonly activated in cancer in response to growth factors or metabolic stress, influencing the decision-making between these pathways, see Fig. \ref{Fig1}. The following section briefly discusses these signalling pathways and their involvement in glucose metabolism.
\subsection{Signalling pathways in glucose metabolism}
\subsubsection{Growth factor signalling pathway}
In response to growth factors, extracellular molecules activate specific receptors on the cell membrane, initiating the intracellular activation of phosphoinositide 3-kinase (PI3K). PI3K then phosphorylates phosphatidylinositol 4,5-bisphosphate (PIP2) into phosphatidylinositol 3,4,5-trisphosphate (PIP3), facilitating the protein kinase B (AKT) activation \citep{danielsen2015portrait,vara2004pi3k,carnero2014pten}.\\

Upon AKT activation, a cascade of events occurs, ultimately activating the mechanistic target of rapamycin (mTOR) \citep{dan2014akt,inoki2002tsc2}, which subsequently promotes the synthesis of proteins required for cell growth and proliferation. One of these proteins is HIF1$\alpha$ \citep{duvel2010activation,laughner2001her2,hudson2002regulation,treins2005regulation}, a transcription factor that drives the expression of genes essential for glycolytic metabolism. These genes include glucose transporter-1 and -3 (GLUT1) and (GLUT3), respectively \citep{ancey2018glucose}, lactate dehydrogenase (LDH) \citep{valvona2016regulation}, and pyruvate dehydrogenase kinase-1 and -3 (PDK1) and (PDK3), respectively \citep{anwar2021targeting,lu2011overexpression,wang2021pyruvate,kim2006hif}.
\subsubsection{Metabolic stress signalling pathway}
Different forms of metabolic stress activate AMPK, a crucial enzyme that restores cellular energy balance by suppressing ATP-consuming processes and promoting ATP production \citep{hardie2011sensing,hardie2012ampk,faubert2015amp,li2015targeting}. Through this activation, AMPK restrains cellular growth and increases ATP production efficiency primarily by repressing mTOR activation and stimulating the transcriptional activity of p53 \citep{faubert2015amp,li2015targeting,inoki2003tsc2}.\\

By phosphorating p53, AMPK disrupts the interaction between p53 and its negative regulator, murine double minute 2 (MDM2), leading to p53 stabilisation \citep{jones2005amp,imamura2001cell}. Consequently, p53 accumulates and translocates to the nucleus, triggering the promoter activity of its target genes.
\subsubsection{p53 transcriptional targets and feedback mechanisms}
p53 exerts its effects through the transcriptional regulation of a wide array of target genes involved in various cellular processes. Some p53 targets form a negative feedback loop that dampens its stability, such as MDM2 and wild-type p53-induced phosphatase 1 (WIP1) \citep{barak1993mdm2,batchelor2011stimulus}. WIP1 dephosphorylates p53, increasing its susceptibility to MDM2-mediated degradation \citep{batchelor2011stimulus}.\\ 

Conversely, p53 activates genes that promote its activation and simultaneously inhibit the glycolytic regulator HIF1. For instance, p53 induces the expression of proteins that activate AMPK, which in turn inhibits mTOR activity and its downstream target HIF1 \citep{budanov2008p53,sanli2012sestrin2,feng2010regulation}. This process forms a positive feedback loop, as active AMPK phosphorylates p53, further enhancing its stabilization \citep{feng2010regulation}. Another crucial target of p53 is phosphatase and tensin homolog (PTEN) \citep{stambolic2001regulation}, which attenuates the PI3K/AKT/mTOR pathway by converting PIP3 back to PIP2 \citep{mayo2002pten,carnero2014pten,feng2010regulation}. The induction of PTEN also supports p53 by hindering the AKT-dependent translocation of MDM2 to the nucleus, thereby boosting the transcriptional activity of p53 \citep{mayo2002pten,mayo2005phosphorylation}.\\

In addition, p53 regulates genes directly involved in glucose metabolic pathways. This includes diminishing the protein synthesis of GLUT1, GLUT3 \citep{ancey2018glucose,schwartzenberg2004tumor,kawauchi2008p53}, and PDK2 \citep{anwar2021targeting,liang2020dichloroacetate}, while stimulating the expression of other molecules, such as the TP53-inducible glycolysis and apoptosis regulator (TIGAR) \citep{bensaad2006tigar,lee2015p53}, and the synthesis of cytochrome c oxidase 2 (SCO2) \citep{matoba2006p53}.
\subsubsection{Glucose metabolism pathways}
Glucose metabolism comprises three main stages: starting with glycolysis, advancing to the tricarboxylic acid (TCA) cycle, and finishing with the electron transport chain (ETC).\\
 \bmhead{Glycolysis}
 Glucose is transported to the cells by specific glucose transporters located in the cell membrane \citep{schwartzenberg2004tumor,szablewski2013expression,ancey2018glucose,mamun2020hypoxia}, namely GLUT1 and GLUT3 in our model, which are negatively regulated by p53 and positively by HIF1, controlling glucose uptake rate \citep{ancey2018glucose,schwartzenberg2004tumor,kawauchi2008p53}. Once inside the cell, glucose undergoes an irreversible conversion to glucose-6-phosphate (G6P), consuming one ATP molecule \citep{golias2019microenvironmental}. G6P can then either proceed through glycolysis or enter the pentose phosphate pathway (PPP) \citep{jiang2014regulation}, a decision influenced by the p53 target TIGAR, which inhibits the enzyme catalysing the third step in the glycolysis pathway, diminishing glycolysis flux \citep{bensaad2006tigar,lee2015p53}. Continuing with glycolysis, a molecule of G6P is converted into two molecules of pyruvate, yielding three net ATP and two reduced nicotinamide adenine dinucleotide (NADH) molecules \citep{valvona2016regulation,golias2019microenvironmental}.\\
 
 In the presence of oxygen, pyruvate derived from glycolysis typically enters the mitochondria for further energy production through oxidative phosphorylation. However, increased levels of LDH enzyme, driven by HIF1 activation, redirect pyruvate away from the mitochondria by catalysing its conversion to lactate and oxidising NADH back to NAD$^{+}$ \citep{valvona2016regulation,golias2019microenvironmental}. This shift allows cells to produce ATP less efficiently through glycolysis while consuming more glucose \citep{valvona2016regulation}.\\
\bmhead{TCA cycle}
 Within the mitochondria, the pyruvate dehydrogenase (PDH) complex facilitates the first step towards glucose respiration by catalysing the oxidative decarboxylation of pyruvate to acetyl-CoA, concurrently reducing one NAD$^{+}$ into NADH \citep{wang2021pyruvate,woolbright2019metabolic,rodrigues2015dichloroacetate}. Acetyl-CoA subsequently enters the TCA cycle, undergoing a series of chemical reactions that generate one energy molecule, three NADH, and one reduced flavin adenine dinucleotide (FADH2) molecule \citep{martinez2020mitochondrial}.\\
 
 The function of the PDH enzyme is controlled by the PDK family, which phosphorylates the E1-$\alpha$ subunit of PDH, blocking its decarboxylation activity \citep{wang2021pyruvate,woolbright2019metabolic,rodrigues2015dichloroacetate}. HIF1 enhances the promoter activities of two PDK family members, namely PDK1 and PDK3 \citep{anwar2021targeting,lu2011overexpression,wang2021pyruvate,kim2006hif}, while p53 suppresses the expression of another PDK member called PDK2 \citep{anwar2021targeting,liang2020dichloroacetate}.\\
\bmhead{ETC} 
 In the last stage, NADH and FADH2 are oxidised back into NAD$^{+}$ and FAD via protein complexes in the inner mitochondrial membrane \citep{ahmad2018biochemistry}. This oxidation process involves electron transfer from NADH and FADH2 across these complexes, during which protons are pumped from the mitochondrial matrix to the intermembrane space. This action creates a proton gradient that drives protons back into the matrix, facilitating ATP production \citep{ahmad2018biochemistry}.\\
 
 Among these complexes, Complex IV acts as the final electron acceptor, channelling the electrons to molecular oxygen by using 0.5 oxygen (O$_{2}$) molecules for each pair of electrons received from NADH or FADH2 \citep{ahmad2018biochemistry}. However, the activity of Complex IV is regulated by SCO2, a p53-regulated gene essential for its proper assembly and maturation. Thus, deficiency of SCO2 can impair Complex IV functionality and disrupt the electron flow within the ETC \citep{matoba2006p53,wanka2012synthesis}.
\subsection{Mathematical framework}
Our model encompasses the entire glucose oxidation network, spanning three compartments—cytoplasm, nucleus, and mitochondria. The directional fluxes within these compartments drive the temporal changes in the concentrations of 33 molecules involved in glucose oxidation pathways, as illustrated in Fig. \ref{Fig1}. Consequently, the system is governed by 33 differential equations where cytoplasmic, nuclear, and mitochondrial molecular species are represented by the subscripts `c', `n', and `m', respectively, while the `$*$' superscript symbol indicates active species for those existing in both active and inactive states. Each term in the model represented by '$\textbf{\textit{v}}$' notation corresponds to one of the reactions detailed in the model reactions section in the Appendix (\ref{secA1}), where $\textbf{\textit{v}}= v/V_{max}$ or $v/k$.
\begin{align}
\frac{dP53_{c}}{dt}&= k_{1}-k_{2}Ampk_{c}^*[\textbf{\textit{v}}_{1(P53_{c})}]-k_{3}Mdm2_{c}[\textbf{\textit{v}}_{1(P53_{c})}]+k_{4}Wip1_{n}[\textbf{\textit{v}}_{1(P53_{n})}]\notag\\
&-k_{5}P53_{c},\label{1}\\
\frac{dP53_{n}}{dt}&=k_{2}Ampk_{c}^*[\textbf{\textit{v}}_{1(P53_{c})}]-k_{6}Mdm2_{n}[\textbf{\textit{v}}_{1(P53_{n})}]-k_{4}Wip1_{n}[\textbf{\textit{v}}_{1(P53_{n})}],\label{2}\\
\frac{dMdm2_{c}}{dt}&= k_{7}+k_{8}[\textbf{\textit{v}}_{2(P53_{n})}^{+}]-k_{9}Akt_{c}^*[\textbf{\textit{v}}_{1(Mdm2_{c})}]
+k_{10}[\textbf{\textit{v}}_{1(Mdm2_{n})}]\notag\\
&-k_{11}Mdm2_{c},\label{3}\\
\frac{dMdm2_{n}}{dt}&=k_{9}Akt_{c}^*[\textbf{\textit{v}}_{1(Mdm2_{c})}]-k_{10}[\textbf{\textit{v}}_{1(Mdm2_{n})}]
-k_{11}Mdm2_{n},\label{4}\\
\frac{dWip1_{n}}{dt}&= k_{12}+k_{13}[\textbf{\textit{v}}_{2(P53_{n})}^{+}]-k_{14}Wip1_{n},\label{5}\\
\frac{dPten_{c}}{dt}&= k_{15}+k_{16}[\textbf{\textit{v}}_{2(P53_{n})}^{+}]-k_{17}Pten_{c},\label{6}\\
\frac{dSco2_{m}}{dt}&= k_{18}+k_{19}[\textbf{\textit{v}}_{2(P53_{n})}^{+}]-k_{20}Sco2_{m},\label{7}\\
\frac{dTigar_{c}}{dt}&= k_{21}+k_{22}[\textbf{\textit{v}}_{2(P53_{n})}^{+}]-k_{23}Tigar_{c},\label{8}\\
\frac{dAmpk_{c}^*}{dt}&= k_{24}[\textbf{\textit{v}}_{1(Ampk_{c})}]+k_{25}[\textbf{\textit{v}}_{2(P53_{n})}^{+}][\textbf{\textit{v}}_{1(Ampk_{c})}]-k_{26}[\textbf{\textit{v}}_{1(Ampk_{c}^*)}],\label{9}\\
\frac{dPip3_{c}}{dt}&= k_{27}[\textbf{\textit{v}}_{1(Pip2_{c})}]-k_{28}Pten_{c}[\textbf{\textit{v}}_{1(Pip3_{c})}],\label{10}\\
\frac{dAkt_{c}^*}{dt}&= k_{29}Pip3_{c}[\textbf{\textit{v}}_{1(Akt_{c})}]-k_{30}[\textbf{\textit{v}}_{1(Akt_{c}^*)}],\label{11}\\
\frac{dMtor_{c}^*}{dt}&= k_{31}Akt_{c}^*[\textbf{\textit{v}}_{1(Mtor_{c})}]-k_{32}Ampk_{c}^*[\textbf{\textit{v}}_{1(Mtor_{c}^*)}]-k_{33}[\textbf{\textit{v}}_{1(Mtor_{c}^*)}],\label{12}\\
\frac{dHif1\alpha_{c}}{dt} &=k_{34}+k_{35}Mtor_{c}^*-k_{36}Mtor_{c}^*Hif1\alpha_{c}-k_{37}Hif1\alpha_{c},\label{13}\\
\frac{dHif1\alpha_{n}}{dt} &=k_{36}Mtor_{c}^*Hif1\alpha_{c}-k_{37}Hif1\alpha_{n},\label{14}\\
\frac{dGlut1_{c}}{dt} &=k_{38}[\textbf{\textit{v}}_{2(P53_{n})}^{-}]+k_{39}Hif1\alpha_{n}-k_{40}Glut1_{c},\label{15}\\
\frac{dGlut3_{c}}{dt} &=k_{41}[\textbf{\textit{v}}_{2(P53_{n})}^{-}]+k_{42}Hif1\alpha_{n}-k_{43}Glut3_{c},\label{16}\\
\frac{dPdk13_{m}}{dt} &=k_{44}+k_{45}Hif1\alpha_{n}-k_{46}Pdk13_{m},\label{17}\\
\frac{dPdk2_{m}}{dt} &=k_{47}[\textbf{\textit{v}}_{2(P53_{n})}^{-}]-k_{46}Pdk2_{m},\label{18}\\
\frac{dLdh_{c}}{dt} &=k_{48}+k_{49}Hif1\alpha_{n}-k_{50}Ldh_{c},\label{19}\\
\frac{dPdh_{m}^*}{dt} &=k_{51}-k_{52}\big(Pdk13_{m}+Pdk2_{m}\big)[\textbf{\textit{v}}_{1(Pdh_{m}^*)}]+k_{53}[\textbf{\textit{v}}_{1(Pdh_{m})}]\notag\\&-k_{54}Pdh_{m}^*,\label{20}\\
\frac{dPdh_{m}}{dt} &=k_{52}\big(Pdk13_{m}+Pdk2_{m}\big)[\textbf{\textit{v}}_{1(Pdh_{m}^*)}]-k_{53}[\textbf{\textit{v}}_{1(Pdh_{m})}]-k_{54}Pdh_{m},\label{21}\\
\frac{dGlucose_{c}}{dt} &=k_{55}\big(Glut1_{c}+Glut3_{c}\big)[\textbf{\textit{v}}_{3(Glucose)}]-k_{56}[\textbf{\textit{v}}_{4(Glucose_{c},Atp)}],\label{22}\\
\frac{dG6p_{c}}{dt} &=k_{56}[\textbf{\textit{v}}_{4(Glucose_{c},Atp)}]-k_{57}[\textbf{\textit{v}}_{7(G6p_{c})}]-k_{58}G6p_{c},\label{23}\\
\frac{dPyruvate_{c}}{dt} &=2k_{57}[\textbf{\textit{v}}_{7(G6p_{c})}]-k_{59}[\textbf{\textit{v}}_{2(Ldh_{c})}^{+}][\textbf{\textit{v}}_{4(Pyruvate_{c},Nadh_{c})}]\notag\\
&-k_{60}[\textbf{\textit{v}}_{6(Pyruvate)}],\label{24}\\
\frac{dPyruvate_{m}}{dt} &=k_{60}[\textbf{\textit{v}}_{6(Pyruvate)}]-k_{61}Pdh_{m}^*[\textbf{\textit{v}}_{5(Pyruvate_{m},Nad_{m})}],\label{25}\\
\frac{dAcetyl_{m}}{dt} &=k_{61}Pdh_{m}^*[\textbf{\textit{v}}_{5(Pyruvate_{m},Nad_{m})}]-k_{62}[\textbf{\textit{v}}_{7(Acetyl_{m})}],\label{26}\\
\frac{dNadh_{c}}{dt} &=2k_{57}[\textbf{\textit{v}}_{7(G6p_{c})}]-k_{59}[\textbf{\textit{v}}_{2(Ldh_{c})}^{+}][\textbf{\textit{v}}_{4(Pyruvate_{c},Nadh_{c})}]\notag\\&-k_{63}[\textbf{\textit{v}}_{6(Nadh)}],\label{27}\\
\frac{dNadh_{m}}{dt} &=k_{61}Pdh_{m}^*[\textbf{\textit{v}}_{5(Pyruvate_{m},Nadh_{m})}]+3k_{62}[\textbf{\textit{v}}_{7(Acetyl_{m})}]+k_{63}[\textbf{\textit{v}}_{6(Nadh)}]\notag\\
&-k_{64}Sco2_{m}[\textbf{\textit{v}}_{7(Nadh_{m})}],\label{28}\\
\frac{dFadh_{m}}{dt} &=k_{62}[\textbf{\textit{v}}_{7(Acetyl_{m})}]-k_{64}Sco2_{m}[\textbf{\textit{v}}_{7(Fadh_{m})}],\label{29}\\
\frac{dLactate_{c}}{dt} &=k_{59}[\textbf{\textit{v}}_{2(Ldh_{c})}^{+}][\textbf{\textit{v}}_{4(Pyruvate_{c},Nadh_{c})}]-k_{65}[\textbf{\textit{v}}_{3(Lactate)}],\label{30}\\
\frac{dLactate_{out}}{dt} &=k_{65}[\textbf{\textit{v}}_{3(Lactate)}]-k_{66}Lactate_{out},\label{31}\\
\frac{dAtp}{dt} &=-k_{56}[\textbf{\textit{v}}_{4(Glucose_{c},Atp)}]+3k_{57}[\textbf{\textit{v}}_{7(G6p_{c})}]+k_{62}[\textbf{\textit{v}}_{7(Acetyl_{m})}]\notag\\
&+2.5k_{64}Sco2_{m}[\textbf{\textit{v}}_{7(Nadh_{m})}]+1.5k_{64}Sco2_{m}[\textbf{\textit{v}}_{7(Fadh_{m})}]-k_{67}Atp,\label{32}\\
\frac{dO2_{con}}{dt} &=0.5k_{64}Sco2_{m}[\textbf{\textit{v}}_{7(Nadh_{m})}]+0.5k_{64}Sco2_{m}[\textbf{\textit{v}}_{7(Fadh_{m})}]\label{33}.
\end{align}
This model applies to all cell types, both normal and cancerous, with certain restrictions. Under the assumption that normal cells exist in a healthy environment, metabolic stress and continuous activation of growth factor signals are exclusively attributed to cancer cells. Consequently, $k_{24}$ and $k_{27}$ are set to zero in normal cells. Additionally, to differentiate between cancer cell types, reducing $k_{2}$ to zero in p53-mutated cancer cells can ensure the absence of p53 response in these cells.\\

For a comprehensive understanding, detailed descriptions of the model—including assumptions, reactions, parameter values, experimental justifications, and sensitivity analysis—are provided in the Appendix (\ref{secA1}). Additionally, all numerical simulations in this study were performed using MATLAB's 'ode' routine and Gear's method in the XPPAUT software.
\section{Results}
\subsection{p53 orchestrates the metabolic shift in cancer: enhancing oxidative phosphorylation, suppressing glucose consumption and lactate production}
To examine the normal and cancer cellular metabolism and investigate the influence of p53 deficiency on cancer metabolic pathways seen in many laboratory experiments, we simulated the experiment conducted by Wanka et al. \citep{wanka2012synthesis}, using our mathematical model.\\

\begin{figure}
    \centering
    \includegraphics[width=1\textwidth]{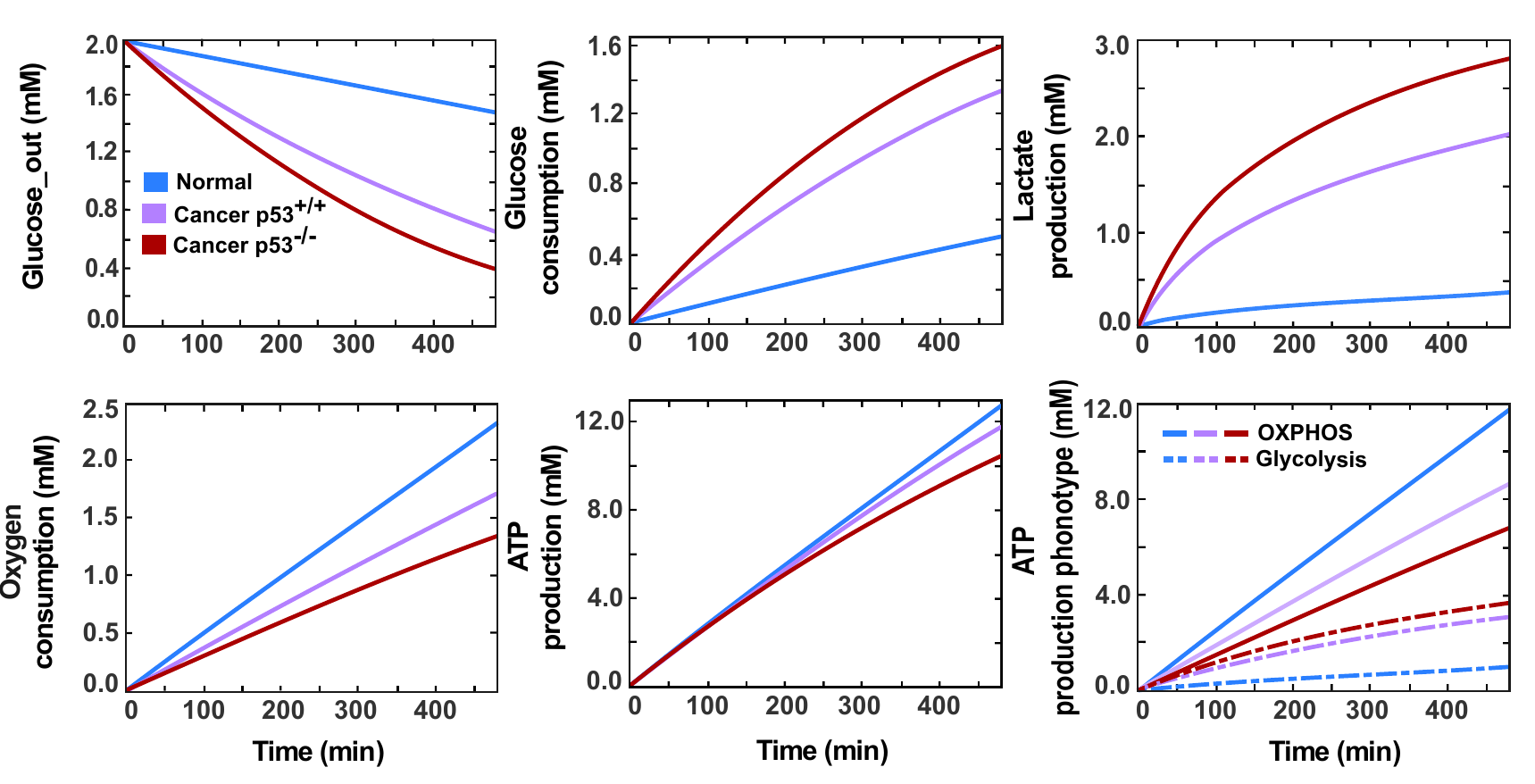}
    \caption{A comparison between normal cells and cancer cells (p53$^{+/+}$, p53$^{-/-}$) regarding their metabolic pathways. It shows the time course of the glucose consumption, lactate production, oxygen consumption, and ATP production by each cell type exposed to 2 mM of glucose over 8 hours}
    \label{Fig2}
\end{figure}

In-silico, different types of cells (normal, cancer p53$^{+/+}$, and cancer p53$^{-/-}$) were exposed to limited glucose (2 mM) for an 8-hour duration. Throughout this timeframe, the glucose consumed, lactate produced, oxygen consumed, and ATP produced were systematically monitored and quantified to provide a comprehensive comparison of metabolic processes across these distinct cell types, as illustrated in Fig. \ref{Fig2}.\\

Our simulations succeeded in clarifying the distinctions in glucose metabolic pathways between cancer and normal cells. Cancer cells exhibited heightened glucose consumption and elevated lactate secretion, signifying their commitment to the aerobic glycolysis phenotype. Conversely, in normal cells, glucose was mainly metabolised by oxidative phosphorylation, accounting for 92\% of the total ATP produced. Moreover, cancer cells in our model displayed high sensitivity to glucose availability, experiencing a notable decline in metabolic activity as glucose levels decreased. However, normal cells maintained relatively stable metabolic levels that were minimally affected by glucose fluctuations.\\

Our finding further confirms the significant influences of losing p53 in cancer metabolism, which caused a high tendency towards the glycolytic pathway reducing the oxygen consumption required for glucose oxidation by 22.5\%, compensating that by increasing glucose consumption and thus producing lactate at higher rates. A comparison of our simulation findings with experimental observations from various studies reveals a good match \citep{wanka2012synthesis,matoba2006p53,wu2016lactic}. Detailed insights are presented in Table \ref{Tab1}, corroborating the consistency between simulated and experimentally observed data.
\begin{table}
\centering
\resizebox{\textwidth}{!}
{
\begin{tabular}{lccccc}
\toprule
\multicolumn{6}{l}{\multirow{4}{*}{\Large\textbf{(a) Normoxic conditions (O$_2$ 21\%)}}}\\
&&&&&\\
&&&&&\\
&&&&&\\
\toprule
\multirow{2}{*}{}&\multicolumn{2}{@{}c@{}}{\multirow{2}{*}{\large\textbf{In experiment}}}&\multicolumn{2}{@{}c@{}}{\multirow{2}{*}{\large\textbf{In simulation}}}&\multirow{2}{*}{}\\
&&&&&\\
\cmidrule{2-3}\cmidrule{4-5}
\multirow{2}{*}{\large\textbf{Cell type}}&\multirow{2}{*}{\large\textbf{HCT116 p53$^{\text{+/+}}$}}&\multirow{2}{*}{\large\textbf{HCT116 p53$^{\text{-/-}}$}}&\multirow{2}{*}{\large\textbf{Cancer p53$^{\text{+/+}}$}}&\multirow{2}{*}{\large\textbf{Cancer p53$^{\text{-/-}}$}}&\multirow{2}{*}{}\\
&&&&&\\
\midrule
&&&&&\\
\large\textbf{Glucose consumption}&\large 1.35 mmol&\large 1.60 mmol&\large 1.34 mmol&\large 1.60 mmol&\large \citep{wanka2012synthesis}\\
&&&&&\\
\large\textbf{Lactate production}&\large 2.00 mmol&\large 2.88 mmol&\large 2.01 mmol&\large 2.82 mmol&\large \citep{wanka2012synthesis}\\
&&&&&\\
\multirow{2}{*}{\large\textbf{Oxygen consumption}}&\multicolumn{2}{c}{\multirow{2}{*}{\large 20 - 25\% less in HCT116 p53$^{\text{\large{-/-}}}$}}&\multicolumn{2}{c}{\multirow{2}{*}{\large 22.5\% less in Cancer p53$^{\text{\large{-/-}}}$}}&\large \citep{wanka2012synthesis},\\
&&&&&\large \citep{matoba2006p53}\\
&&&&&\\
\large\textbf{ATP production phenotype}&\multirow{2}{*}{\large 0.81 $\pm$ 0.12}&\multirow{2}{*}{\large 1.72 $\pm$ 0.16}&\multirow{2}{*}{\large 1.16}&\multirow{2}{*}{\large 2.10}&\multirow{2}{*}{\large \citep{matoba2006p53}}\\
\quad\quad\large (Lactate$_{\text{pro}}$/Oxygen$_{\text{con}}$)&&&&&\\
&&&&&\\
\cmidrule{2-3}\cmidrule{4-5}
\multirow{2}{*}{\large\textbf{Cell type}}&\multirow{2}{*}{\large\textbf{Normal}}&\multirow{2}{*}{\large\textbf{Cancer}}&\multirow{2}{*}{\large\textbf{Normal}}&\multirow{2}{*}{\large\textbf{Cancer}}&\multirow{2}{*}{}\\
&&&&&\\
\midrule
&&&&&\\
\large\textbf{ATP produced by glycolysis}&\large 6 - 13\%&\large 24 - 52\%&\large 8\%&\large 26 - 35\%&\large \citep{wu2016lactic}\\
&&&&&\\
\large\textbf{ATP produced by OXPHOS}&\large 87 - 94\%&\large 48 - 76\%&\large 92\%&\large 65 - 74\%&\large \citep{wu2016lactic}\\
&&&&&\\
\large\textbf{ATP production phenotype}&\multirow{2}{*}{\text{\Huge{-}}}&\multirow{2}{*}{\large 1.57 - 1.80}&\multirow{2}{*}{\text{\Huge{-}}}&\multirow{2}{*}{\large 1.50 - 1.76}&\multirow{2}{*}{\large \citep{wu2016lactic}}\\
\quad\quad\large (Lactate$_{\text{pro}}$/Glucose$_{\text{con}}$)&&&&&\\
&&&&&\\
\toprule
\multicolumn{6}{l}{\multirow{4}{*}{\Large\textbf{(b) Hypoxic conditions (O$_2$ 1\%)}}}\\
&&&&&\\
&&&&&\\
&&&&&\\
\toprule
\multirow{2}{*}{}&\multicolumn{2}{@{}c@{}}{\multirow{2}{*}{\large\textbf{In experiment}}}&\multicolumn{2}{@{}c@{}}{\multirow{2}{*}{\large\textbf{In simulation}}}&\multirow{2}{*}{}\\
&&&&&\\
\cmidrule{2-3}\cmidrule{4-5}
\multirow{2}{*}{\large\textbf{Cell type}}&\multirow{2}{*}{\large\textbf{HCT116 p53$^{\text{+/+}}$}}&\multirow{2}{*}{\large\textbf{HCT116 p53$^{\text{-/-}}$}}&\multirow{2}{*}{\large\textbf{Cancer p53$^{\text{+/+}}$}}&\multirow{2}{*}{\large\textbf{Cancer p53$^{\text{-/-}}$}}&\multirow{2}{*}{}\\
&&&&&\\
\midrule
&&&&&\\
\large\textbf{Glucose consumption}&\large 1.75 mmol&\large 2.00 mmol&\large 1.82 mmol&\large 1.93 mmol&\large \citep{wanka2012synthesis}\\
&&&&&\\
\large\textbf{Lactate production}&\large 2.90 mmol&\large 3.50 mmol&\large 3.86 mmol&\large 4.89 mmol&\large \citep{wanka2012synthesis}\\
&&&&&\\
\botrule
\end{tabular}
}
\caption{Comparing our simulation results with experimental observations under normoxia (a) and hypoxia (b) after 8h. By fitting the glucose uptake rate to that observed in p53-deficient cancer cells (HCT116 p53$^{-/-}$) under normoxic conditions, the detailed results were obtained}
\label{Tab1}
\end{table}
\subsection{The influence of abundant extracellular glucose level on stimulating high-energy production in cancer cells}
Considering the dynamic nature of cellular behaviour within the body, it is crucial to note that experiments may not comprehensively capture the full spectrum of cellular responses. In the experiment we reproduced, cells were subjected to a limited supply of glucose that depletes over time. However, this scenario contrasts with the relatively constant glucose level in the bloodstream that is readily accessible to cells within the body.\\
 
Accordingly, for a more realistic representation of cellular metabolism, we need to bridge the gap between the laboratory settings and the continuous physiological conditions experienced by cells within the body. In pursuit of this goal, we replicated previous simulations but this time assumed a consistent extracellular glucose level, maintaining it at the normal physiological glucose blood concentration (5mM) \citep{grupe1995transgenic}, regardless of the cellular consumption rate.\\

\begin{figure}
    \centering
    \includegraphics[width=1\textwidth]{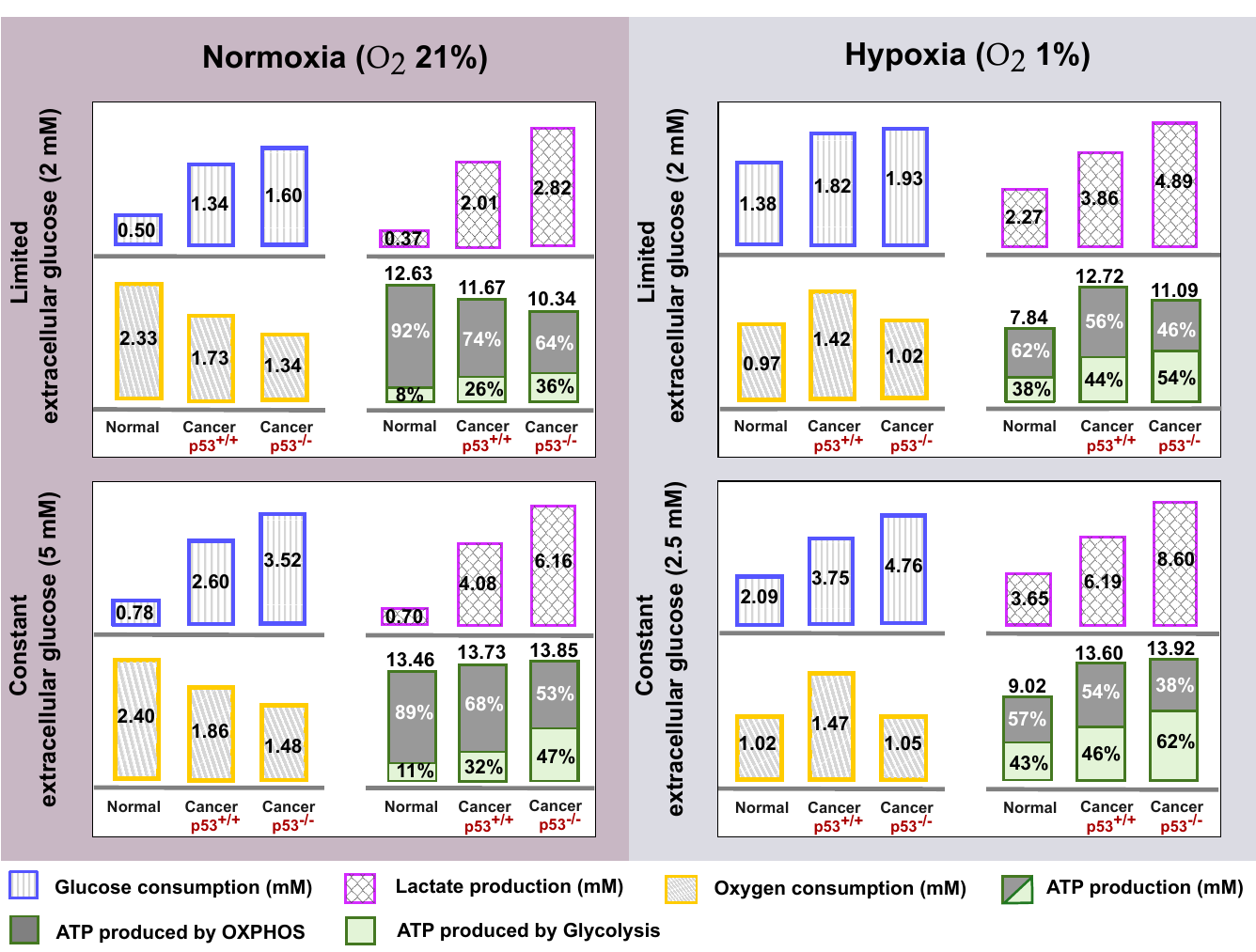}
    \caption{Glucose metabolism under normoxic and hypoxic conditions. A comparison of key outputs of metabolic pathways is shown for normal and cancer (p53$^{+/+}$, p53$^{-/-}$) cells, considering limited extracellular glucose level that depletes over time (Top row) and constant extracellular glucose levels (Bottom row). The glucose consumption, lactate production, oxygen consumption, and ATP production were measured by simulating each cell type for 8h under normoxic/hypoxic conditions}
    \label{Fig3}
\end{figure}

By adopting this methodology, our simulations demonstrated that the maintenance of stable glucose availability prompted both normal and cancer cells to exhibit a sustained rate of glucose consumption throughout the 8-hour duration, which revealed the distinctive ability of cancer cells to produce markedly higher levels of ATP compared to our previous simulations and even more than normal cells (Fig. \ref{Fig3}, Left).\\

This insight suggests that lowering the glucose levels in the bloodstream by following a specific regime could substantially diminish the ATP production in cancer cells, limiting their ability to sustain and spread. In addition, a comparative analysis of the three cell types under both limited and constant glucose levels highlights that the more the cell relies on the glycolytic pathway, the more it is affected by reducing glucose availability.
\subsection{Unravelling hypoxia's metabolism: adaptive strategies, energy production, and mTOR signalling dynamics in cancer progression}
In cancer progression, hypoxia emerges as a vital challenge faced by rapidly proliferating cancer cells due to the formation of regions within the tumour that are deprived of an adequate blood supply. In response, cancer cells exhibit remarkable adaptive strategies. They undergo complex molecular alterations, activating a cascade of signalling pathways that drive angiogenesis (the formation of new blood vessels) to restore oxygen balance and intensify the shift toward glycolytic energy production mode to offset the deficit in respiration \citep{xu2019hypoxia}. These dynamic responses are primarily governed by stabilising HIF1, a regulator suppressed under normal conditions in an oxygen availability-dependent manner \citep{valvona2016regulation,laughner2001her2,xu2019hypoxia}.\\

To investigate the impact of hypoxic conditions on cellular metabolism, we mimicked the hypoxic environment by diminishing the oxygen-dependent degradation rate of HIF-1, factoring in the HIF-1 half-life observed under hypoxic conditions \citep{kubaichuk2023usp10}. In parallel, we attenuated the activity of the electron transport chain by an equivalent rate (50\%), considering the inadequate availability of oxygen to facilitate the oxygen reduction process. Additionally, because hypoxia is often accompanied by a lack of blood supply to cancer cells, we reduced the glucose availability to cells within the body by the same percentage (from 5 mM to 2.5 mM).\\

By employing this approach, we anticipated and indeed observed a notable increase in lactate fermentation by both normal and cancer cells to maximise energy production as mitochondrial capacity diminishes, consequently escalating overall glucose consumption (Fig. \ref{Fig3}, right). This adaptive strategy mirrors the metabolic response seen in normal cells during intense exercise, where lower oxygen availability prompts alternative energy pathways. Furthermore, our simulation closely aligned the observed glucose metabolism outcomes for colon cancer cells HCT116 (p53$^{+/+}$ and p53$^{-/-}$) under hypoxic conditions (O$_{2}$ 1\%) \citep{wanka2012synthesis}, providing a good estimation of the glucose consumption levels with a slight increase in lactate production, as detailed in Table \ref{Tab1}, (b). The discrepancy in lactate production levels may be attributed to the potential conversion of some lactate back to pyruvate, especially in instances of extremely high lactate production not accounted for in our model.\\

On the other hand, our simulations revealed a prominent divergence in the response to hypoxia between normal and cancer cells regarding their energy production ability. While hypoxia led to a reduction in ATP production in normal cells, cancer cells displayed resilience, maintaining their energy productivity close to normal conditions (Fig. \ref{Fig3}, second row). This intriguing observation prompted a thorough investigation into possible factors that may be missed in our signalling network influencing energy production under hypoxic conditions. Our investigation unveiled that hypoxia typically induces the expression of the hypoxia-responsive REDD1 gene (not incorporated in our model), which, in turn, disrupts mTOR activity as a major control point to inhibit energy-intensive processes like protein translation \citep{brugarolas2004regulation,connolly2006hypoxia,deyoung2008hypoxia,horak2010negative}. This cascade leads to a decrease in HIF1 levels and a dampening of the glycolytic pathway \citep{brugarolas2004regulation,horak2010negative}.\\

\begin{figure}
    \centering
    \includegraphics[width=1\textwidth]{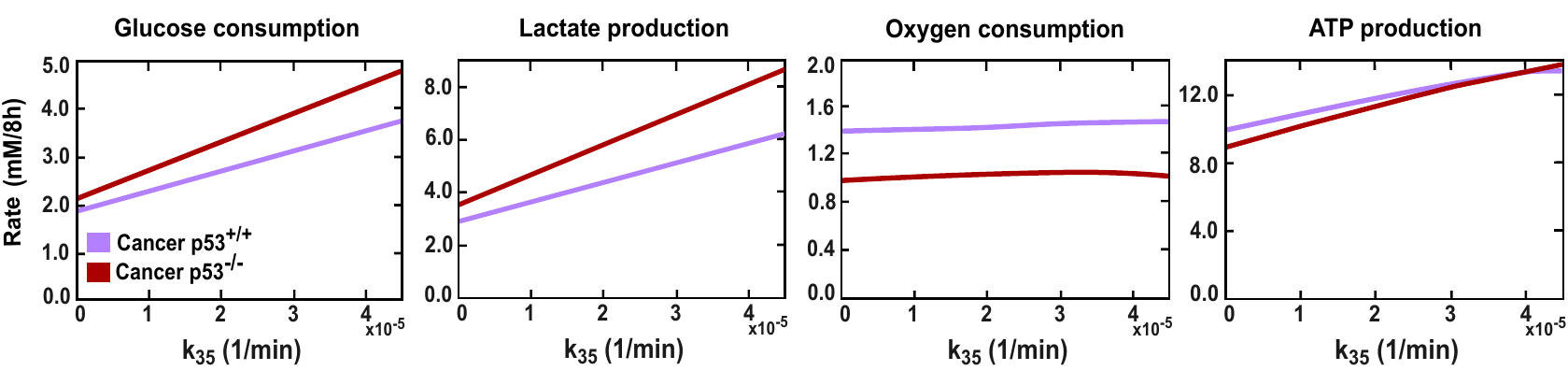}
    \caption{The effect of mTOR on the cellular metabolism and energy production levels in hypoxic cancer cells (p53$^{+/+}$, p53$^{-/-}$). The glucose consumption, lactate production, oxygen consumption, and ATP production were calculated under different mTOR-dependent HIF1 activation rates, $k_{35}$}
    \label{Fig4}
\end{figure}

Motivated by these findings, we studied the impact of inhibiting mTOR activity on metabolic pathways and energy production levels under hypoxic conditions. We constructed diagrams showcasing the metabolic activity of glycolysis and OXPHOS and their contributions to energy production under different $k_{35}$ rates (mTOR-dependent HIF1 synthesis rate), see Fig. \ref{Fig4}. Analysing these diagrams confirms the mTOR involvement in producing high energy levels in hypoxic cancer cells, as impeding its activity drove the cell towards a similar energy level produced in our hypoxic normal cells.\\

Nevertheless, numerous studies have consistently reported resistance of transformed cells to mTOR inhibition under hypoxic conditions \citep{connolly2006hypoxia}. This phenomenon is seen to preserve the protein synthesis rates and promote cell proliferation and growth under hypoxia \citep{brugarolas2004regulation,connolly2006hypoxia,deyoung2008hypoxia}. The engagement in energy-demanding processes, such as protein synthesis and growth, underscores the cell's proficiency in generating ample energy. This concurs with our findings regarding hypoxic cancer cells, where the maintenance of mTOR activity correlated with a remarkable ability to produce energy even in the face of oxygen deficiency.
\subsection{Dual stable steady states in cancer cells, contrasted by singular stability in normal cells}
\begin{figure}
    \centering
    \includegraphics[width=1\textwidth]{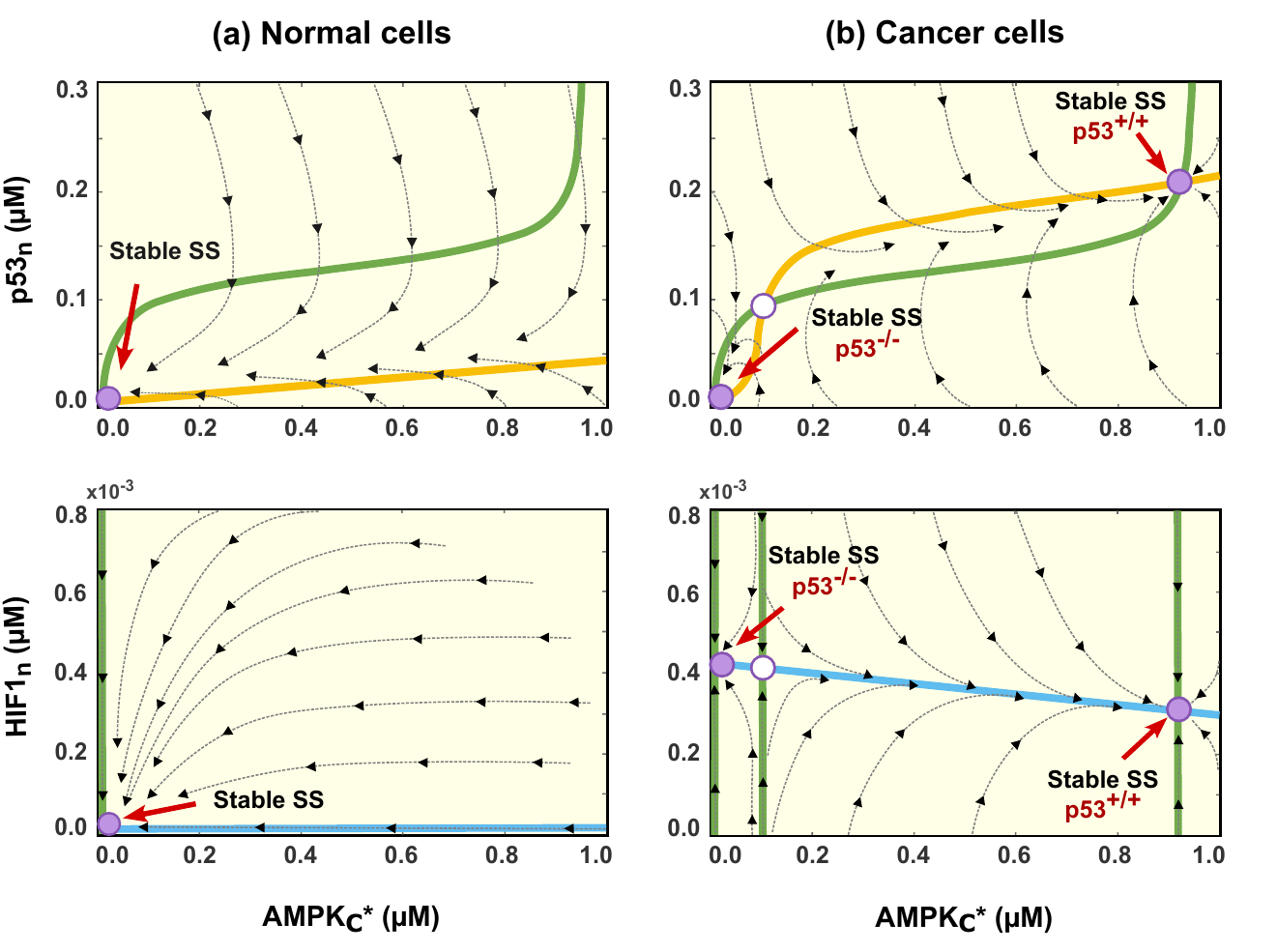}
    \caption{Phase portrait of the system in normal and cancer cells. (Top row) Nullcline corresponding to nuclear p53 (p53$_n$) and active AMPK (AMPK$_{c}^*$). (Bottom row) Nullcline corresponding to nuclear HIF1 (HIF1$_n$) and active AMPK (AMPK$_{c}^*$). The green, yellow, and blue lines represent AMPK$_{c}^*$, p53$_n$, and HIF1$_n$ nullclines, respectively. Solid and hollow magenta dots denote stable and unstable equilibria, respectively. The system exhibits a single stable equilibrium point in normal cells with no p53 and HIF1 activation, while in cancer cells, two stable and one unstable equilibria are observed.  For cancer cells, the stable equilibrium point with low  p53$_n$/high HIF1$_n$ levels represents p53-mutated cancer cells. In contrast, the one with high p53$_n$/low  HIF1$_n$ concentrations indicates p53-wild-type cancer cells}
    \label{Fig5}
\end{figure}

In previous sections, normal and cancerous cells, whether possessing wild-type p53 or mutated p53, manifest distinct metabolic profiles, signifying different stability states. To explore this further, we developed a phase space presenting the nullclines and potential steady states of key players in glucose metabolic pathways, p53, HIF1, and AMPK, across both normal and cancer cells (Fig. \ref{Fig5}).\\

Considering the phase space diagrams, normal cells show a unique stability with no activation of p53 and HIF1, indicative of a healthy environment. Conversely, cancer cells display two stable steady states, with an unstable one in between. The first stable steady state lacks p53 activation but exhibits a high level of HIF1, representing the case when cancer cells have p53 mutations. In contrast, the other stable steady state shows high p53 activation with a lower level of HIF1, indicating the state of cancer cells with wild-type p53.\\

\begin{figure}
    \centering
    \includegraphics[width=1\textwidth]{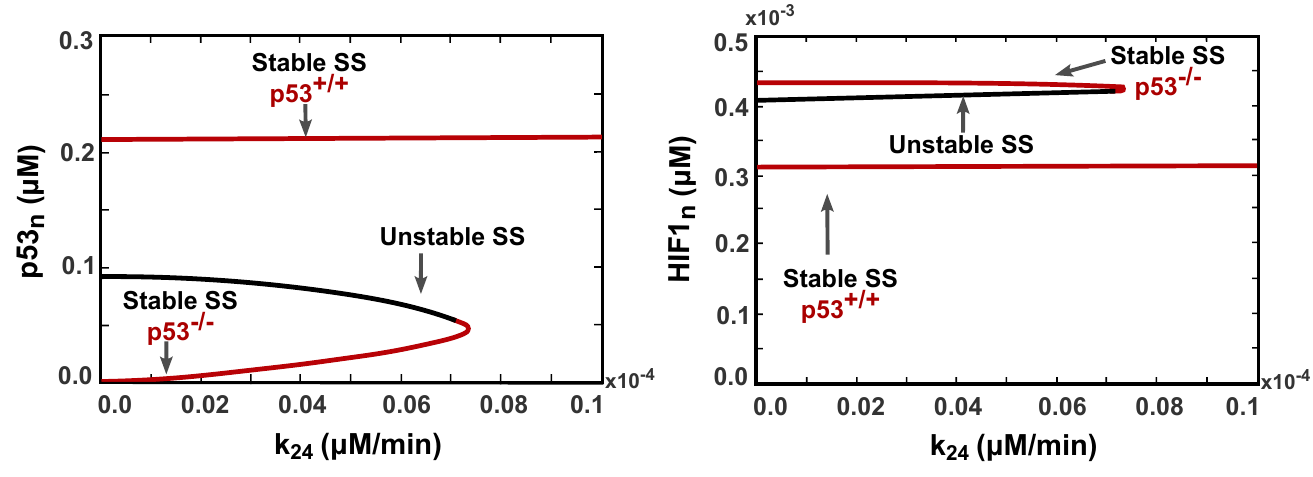}
    \caption{Bifurcation diagrams demonstrate nuclear p53 and HIF1 levels driven by AMPK phosphorylation rate, $k_{24}$, in cancer cells. The diagrams reveal a bistability regime exhibiting low p53/high HIF1 and high p53/low HIF1 levels, which represent the wild-type p53 (p53$^{+/+}$) and mutated p53 (p53$^{-/-}$) states, respectively. However, with high AMPK activation surpassing the bifurcation point, unique stability emerges, transitioning wild-type cancer cells to high p53 activation levels}
    \label{Fig6}
\end{figure}

The transition between these two states in p53 wild-type cells is governed by the phosphorylation levels of AMPK, the protein responsible for instigating the p53-metabolic stress response. This dynamic is further elucidated by the bifurcation diagram, illustrating the levels of p53 and HIF1 under various AMPK phosphorylation rates denoted as $k_{24}$ (Fig. \ref{Fig6}).\\

Under low phosphorylation rates of AMPK, cells exhibit two stable steady states: high activation of p53 (Stable SS p53$^{+/+}$)  and no activation of p53 (Stable SS p53$^{-/-}$). However, exceeding the bifurcation point by increasing the phosphorylation rate ($k_{24}$) induces p53 activation and shifts the cell into a unique stability regime, representing the p53-wild-type state.
\subsection{Restoring normal metabolism in cancer cells by increasing the p53 activation levels}
\begin{figure}
    \centering
    \includegraphics[width=1\textwidth]{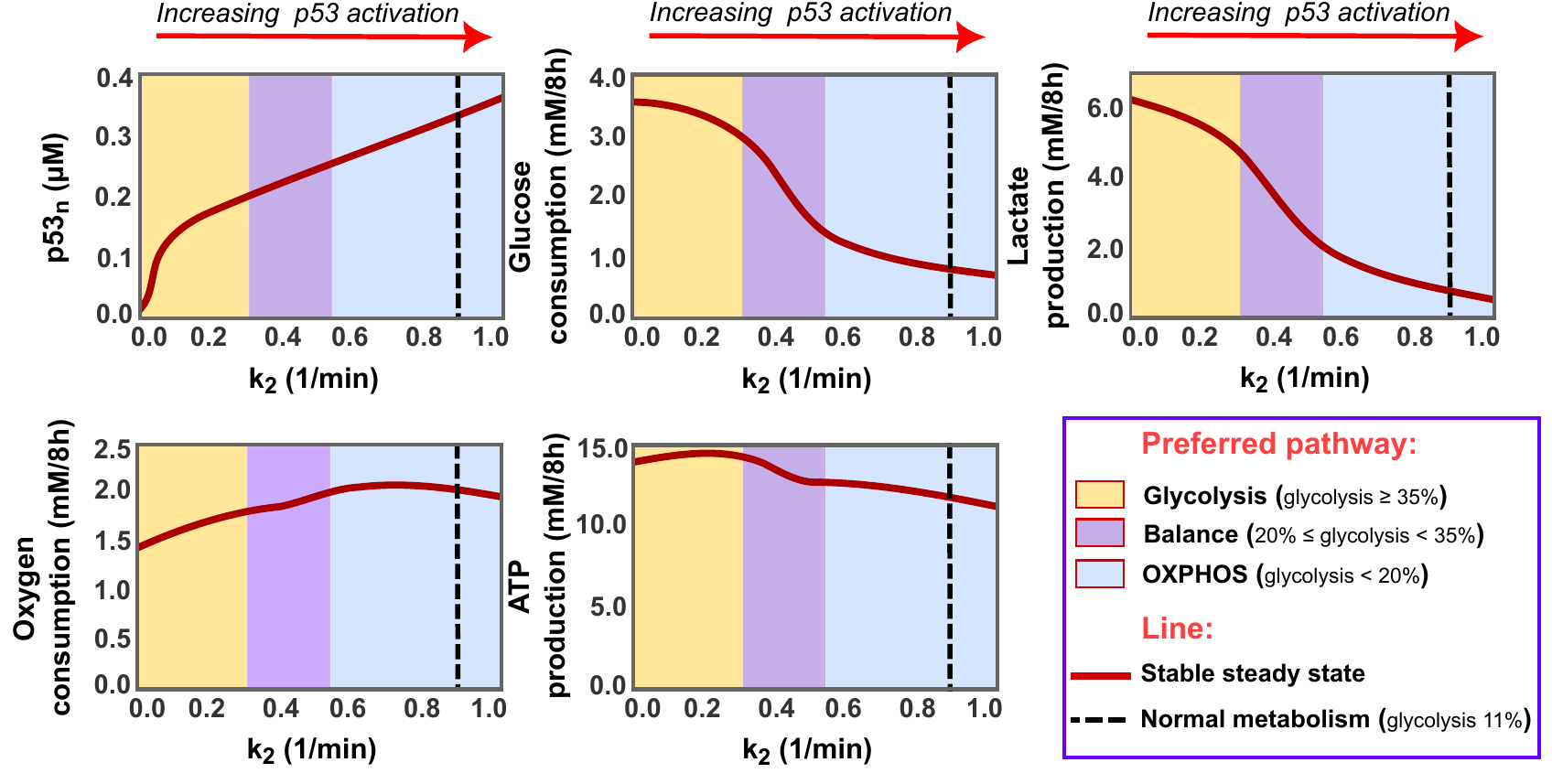}
    \caption{The effect of p53 activation on cancer metabolism. These diagrams show the steady state levels of nuclear p53 and four key metabolic indicators: glucose consumption, lactate production, oxygen consumption, and ATP generation, under varying rates of p53 phosphorylation ($k_{2}$) in cancer cells. Each diagram is divided into three distinct regions: the yellow region, where glycolysis dominates, contributing to 47\%-35\% of ATP production; the magenta region, indicating a balanced state between glycolysis and oxidative phosphorylation, with glycolysis contributing to 20-34\% of total ATP; and the blue region, where OXPHOS becomes dominant, accounting for more than 80\% of ATP production. A black dashed line within the diagrams marks the targeted normal cellular metabolism}
    \label{Fig7}
\end{figure}

Beyond its traditional roles in DNA repair and apoptosis initiation, our study highlights the enhanced activation potential of p53 to counter the Warburg effect, restoring cancer cells to a more normal metabolic state. This transformative impact unfolds across three distinct phases, depicted in Figure \ref{Fig7}.\\

During the first phase (yellow area, Fig. \ref{Fig7}), the elevation of nuclear p53 levels leads to a modest reduction in glucose consumption and lactate production. Nevertheless, energy production levels remain high due to improved glucose respiration, explaining the increase in oxygen consumption despite the lower amount of glucose consumed. Glycolysis maintains dominance in this phase, contributing to 47\%-35\% of the overall energy produced.\\

Advancing the p53 activation will shift the cells towards the next phase (magenta area, Fig. \ref{Fig7}), further reducing glucose uptake and lactate formation. However, this time, the ATP production is negatively impacted as a balanced state between glycolysis and oxidative phosphorylation is achieved, with glycolysis responsible for 20-34\% of ATP output.\\

In the third phase (blue area, Fig. \ref{Fig7}), oxidative phosphorylation overcomes glycolysis, intensifying oxygen consumption while consistently diminishing glucose utilisation and lactate production. This transition guides the cell towards achieving the standards of normal cellular metabolism, represented by the dashed black line around $k_{2}=0.9$. Along this line, glycolysis and OXPHOS are involved in producing energy with the same percentage seen in our normal cells, attaining standard rates of glucose consumption and lactate production. However, with high activation of TIGAR, glycolysis flux is lower than that of normal cells, resulting in reduced pyruvate production and overall ATP synthesis compared to the normal cellular state.\\

This finding sheds light on the crucial role of p53 activation levels in the cellular outcomes following chemotherapy that activates p53 to trigger apoptosis. Unlike cells with p53 mutation, those with intact p53 can manipulate cancer cells' metabolism in response to chemotherapy, restraining the glycolytic pathway and decreasing the intracellular ATP levels, thereby boosting cells' sensitivity to drugs.
\subsection{Targeting PI3K as an alternative player to p53 in modulating the metabolism of p53-mutated cancer cells}
\begin{figure}
    \centering
    \includegraphics[width=1\textwidth]{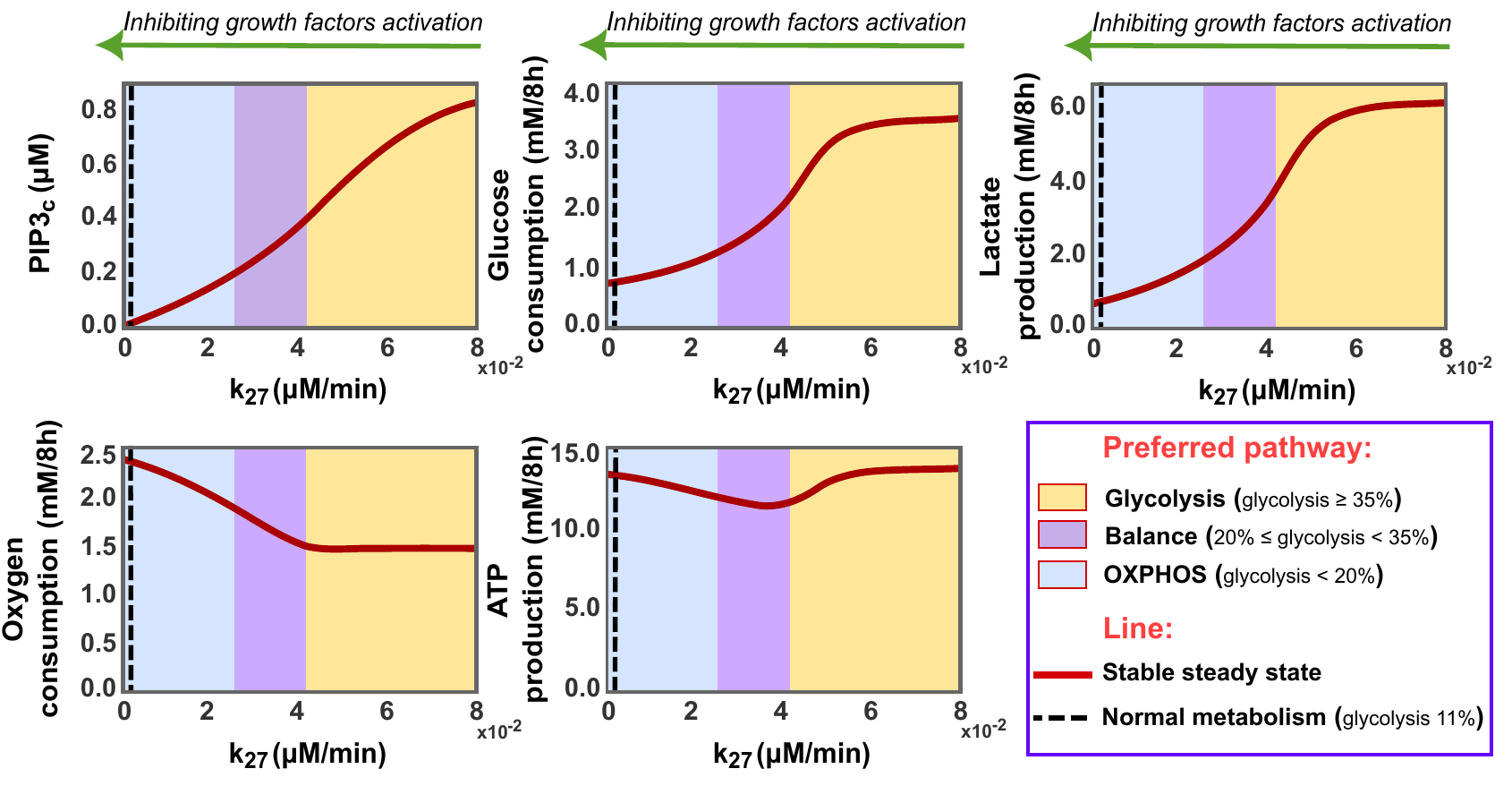}
    \caption{The impact of disrupting the growth factors signalling pathway on p53-mutated cancer cells metabolism. These diagrams depict the steady state levels of PIP3 alongside key metabolic metrics—glucose consumption, lactate production, oxygen consumption, and ATP production—at different PIP2 phosphorylation rates ($k_{27}$) in cancer cells (p53$^{-/-}$). The diagrams are categorised into three zones: yellow for glycolytic predominance (accounting for over 35\% of ATP production), magenta for a metabolic balance between glycolysis and oxidative phosphorylation (20-34\% of ATP from glycolysis), and blue for oxidative phosphorylation supremacy (exceeding 80\% of energy output). A black dashed line marks the standard for normal cellular metabolism, with energy contributions of 11\% from glycolysis and 89\% from OXPHOS}
    \label{Fig8}
\end{figure}

In the context of addressing cancer metabolism in cells harbouring p53 mutations, our analysis suggests an alternative strategy by targeting the growth factors signalling pathway. This critical pathway plays a central role in instigating the HIF1 and its associated targets that mainly support the aerobic glycolysis of cancer \citep{lien2016metabolic}. The initiation of this pathway involves the activation of PI3K, leading to the transformation of PIP2 into PIP3 \citep{danielsen2015portrait,vara2004pi3k,lien2016metabolic}. Thus, our investigation has focused on perturbing this pathway by simulating methodologies such as triggering PTEN or blocking PI3K activation with specific inhibitors, like idelalisib or copanlisib  \citep{lannutti2011cal,liu2013bay}. The outcomes reveal a profound and systematic influence on cellular metabolism, manifesting across three phases (Fig. \ref{Fig8}).\\

In the initial phase (yellow area, Fig. \ref{Fig8}), the emphasis is placed on inhibiting the glycolysis pathway while leaving the oxidative phosphorylation unaffected, leading to a decrease in total energy production following a loss of more than 20\% in PIP3 concentration. Despite the reduction in glycolysis activity, it is considered predominant in this phase, accounting for over 35\% of cellular energy.\\

As PIP3 levels decrease, the glycolytic pathway continues to diminish, indirectly prompting the OXPHOS pathway to regain its functionality and bringing the two pathways into a balanced state (magenta area, Fig. \ref{Fig8}). Restricting pyruvate flux to lactate is expected to elevate cytosolic pyruvate concentrations, redirecting them towards mitochondria and thus promoting pyruvate oxidation. This shift is reflected in the notable oxygen consumption boost and sustained ATP production levels despite lower glucose utilisation in this phase.\\

In the last phase (blue area, Fig. \ref{Fig8}), glycolysis experiences a significant decline, allowing OXPHOS to overcome it, thus improving energy production efficiency. In this stage, glucose respiration becomes the preferred cellular pathway responsible for 80-90\% of the total energy output. The black dashed line in this phase signifies the targeted normal cellular metabolism, with 11\% of energy production attributed to glycolysis and 89\% to OXPHOS.\\

Our simulations reveal the efficacy of targeting the growth factors signalling pathway and highlight the potency of PI3K inhibitors in disrupting the aerobic glycolysis in p53-mutated cancer cells, enhancing therapeutic outcomes.
\subsection{SCO2: a critical component in boosting the OXPHOS, yet alone insufficient for reversing the Warburg effect}
Numerous studies have emphasised the crucial role of SCO2, a p53 target, in the efficient functioning of the mitochondrial respiratory chain and cellular energy production \citep{matoba2006p53,wanka2012synthesis}. SCO2 is essential for the proper assembly and function of cytochrome c oxidase (Complex IV in ETC), which catalyses electron transfer to molecular oxygen in the inner mitochondrial membrane \citep{matoba2006p53,wanka2012synthesis}. Given its significance in cellular respiration, some studies have proposed targeting it as a potential strategy to rescue oxygen consumption in p53-deficient cells and modulate the Warburg effect \citep{matoba2006p53,wanka2012synthesis}.\\

\begin{figure}
    \centering
    \includegraphics[width=1\textwidth]{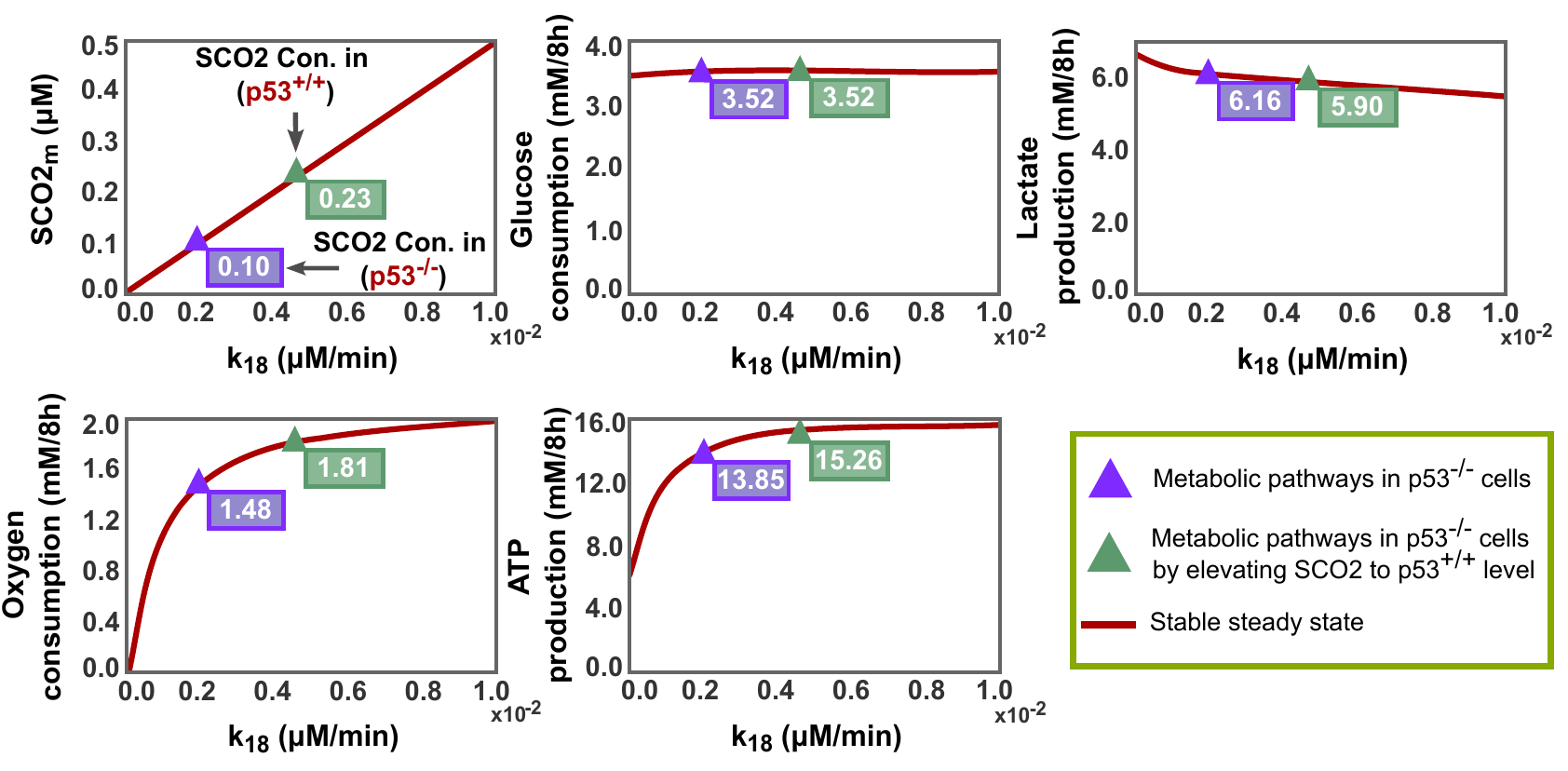}
    \caption{The role of SCO2 in the metabolism of p53-deficient cancer cells. The diagrams represent steady state levels of SCO2 alongside key metabolic metrics—glucose consumption, lactate production, oxygen consumption, and ATP production—influenced by various SCO2 basal production rates ($k_{18}$) in p53$^{-/-}$ cancer cells. Magenta triangles denote the baseline scenario of p53$^{-/-}$ cancer cells, whereas green triangles signify the altered state after increasing SCO2 concentration to match levels observed in p53$^{+/+}$ cells}
    \label{Fig9}
\end{figure}

Inspired by these insights, we delved into the impact of boosting SCO2 levels on the metabolic phenotypes of cancer cells, particularly those with p53 mutations. In-silico, we elevated the SCO2 expression levels of p53$^{-/-}$ cells by increasing its basal production rate, $k_{18}$ (Fig. \ref{Fig9}). \\

Indeed, our simulations agreed with those studies' observations \citep{matoba2006p53,wanka2012synthesis}, revealing a substantial activation of aerobic respiration in a SCO2 level-dependent manner. Additionally, we noticed that when SCO2 concentration achieves its level in wild-type p53 cells, the oxygen consumption activity of p53$^{-/-}$ cells rises at a rate comparable to that in wild-type p53 cells. This is completely consistent with what was observed in Matoba’s study, which noted that the amount of SCO2 protein needed to rescue the deficit in mitochondrial respiration of the p53$^{-/-}$ cells corresponded well to the physiological levels observed in the p53$^{+/+}$ cells \citep{matoba2006p53}.\\

On the other hand, our findings also indicate that increasing SCO2 alone is insufficient to eliminate or reverse the Warburg effect. Enhancing oxidative phosphorylation does not necessarily lead to efficient suppression of the glycolysis pathway, especially with continued incentives to consume large amounts of glucose and high activation of glycolysis enzymes. This clearly explains our results, which show a slight decline in glycolysis despite a striking increase in the oxidative phosphorylation pathway (Fig. \ref{9}). Consequently, solely targeting SCO2 may elevate energy production levels, as shown in our simulations, potentially promoting the proliferative capacity of cancer cells.\\

In brief, our results demonstrate that SCO2 may indeed play a robust role in transforming cancer cell metabolism, but in conjunction with targeting enzymes stimulating the glycolysis pathway.
\section{Discussion}
The Warburg effect is a hallmark of cancer metabolism, granting cancer cells exceptional metabolic flexibility that enables their rapid adaptation and survival in hostile microenvironments. This phenomenon is pivotal in cancer research, with particular interest in the regulatory mechanisms that govern metabolic pathways. At the forefront of these is the tumour suppressor gene p53, whose role extends beyond cell cycle control and apoptosis to include metabolic processes. Our investigation delves into the critical role of p53 in modulating cancer cell metabolism, offering novel insights into its capacity to counteract the Warburg effect phenomenon.\\

In terms of existing research, Linglin et al. have made good strides in elucidating the influence of genetic regulation on glycolysis and oxidative phosphorylation, identifying a hybrid metabolic state unique to cancer cells \citep{yu2017modeling}. However, this study did not consider the vital influence of p53, nor did it conduct a quantitative analysis of how genetic factors impact metabolic outcomes or potential strategies to mitigate the Warburg effect. Our research bridges these gaps by constructing a comprehensive mathematical framework that dissects the mechanisms through which p53, alongside other genetic regulations, influences glycolysis and OXPHOS and quantitatively explores their impact on these pathways under various cellular conditions.\\

Our model analysis reveals distinct metabolic profiles characterised by different stability regimes, delineating clear metabolic distinctions between normal and cancer cells with or without p53 mutations. Importantly, our model successfully replicated experimental observations on glucose metabolism in both p53-mutated and wild-type colon cancer cells, underscoring its validity.\\

By exploring various scenarios, our study uncovers the mechanism of how diminished glucose availability massively curtails cancer cell proliferation and viability. We further identify adaptive tactics cancer cells employ under low-oxygen conditions to maintain energy production and growth, particularly emphasising the crucial role of mTOR activation. This adaptation starkly contrasts with the energy production downturn observed in normal cells under similar hypoxic conditions, highlighting the unique metabolic resilience of cancer cells.\\

Interestingly, we detect a novel aspect of chemotherapy resistance linked to insufficient p53 activation levels, suggesting that beyond apoptosis evasion, inadequate p53 activity also impedes the reversal of the Warburg effect, enhancing cellular resistance.\\

Moreover, this study discusses strategies to combat the Warburg effect in p53-mutated contexts, evaluating the efficacy of augmenting cellular respiration by increasing the SCO2 expression levels. While this approach indeed elevates mitochondrial respiration, it does so without a noticeable reduction in the glycolysis pathway, thereby boosting the overall ATP production and potentially supporting cancer cells even further. Alternatively, we suggest inhibiting the glycolysis pathway using a PI3K inhibitor, which has shown promising results in our simulations.\\

While our model has shown considerable success and offered valuable insights, it is important to acknowledge its limitations. Our model does not incorporate the competitive dynamics between p53 and HIF1 over transcriptional coactivators. Transcription factors like p53 and HIF1 depend on coactivators such as p300/CBP for gene regulation, which involves acetylating histones at specific gene promoters to facilitate the recruitment of the transcriptional machinery \citep{grossman2001p300,freedman2002structural}. Given the finite availability of these coactivators, competition for access between p53 and HIF1 emerges, affecting their transcriptional activities \citep{schmid2004p300}.\\

Furthermore, our current work concentrated exclusively on glucose metabolism, yet cells can utilise additional energy sources, such as glutamine and fatty acids. Integrating these energy sources and the p53 influence on their respective metabolic pathways might give a more comprehensive overview of the metabolism outcomes and analyse the p53 role much deeper. Future research aims to expand our signalling network to include these pathways, providing a more holistic view of p53 impact on cancer metabolism.\\

In conclusion, this study broadens our understanding of the Warburg effect through the lens of p53 regulatory mechanisms, introducing, for the first time, a mathematical model that captures the observed impact of p53 deficiency on cancer metabolism. This pioneering model unravels the metabolic underpinnings of cancer, thoroughly scrutinising glucose metabolic pathways across different scenarios. Additionally, model findings propose fresh perspectives to improve therapeutic approaches, significantly highlighting the importance of optimal p53 activation for reversing the Warburg effect and the efficacy of PI3K inhibitors in overcoming metabolic adaptations in p53-mutated cancer cells.\\
\bmhead{Acknowledgements}
The authors would like to thank the Saudi Arabian Cultural Bureau (SACB) for funding \textbf{R.A.}, Cancer Research UK for supporting \textbf{D.T.} and \textbf{E.V.-S.} (C42109/A26982 and C42109/A24747), and the UKRI Future Leaders Fellowship (MR/T043571/1) for backing \textbf{F.S.}. This work has also benefitted from the support and resources provided by the Birmingham Metabolic Tracer Analysis Core (MTAC), alongside financial contributions from the University of Birmingham Dynamic Investment Fund and the EPSRC via grant no. (EP/N014391/2).\\
\bmhead{Author Contributions}
\textbf{R.A.} conceived the project and the mathematical model, performed in silico experimentation and analysis, and wrote the manuscript. \textbf{E.V.-S.} provided technical advice concerning the in-silico modelling and critically revised the manuscript. \textbf{D.T.} offered technical advice on the biology discussed in this article and the validity of the model’s results. \textbf{F.S.} provided technical advice concerning the in-silico modelling and analysis, organised, supervised and managed the study.\\
\bmhead{Data Availability}
The codes used in this article are publicly available on the Gitlab webpage \url{https://gitlab.bham.ac.uk/spillf-systems-mechanobiology-health-disease/p53-and-metabolism}.
\section*{Declarations}
\bmhead{Conflict of interest} 
The authors declare no competing interests influence the work reported in this article.\\
\bmhead{Use of AI} 
We used Grammarly and ChatGPT to refine and enhance the clarity of the manuscript's text.
\bibliography{References}


\begin{thebibliography}{140}
\ifx \bisbn   \undefined \def \bisbn  #1{ISBN #1}\fi
\ifx \binits  \undefined \def \binits#1{#1}\fi
\ifx \bauthor  \undefined \def \bauthor#1{#1}\fi
\ifx \batitle  \undefined \def \batitle#1{#1}\fi
\ifx \bjtitle  \undefined \def \bjtitle#1{#1}\fi
\ifx \bvolume  \undefined \def \bvolume#1{\textbf{#1}}\fi
\ifx \byear  \undefined \def \byear#1{#1}\fi
\ifx \bissue  \undefined \def \bissue#1{#1}\fi
\ifx \bfpage  \undefined \def \bfpage#1{#1}\fi
\ifx \blpage  \undefined \def \blpage #1{#1}\fi
\ifx \burl  \undefined \def \burl#1{\textsf{#1}}\fi
\ifx \doiurl  \undefined \def \doiurl#1{\url{https://doi.org/#1}}\fi
\ifx \betal  \undefined \def \betal{\textit{et al.}}\fi
\ifx \binstitute  \undefined \def \binstitute#1{#1}\fi
\ifx \binstitutionaled  \undefined \def \binstitutionaled#1{#1}\fi
\ifx \bctitle  \undefined \def \bctitle#1{#1}\fi
\ifx \beditor  \undefined \def \beditor#1{#1}\fi
\ifx \bpublisher  \undefined \def \bpublisher#1{#1}\fi
\ifx \bbtitle  \undefined \def \bbtitle#1{#1}\fi
\ifx \bedition  \undefined \def \bedition#1{#1}\fi
\ifx \bseriesno  \undefined \def \bseriesno#1{#1}\fi
\ifx \blocation  \undefined \def \blocation#1{#1}\fi
\ifx \bsertitle  \undefined \def \bsertitle#1{#1}\fi
\ifx \bsnm \undefined \def \bsnm#1{#1}\fi
\ifx \bsuffix \undefined \def \bsuffix#1{#1}\fi
\ifx \bparticle \undefined \def \bparticle#1{#1}\fi
\ifx \barticle \undefined \def \barticle#1{#1}\fi
\bibcommenthead
\ifx \bconfdate \undefined \def \bconfdate #1{#1}\fi
\ifx \botherref \undefined \def \botherref #1{#1}\fi
\ifx \url \undefined \def \url#1{\textsf{#1}}\fi
\ifx \bchapter \undefined \def \bchapter#1{#1}\fi
\ifx \bbook \undefined \def \bbook#1{#1}\fi
\ifx \bcomment \undefined \def \bcomment#1{#1}\fi
\ifx \oauthor \undefined \def \oauthor#1{#1}\fi
\ifx \citeauthoryear \undefined \def \citeauthoryear#1{#1}\fi
\ifx \endbibitem  \undefined \def \endbibitem {}\fi
\ifx \bconflocation  \undefined \def \bconflocation#1{#1}\fi
\ifx \arxivurl  \undefined \def \arxivurl#1{\textsf{#1}}\fi
\csname PreBibitemsHook\endcsname

\bibitem[\protect\citeauthoryear{Ancey et~al.}{2018}]{ancey2018glucose}
\begin{barticle}
\bauthor{\bsnm{Ancey}, \binits{P.-B.}},
\bauthor{\bsnm{Contat}, \binits{C.}},
\bauthor{\bsnm{Meylan}, \binits{E.}}:
\batitle{Glucose transporters in cancer--from tumor cells to the tumor microenvironment}.
\bjtitle{The FEBS journal}
\bvolume{285}(\bissue{16}),
\bfpage{2926}--\blpage{2943}
(\byear{2018})
\end{barticle}
\endbibitem

\bibitem[\protect\citeauthoryear{Anad{\'o}n et~al.}{2014}]{anadon2014biomarkers}
\begin{botherref}
\oauthor{\bsnm{Anad{\'o}n}, \binits{A.}},
\oauthor{\bsnm{Castellano}, \binits{V.}},
\oauthor{\bsnm{Mart{\'\i}nez-Larra{\~n}aga}, \binits{M.R.}}:
Biomarkers in drug safety evaluation. in: Biomarkers in toxicolog.
Elsevier,
923--945
(2014)
\end{botherref}
\endbibitem

\bibitem[\protect\citeauthoryear{Al-Khayal et~al.}{2016}]{al2016identification}
\begin{barticle}
\bauthor{\bsnm{Al-Khayal}, \binits{K.}},
\bauthor{\bsnm{Abdulla}, \binits{M.}},
\bauthor{\bsnm{Al-Obeed}, \binits{O.}},
\bauthor{\bsnm{Al~Kattan}, \binits{W.}},
\bauthor{\bsnm{Zubaidi}, \binits{A.}},
\bauthor{\bsnm{Vaali-Mohammed}, \binits{M.-A.}},
\bauthor{\bsnm{Alsheikh}, \binits{A.}},
\bauthor{\bsnm{Ahmad}, \binits{R.}}:
\batitle{Identification of the tp53-induced glycolysis and apoptosis regulator in various stages of colorectal cancer patients}.
\bjtitle{Oncology Reports}
\bvolume{35}(\bissue{3}),
\bfpage{1281}--\blpage{1286}
(\byear{2016})
\end{barticle}
\endbibitem

\bibitem[\protect\citeauthoryear{Anwar et~al.}{2021}]{anwar2021targeting}
\begin{barticle}
\bauthor{\bsnm{Anwar}, \binits{S.}},
\bauthor{\bsnm{Shamsi}, \binits{A.}},
\bauthor{\bsnm{Mohammad}, \binits{T.}},
\bauthor{\bsnm{Islam}, \binits{A.}},
\bauthor{\bsnm{Hassan}, \binits{M.I.}}:
\batitle{Targeting pyruvate dehydrogenase kinase signaling in the development of effective cancer therapy}.
\bjtitle{Biochimica et Biophysica Acta (BBA)-Reviews on Cancer}
\bvolume{1876}(\bissue{1}),
\bfpage{188568}
(\byear{2021})
\end{barticle}
\endbibitem

\bibitem[\protect\citeauthoryear{Ataullakhanov and Vitvitsky}{2002}]{ataullakhanov2002determines}
\begin{barticle}
\bauthor{\bsnm{Ataullakhanov}, \binits{F.I.}},
\bauthor{\bsnm{Vitvitsky}, \binits{V.M.}}:
\batitle{What determines the intracellular atp concentration}.
\bjtitle{Bioscience reports}
\bvolume{22}(\bissue{5-6}),
\bfpage{501}--\blpage{511}
(\byear{2002})
\end{barticle}
\endbibitem

\bibitem[\protect\citeauthoryear{Abukwaik et~al.}{2023}]{abukwaik2023interplay}
\begin{botherref}
\oauthor{\bsnm{Abukwaik}, \binits{R.}},
\oauthor{\bsnm{Vera-Siguenza}, \binits{E.}},
\oauthor{\bsnm{Tennant}, \binits{D.A.}},
\oauthor{\bsnm{Spill}, \binits{F.}}:
Interplay of p53 and xiap protein dynamics orchestrates cell fate in response to chemotherapy.
Journal of Theoretical Biology,
111562
(2023)
\end{botherref}
\endbibitem

\bibitem[\protect\citeauthoryear{Ahmad et~al.}{2018}]{ahmad2018biochemistry}
\begin{botherref}
\oauthor{\bsnm{Ahmad}, \binits{M.}},
\oauthor{\bsnm{Wolberg}, \binits{A.}},
\oauthor{\bsnm{Kahwaji}, \binits{C.I.}}:
Biochemistry, electron transport chain
(2018)
\end{botherref}
\endbibitem

\bibitem[\protect\citeauthoryear{Bhagavan}{2002}]{bhagavan2002medical}
\begin{botherref}
\oauthor{\bsnm{Bhagavan}, \binits{N.V.}}:
Medical biochemistry.
Academic press
(2002)
\end{botherref}
\endbibitem

\bibitem[\protect\citeauthoryear{Barak et~al.}{1993}]{barak1993mdm2}
\begin{barticle}
\bauthor{\bsnm{Barak}, \binits{Y.}},
\bauthor{\bsnm{Juven}, \binits{T.}},
\bauthor{\bsnm{Haffner}, \binits{R.}},
\bauthor{\bsnm{Oren}, \binits{M.}}:
\batitle{mdm2 expression is induced by wild type p53 activity.}
\bjtitle{The EMBO journal}
\bvolume{12}(\bissue{2}),
\bfpage{461}--\blpage{468}
(\byear{1993})
\end{barticle}
\endbibitem

\bibitem[\protect\citeauthoryear{Budanov and Karin}{2008}]{budanov2008p53}
\begin{barticle}
\bauthor{\bsnm{Budanov}, \binits{A.V.}},
\bauthor{\bsnm{Karin}, \binits{M.}}:
\batitle{p53 target genes sestrin1 and sestrin2 connect genotoxic stress and mtor signaling}.
\bjtitle{Cell}
\bvolume{134}(\bissue{3}),
\bfpage{451}--\blpage{460}
(\byear{2008})
\end{barticle}
\endbibitem

\bibitem[\protect\citeauthoryear{Brugarolas et~al.}{2004}]{brugarolas2004regulation}
\begin{barticle}
\bauthor{\bsnm{Brugarolas}, \binits{J.}},
\bauthor{\bsnm{Lei}, \binits{K.}},
\bauthor{\bsnm{Hurley}, \binits{R.L.}},
\bauthor{\bsnm{Manning}, \binits{B.D.}},
\bauthor{\bsnm{Reiling}, \binits{J.H.}},
\bauthor{\bsnm{Hafen}, \binits{E.}},
\bauthor{\bsnm{Witters}, \binits{L.A.}},
\bauthor{\bsnm{Ellisen}, \binits{L.W.}},
\bauthor{\bsnm{Kaelin}, \binits{W.G.}}:
\batitle{Regulation of mtor function in response to hypoxia by redd1 and the tsc1/tsc2 tumor suppressor complex}.
\bjtitle{Genes \& development}
\bvolume{18}(\bissue{23}),
\bfpage{2893}--\blpage{2904}
(\byear{2004})
\end{barticle}
\endbibitem

\bibitem[\protect\citeauthoryear{Batchelor et~al.}{2011}]{batchelor2011stimulus}
\begin{barticle}
\bauthor{\bsnm{Batchelor}, \binits{E.}},
\bauthor{\bsnm{Loewer}, \binits{A.}},
\bauthor{\bsnm{Mock}, \binits{C.}},
\bauthor{\bsnm{Lahav}, \binits{G.}}:
\batitle{Stimulus-dependent dynamics of p53 in single cells}.
\bjtitle{Molecular systems biology}
\bvolume{7}(\bissue{1}),
\bfpage{488}
(\byear{2011})
\end{barticle}
\endbibitem

\bibitem[\protect\citeauthoryear{Bensaad et~al.}{2006}]{bensaad2006tigar}
\begin{barticle}
\bauthor{\bsnm{Bensaad}, \binits{K.}},
\bauthor{\bsnm{Tsuruta}, \binits{A.}},
\bauthor{\bsnm{Selak}, \binits{M.A.}},
\bauthor{\bsnm{Vidal}, \binits{M.N.C.}},
\bauthor{\bsnm{Nakano}, \binits{K.}},
\bauthor{\bsnm{Bartrons}, \binits{R.}},
\bauthor{\bsnm{Gottlieb}, \binits{E.}},
\bauthor{\bsnm{Vousden}, \binits{K.H.}}:
\batitle{Tigar, a p53-inducible regulator of glycolysis and apoptosis}.
\bjtitle{Cell}
\bvolume{126}(\bissue{1}),
\bfpage{107}--\blpage{120}
(\byear{2006})
\end{barticle}
\endbibitem

\bibitem[\protect\citeauthoryear{Connolly et~al.}{2006}]{connolly2006hypoxia}
\begin{barticle}
\bauthor{\bsnm{Connolly}, \binits{E.}},
\bauthor{\bsnm{Braunstein}, \binits{S.}},
\bauthor{\bsnm{Formenti}, \binits{S.}},
\bauthor{\bsnm{Schneider}, \binits{R.J.}}:
\batitle{Hypoxia inhibits protein synthesis through a 4e-bp1 and elongation factor 2 kinase pathway controlled by mtor and uncoupled in breast cancer cells}.
\bjtitle{Molecular and cellular biology}
\bvolume{26}(\bissue{10}),
\bfpage{3955}--\blpage{3965}
(\byear{2006})
\end{barticle}
\endbibitem

\bibitem[\protect\citeauthoryear{Castillo et~al.}{2018}]{castillo2018effect}
\begin{barticle}
\bauthor{\bsnm{Castillo}, \binits{A.}},
\bauthor{\bsnm{Callejas}, \binits{L.}},
\bauthor{\bsnm{Alvarez-Gonz{\'a}lez}, \binits{C.A.}},
\bauthor{\bsnm{Maldonado}, \binits{C.}},
\bauthor{\bsnm{Cuzon}, \binits{G.}},
\bauthor{\bsnm{Gaxiola}, \binits{G.}}:
\batitle{Effect of native and modified starches on nutritional and hysiological performance of wild juveniles of red grouper (epinephelus morio)}.
\bjtitle{Ecosistemas y recursos agropecuarios}
\bvolume{5}(\bissue{15}),
\bfpage{491}--\blpage{500}
(\byear{2018})
\end{barticle}
\endbibitem

\bibitem[\protect\citeauthoryear{Chang et~al.}{1991}]{chang1991kinetic}
\begin{barticle}
\bauthor{\bsnm{Chang}, \binits{G.-G.}},
\bauthor{\bsnm{Huang}, \binits{S.-M.}},
\bauthor{\bsnm{Chiou}, \binits{S.-H.}}:
\batitle{Kinetic mechanism of the endogenous lactate dehydrogenase activity of duck $\epsilon$-crystallin}.
\bjtitle{Archives of biochemistry and biophysics}
\bvolume{284}(\bissue{2}),
\bfpage{285}--\blpage{291}
(\byear{1991})
\end{barticle}
\endbibitem

\bibitem[\protect\citeauthoryear{Cairns et~al.}{2011}]{cairns2011regulation}
\begin{barticle}
\bauthor{\bsnm{Cairns}, \binits{R.A.}},
\bauthor{\bsnm{Harris}, \binits{I.S.}},
\bauthor{\bsnm{Mak}, \binits{T.W.}}:
\batitle{Regulation of cancer cell metabolism}.
\bjtitle{Nature Reviews Cancer}
\bvolume{11}(\bissue{2}),
\bfpage{85}--\blpage{95}
(\byear{2011})
\end{barticle}
\endbibitem

\bibitem[\protect\citeauthoryear{Carnero and Paramio}{2014}]{carnero2014pten}
\begin{barticle}
\bauthor{\bsnm{Carnero}, \binits{A.}},
\bauthor{\bsnm{Paramio}, \binits{J.M.}}:
\batitle{The pten/pi3k/akt pathway in vivo, cancer mouse models}.
\bjtitle{Frontiers in oncology}
\bvolume{4},
\bfpage{252}
(\byear{2014})
\end{barticle}
\endbibitem

\bibitem[\protect\citeauthoryear{Crewe et~al.}{2017}]{crewe2017regulation}
\begin{barticle}
\bauthor{\bsnm{Crewe}, \binits{C.}},
\bauthor{\bsnm{Schafer}, \binits{C.}},
\bauthor{\bsnm{Lee}, \binits{I.}},
\bauthor{\bsnm{Kinter}, \binits{M.}},
\bauthor{\bsnm{Szweda}, \binits{L.I.}}:
\batitle{Regulation of pyruvate dehydrogenase kinase 4 in the heart through degradation by the lon protease in response to mitochondrial substrate availability}.
\bjtitle{Journal of Biological Chemistry}
\bvolume{292}(\bissue{1}),
\bfpage{305}--\blpage{312}
(\byear{2017})
\end{barticle}
\endbibitem

\bibitem[\protect\citeauthoryear{Day et~al.}{2013}]{day2013factors}
\begin{barticle}
\bauthor{\bsnm{Day}, \binits{P.}},
\bauthor{\bsnm{Cleal}, \binits{J.}},
\bauthor{\bsnm{Lofthouse}, \binits{E.}},
\bauthor{\bsnm{Hanson}, \binits{M.}},
\bauthor{\bsnm{Lewis}, \binits{R.}}:
\batitle{What factors determine placental glucose transfer kinetics?}
\bjtitle{Placenta}
\bvolume{34}(\bissue{10}),
\bfpage{953}--\blpage{958}
(\byear{2013})
\end{barticle}
\endbibitem

\bibitem[\protect\citeauthoryear{Danielsen et~al.}{2015}]{danielsen2015portrait}
\begin{barticle}
\bauthor{\bsnm{Danielsen}, \binits{S.A.}},
\bauthor{\bsnm{Eide}, \binits{P.W.}},
\bauthor{\bsnm{Nesbakken}, \binits{A.}},
\bauthor{\bsnm{Guren}, \binits{T.}},
\bauthor{\bsnm{Leithe}, \binits{E.}},
\bauthor{\bsnm{Lothe}, \binits{R.A.}}:
\batitle{Portrait of the pi3k/akt pathway in colorectal cancer}.
\bjtitle{Biochimica et Biophysica Acta (BBA)-Reviews on Cancer}
\bvolume{1855}(\bissue{1}),
\bfpage{104}--\blpage{121}
(\byear{2015})
\end{barticle}
\endbibitem

\bibitem[\protect\citeauthoryear{Dan et~al.}{2014}]{dan2014akt}
\begin{barticle}
\bauthor{\bsnm{Dan}, \binits{H.C.}},
\bauthor{\bsnm{Ebbs}, \binits{A.}},
\bauthor{\bsnm{Pasparakis}, \binits{M.}},
\bauthor{\bsnm{Van~Dyke}, \binits{T.}},
\bauthor{\bsnm{Basseres}, \binits{D.S.}},
\bauthor{\bsnm{Baldwin}, \binits{A.S.}}:
\batitle{Akt-dependent activation of mtorc1 complex involves phosphorylation of mtor (mammalian target of rapamycin) by i$\kappa$b kinase $\alpha$ (ikk$\alpha$)}.
\bjtitle{Journal of Biological Chemistry}
\bvolume{289}(\bissue{36}),
\bfpage{25227}--\blpage{25240}
(\byear{2014})
\end{barticle}
\endbibitem

\bibitem[\protect\citeauthoryear{Devic}{2016}]{devic2016warburg}
\begin{barticle}
\bauthor{\bsnm{Devic}, \binits{S.}}:
\batitle{Warburg effect-a consequence or the cause of carcinogenesis?}
\bjtitle{Journal of Cancer}
\bvolume{7}(\bissue{7}),
\bfpage{817}
(\byear{2016})
\end{barticle}
\endbibitem

\bibitem[\protect\citeauthoryear{DeYoung et~al.}{2008}]{deyoung2008hypoxia}
\begin{barticle}
\bauthor{\bsnm{DeYoung}, \binits{M.P.}},
\bauthor{\bsnm{Horak}, \binits{P.}},
\bauthor{\bsnm{Sofer}, \binits{A.}},
\bauthor{\bsnm{Sgroi}, \binits{D.}},
\bauthor{\bsnm{Ellisen}, \binits{L.W.}}:
\batitle{Hypoxia regulates tsc1/2--mtor signaling and tumor suppression through redd1-mediated 14--3--3 shuttling}.
\bjtitle{Genes \& development}
\bvolume{22}(\bissue{2}),
\bfpage{239}--\blpage{251}
(\byear{2008})
\end{barticle}
\endbibitem

\bibitem[\protect\citeauthoryear{Dai et~al.}{2020}]{dai2020glut3}
\begin{barticle}
\bauthor{\bsnm{Dai}, \binits{W.}},
\bauthor{\bsnm{Xu}, \binits{Y.}},
\bauthor{\bsnm{Mo}, \binits{S.}},
\bauthor{\bsnm{Li}, \binits{Q.}},
\bauthor{\bsnm{Yu}, \binits{J.}},
\bauthor{\bsnm{Wang}, \binits{R.}},
\bauthor{\bsnm{Ma}, \binits{Y.}},
\bauthor{\bsnm{Ni}, \binits{Y.}},
\bauthor{\bsnm{Xiang}, \binits{W.}},
\bauthor{\bsnm{Han}, \binits{L.}}, \betal:
\batitle{Glut3 induced by ampk/creb1 axis is key for withstanding energy stress and augments the efficacy of current colorectal cancer therapies}.
\bjtitle{Signal transduction and targeted therapy}
\bvolume{5}(\bissue{1}),
\bfpage{177}
(\byear{2020})
\end{barticle}
\endbibitem

\bibitem[\protect\citeauthoryear{D{\"u}vel et~al.}{2010}]{duvel2010activation}
\begin{barticle}
\bauthor{\bsnm{D{\"u}vel}, \binits{K.}},
\bauthor{\bsnm{Yecies}, \binits{J.L.}},
\bauthor{\bsnm{Menon}, \binits{S.}},
\bauthor{\bsnm{Raman}, \binits{P.}},
\bauthor{\bsnm{Lipovsky}, \binits{A.I.}},
\bauthor{\bsnm{Souza}, \binits{A.L.}},
\bauthor{\bsnm{Triantafellow}, \binits{E.}},
\bauthor{\bsnm{Ma}, \binits{Q.}},
\bauthor{\bsnm{Gorski}, \binits{R.}},
\bauthor{\bsnm{Cleaver}, \binits{S.}}, \betal:
\batitle{Activation of a metabolic gene regulatory network downstream of mtor complex 1}.
\bjtitle{Molecular cell}
\bvolume{39}(\bissue{2}),
\bfpage{171}--\blpage{183}
(\byear{2010})
\end{barticle}
\endbibitem

\bibitem[\protect\citeauthoryear{Fan et~al.}{2011}]{fan2011tyrosine}
\begin{barticle}
\bauthor{\bsnm{Fan}, \binits{J.}},
\bauthor{\bsnm{Hitosugi}, \binits{T.}},
\bauthor{\bsnm{Chung}, \binits{T.-W.}},
\bauthor{\bsnm{Xie}, \binits{J.}},
\bauthor{\bsnm{Ge}, \binits{Q.}},
\bauthor{\bsnm{Gu}, \binits{T.-L.}},
\bauthor{\bsnm{Polakiewicz}, \binits{R.D.}},
\bauthor{\bsnm{Chen}, \binits{G.Z.}},
\bauthor{\bsnm{Boggon}, \binits{T.J.}},
\bauthor{\bsnm{Lonial}, \binits{S.}}, \betal:
\batitle{Tyrosine phosphorylation of lactate dehydrogenase a is important for nadh/nad+ redox homeostasis in cancer cells}.
\bjtitle{Molecular and cellular biology}
\bvolume{31}(\bissue{24}),
\bfpage{4938}--\blpage{4950}
(\byear{2011})
\end{barticle}
\endbibitem

\bibitem[\protect\citeauthoryear{Fl{\"o}ter et~al.}{2017}]{floter2017regulation}
\begin{barticle}
\bauthor{\bsnm{Fl{\"o}ter}, \binits{J.}},
\bauthor{\bsnm{Kaymak}, \binits{I.}},
\bauthor{\bsnm{Schulze}, \binits{A.}}:
\batitle{Regulation of metabolic activity by p53}.
\bjtitle{Metabolites}
\bvolume{7}(\bissue{2}),
\bfpage{21}
(\byear{2017})
\end{barticle}
\endbibitem

\bibitem[\protect\citeauthoryear{Feng and Levine}{2010}]{feng2010regulation}
\begin{barticle}
\bauthor{\bsnm{Feng}, \binits{Z.}},
\bauthor{\bsnm{Levine}, \binits{A.J.}}:
\batitle{The regulation of energy metabolism and the igf-1/mtor pathways by the p53 protein}.
\bjtitle{Trends in cell biology}
\bvolume{20}(\bissue{7}),
\bfpage{427}--\blpage{434}
(\byear{2010})
\end{barticle}
\endbibitem

\bibitem[\protect\citeauthoryear{Freedman et~al.}{2002}]{freedman2002structural}
\begin{barticle}
\bauthor{\bsnm{Freedman}, \binits{S.J.}},
\bauthor{\bsnm{Sun}, \binits{Z.-Y.J.}},
\bauthor{\bsnm{Poy}, \binits{F.}},
\bauthor{\bsnm{Kung}, \binits{A.L.}},
\bauthor{\bsnm{Livingston}, \binits{D.M.}},
\bauthor{\bsnm{Wagner}, \binits{G.}},
\bauthor{\bsnm{Eck}, \binits{M.J.}}:
\batitle{Structural basis for recruitment of cbp/p300 by hypoxia-inducible factor-1$\alpha$}.
\bjtitle{Proceedings of the National Academy of Sciences}
\bvolume{99}(\bissue{8}),
\bfpage{5367}--\blpage{5372}
(\byear{2002})
\end{barticle}
\endbibitem

\bibitem[\protect\citeauthoryear{Faubert et~al.}{2015}]{faubert2015amp}
\begin{barticle}
\bauthor{\bsnm{Faubert}, \binits{B.}},
\bauthor{\bsnm{Vincent}, \binits{E.E.}},
\bauthor{\bsnm{Poffenberger}, \binits{M.C.}},
\bauthor{\bsnm{Jones}, \binits{R.G.}}:
\batitle{The amp-activated protein kinase (ampk) and cancer: many faces of a metabolic regulator}.
\bjtitle{Cancer letters}
\bvolume{356}(\bissue{2}),
\bfpage{165}--\blpage{170}
(\byear{2015})
\end{barticle}
\endbibitem

\bibitem[\protect\citeauthoryear{Fiscella et~al.}{1997}]{fiscella1997wip1}
\begin{barticle}
\bauthor{\bsnm{Fiscella}, \binits{M.}},
\bauthor{\bsnm{Zhang}, \binits{H.}},
\bauthor{\bsnm{Fan}, \binits{S.}},
\bauthor{\bsnm{Sakaguchi}, \binits{K.}},
\bauthor{\bsnm{Shen}, \binits{S.}},
\bauthor{\bsnm{Mercer}, \binits{W.E.}},
\bauthor{\bsnm{Vande~Woude}, \binits{G.F.}},
\bauthor{\bsnm{O’Connor}, \binits{P.M.}},
\bauthor{\bsnm{Appella}, \binits{E.}}:
\batitle{Wip1, a novel human protein phosphatase that is induced in response to ionizing radiation in a p53-dependent manner}.
\bjtitle{Proceedings of the National Academy of Sciences}
\bvolume{94}(\bissue{12}),
\bfpage{6048}--\blpage{6053}
(\byear{1997})
\end{barticle}
\endbibitem

\bibitem[\protect\citeauthoryear{Garc{\'\i}a-Aguilar et~al.}{2019}]{garcia2019changes}
\begin{barticle}
\bauthor{\bsnm{Garc{\'\i}a-Aguilar}, \binits{A.}},
\bauthor{\bsnm{Mart{\'\i}nez-Reyes}, \binits{I.}},
\bauthor{\bsnm{Cuezva}, \binits{J.M.}}:
\batitle{Changes in the turnover of the cellular proteome during metabolic reprogramming: a role for mtros in proteostasis}.
\bjtitle{Journal of proteome research}
\bvolume{18}(\bissue{8}),
\bfpage{3142}--\blpage{3155}
(\byear{2019})
\end{barticle}
\endbibitem

\bibitem[\protect\citeauthoryear{Grupe et~al.}{1995}]{grupe1995transgenic}
\begin{barticle}
\bauthor{\bsnm{Grupe}, \binits{A.}},
\bauthor{\bsnm{Hultgren}, \binits{B.}},
\bauthor{\bsnm{Ryan}, \binits{A.}},
\bauthor{\bsnm{Ma}, \binits{Y.H.}},
\bauthor{\bsnm{Bauer}, \binits{M.}},
\bauthor{\bsnm{Stewart}, \binits{T.A.}}:
\batitle{Transgenic knockouts reveal a critical requirement for pancreatic $\beta$ cell glucokinase in maintaining glucose homeostasis}.
\bjtitle{Cell}
\bvolume{83}(\bissue{1}),
\bfpage{69}--\blpage{78}
(\byear{1995})
\end{barticle}
\endbibitem

\bibitem[\protect\citeauthoryear{Golias et~al.}{2019}]{golias2019microenvironmental}
\begin{barticle}
\bauthor{\bsnm{Golias}, \binits{T.}},
\bauthor{\bsnm{Kery}, \binits{M.}},
\bauthor{\bsnm{Radenkovic}, \binits{S.}},
\bauthor{\bsnm{Papandreou}, \binits{I.}}:
\batitle{Microenvironmental control of glucose metabolism in tumors by regulation of pyruvate dehydrogenase}.
\bjtitle{International journal of cancer}
\bvolume{144}(\bissue{4}),
\bfpage{674}--\blpage{686}
(\byear{2019})
\end{barticle}
\endbibitem

\bibitem[\protect\citeauthoryear{Grossman}{2001}]{grossman2001p300}
\begin{barticle}
\bauthor{\bsnm{Grossman}, \binits{S.R.}}:
\batitle{p300/cbp/p53 interaction and regulation of the p53 response}.
\bjtitle{European journal of biochemistry}
\bvolume{268}(\bissue{10}),
\bfpage{2773}--\blpage{2778}
(\byear{2001})
\end{barticle}
\endbibitem

\bibitem[\protect\citeauthoryear{Hardie}{2011}]{hardie2011sensing}
\begin{barticle}
\bauthor{\bsnm{Hardie}, \binits{D.G.}}:
\batitle{Sensing of energy and nutrients by amp-activated protein kinase}.
\bjtitle{The American journal of clinical nutrition}
\bvolume{93}(\bissue{4}),
\bfpage{891}--\blpage{896}
(\byear{2011})
\end{barticle}
\endbibitem

\bibitem[\protect\citeauthoryear{Horak et~al.}{2010}]{horak2010negative}
\begin{barticle}
\bauthor{\bsnm{Horak}, \binits{P.}},
\bauthor{\bsnm{Crawford}, \binits{A.R.}},
\bauthor{\bsnm{Vadysirisack}, \binits{D.D.}},
\bauthor{\bsnm{Nash}, \binits{Z.M.}},
\bauthor{\bsnm{DeYoung}, \binits{M.P.}},
\bauthor{\bsnm{Sgroi}, \binits{D.}},
\bauthor{\bsnm{Ellisen}, \binits{L.W.}}:
\batitle{Negative feedback control of hif-1 through redd1-regulated ros suppresses tumorigenesis}.
\bjtitle{Proceedings of the National Academy of Sciences}
\bvolume{107}(\bissue{10}),
\bfpage{4675}--\blpage{4680}
(\byear{2010})
\end{barticle}
\endbibitem

\bibitem[\protect\citeauthoryear{Hu et~al.}{2006}]{hu2006differential}
\begin{barticle}
\bauthor{\bsnm{Hu}, \binits{C.-J.}},
\bauthor{\bsnm{Iyer}, \binits{S.}},
\bauthor{\bsnm{Sataur}, \binits{A.}},
\bauthor{\bsnm{Covello}, \binits{K.L.}},
\bauthor{\bsnm{Chodosh}, \binits{L.A.}},
\bauthor{\bsnm{Simon}, \binits{M.C.}}:
\batitle{Differential regulation of the transcriptional activities of hypoxia-inducible factor 1 alpha (hif-1$\alpha$) and hif-2$\alpha$ in stem cells}.
\bjtitle{Molecular and cellular biology}
\bvolume{26}(\bissue{9}),
\bfpage{3514}--\blpage{3526}
(\byear{2006})
\end{barticle}
\endbibitem

\bibitem[\protect\citeauthoryear{Han et~al.}{2013}]{han2013does}
\begin{barticle}
\bauthor{\bsnm{Han}, \binits{T.}},
\bauthor{\bsnm{Kang}, \binits{D.}},
\bauthor{\bsnm{Ji}, \binits{D.}},
\bauthor{\bsnm{Wang}, \binits{X.}},
\bauthor{\bsnm{Zhan}, \binits{W.}},
\bauthor{\bsnm{Fu}, \binits{M.}},
\bauthor{\bsnm{Xin}, \binits{H.-B.}},
\bauthor{\bsnm{Wang}, \binits{J.-B.}}:
\batitle{How does cancer cell metabolism affect tumor migration and invasion?}
\bjtitle{Cell adhesion \& migration}
\bvolume{7}(\bissue{5}),
\bfpage{395}--\blpage{403}
(\byear{2013})
\end{barticle}
\endbibitem

\bibitem[\protect\citeauthoryear{Hudson et~al.}{2002}]{hudson2002regulation}
\begin{barticle}
\bauthor{\bsnm{Hudson}, \binits{C.C.}},
\bauthor{\bsnm{Liu}, \binits{M.}},
\bauthor{\bsnm{Chiang}, \binits{G.G.}},
\bauthor{\bsnm{Otterness}, \binits{D.M.}},
\bauthor{\bsnm{Loomis}, \binits{D.C.}},
\bauthor{\bsnm{Kaper}, \binits{F.}},
\bauthor{\bsnm{Giaccia}, \binits{A.J.}},
\bauthor{\bsnm{Abraham}, \binits{R.T.}}:
\batitle{Regulation of hypoxia-inducible factor 1$\alpha$ expression and function by the mammalian target of rapamycin}.
\bjtitle{Molecular and cellular biology}
\bvolume{22}(\bissue{20}),
\bfpage{7004}--\blpage{7014}
(\byear{2002})
\end{barticle}
\endbibitem

\bibitem[\protect\citeauthoryear{Haupt et~al.}{1997}]{haupt1997mdm2}
\begin{barticle}
\bauthor{\bsnm{Haupt}, \binits{Y.}},
\bauthor{\bsnm{Maya}, \binits{R.}},
\bauthor{\bsnm{Kazaz}, \binits{A.}},
\bauthor{\bsnm{Oren}, \binits{M.}}:
\batitle{Mdm2 promotes the rapid degradation of p53}.
\bjtitle{Nature}
\bvolume{387}(\bissue{6630}),
\bfpage{296}--\blpage{299}
(\byear{1997})
\end{barticle}
\endbibitem

\bibitem[\protect\citeauthoryear{Hardie et~al.}{2012}]{hardie2012ampk}
\begin{barticle}
\bauthor{\bsnm{Hardie}, \binits{D.G.}},
\bauthor{\bsnm{Ross}, \binits{F.A.}},
\bauthor{\bsnm{Hawley}, \binits{S.A.}}:
\batitle{Ampk: a nutrient and energy sensor that maintains energy homeostasis}.
\bjtitle{Nature reviews Molecular cell biology}
\bvolume{13}(\bissue{4}),
\bfpage{251}--\blpage{262}
(\byear{2012})
\end{barticle}
\endbibitem

\bibitem[\protect\citeauthoryear{Hennessy et~al.}{2005}]{hennessy2005exploiting}
\begin{barticle}
\bauthor{\bsnm{Hennessy}, \binits{B.T.}},
\bauthor{\bsnm{Smith}, \binits{D.L.}},
\bauthor{\bsnm{Ram}, \binits{P.T.}},
\bauthor{\bsnm{Lu}, \binits{Y.}},
\bauthor{\bsnm{Mills}, \binits{G.B.}}:
\batitle{Exploiting the pi3k/akt pathway for cancer drug discovery}.
\bjtitle{Nature reviews Drug discovery}
\bvolume{4}(\bissue{12}),
\bfpage{988}--\blpage{1004}
(\byear{2005})
\end{barticle}
\endbibitem

\bibitem[\protect\citeauthoryear{Hu et~al.}{1983}]{hu1983induction}
\begin{barticle}
\bauthor{\bsnm{Hu}, \binits{C.}},
\bauthor{\bsnm{Utter}, \binits{M.}},
\bauthor{\bsnm{Patel}, \binits{M.}}:
\batitle{Induction of pyruvate dehydrogenase in 3t3-l1 cells during differentiation.}
\bjtitle{Journal of Biological Chemistry}
\bvolume{258}(\bissue{4}),
\bfpage{2315}--\blpage{2320}
(\byear{1983})
\end{barticle}
\endbibitem

\bibitem[\protect\citeauthoryear{Hanahan and Weinberg}{2011}]{hanahan2011hallmarks}
\begin{barticle}
\bauthor{\bsnm{Hanahan}, \binits{D.}},
\bauthor{\bsnm{Weinberg}, \binits{R.A.}}:
\batitle{Hallmarks of cancer: the next generation}.
\bjtitle{cell}
\bvolume{144}(\bissue{5}),
\bfpage{646}--\blpage{674}
(\byear{2011})
\end{barticle}
\endbibitem

\bibitem[\protect\citeauthoryear{Huang et~al.}{2002}]{huang2002regulation}
\begin{barticle}
\bauthor{\bsnm{Huang}, \binits{B.}},
\bauthor{\bsnm{Wu}, \binits{P.}},
\bauthor{\bsnm{Bowker-Kinley}, \binits{M.M.}},
\bauthor{\bsnm{Harris}, \binits{R.A.}}:
\batitle{Regulation of pyruvate dehydrogenase kinase expression by peroxisome proliferator--activated receptor-$\alpha$ ligands, glucocorticoids, and insulin}.
\bjtitle{diabetes}
\bvolume{51}(\bissue{2}),
\bfpage{276}--\blpage{283}
(\byear{2002})
\end{barticle}
\endbibitem

\bibitem[\protect\citeauthoryear{Iijima et~al.}{2004}]{iijima2004novel}
\begin{barticle}
\bauthor{\bsnm{Iijima}, \binits{M.}},
\bauthor{\bsnm{Huang}, \binits{Y.E.}},
\bauthor{\bsnm{Luo}, \binits{H.R.}},
\bauthor{\bsnm{Vazquez}, \binits{F.}},
\bauthor{\bsnm{Devreotes}, \binits{P.N.}}:
\batitle{Novel mechanism of pten regulation by its phosphatidylinositol 4, 5-bisphosphate binding motif is critical for chemotaxis}.
\bjtitle{Journal of Biological Chemistry}
\bvolume{279}(\bissue{16}),
\bfpage{16606}--\blpage{16613}
(\byear{2004})
\end{barticle}
\endbibitem

\bibitem[\protect\citeauthoryear{Inoki et~al.}{2002}]{inoki2002tsc2}
\begin{barticle}
\bauthor{\bsnm{Inoki}, \binits{K.}},
\bauthor{\bsnm{Li}, \binits{Y.}},
\bauthor{\bsnm{Zhu}, \binits{T.}},
\bauthor{\bsnm{Wu}, \binits{J.}},
\bauthor{\bsnm{Guan}, \binits{K.-L.}}:
\batitle{Tsc2 is phosphorylated and inhibited by akt and suppresses mtor signalling}.
\bjtitle{Nature cell biology}
\bvolume{4}(\bissue{9}),
\bfpage{648}--\blpage{657}
(\byear{2002})
\end{barticle}
\endbibitem

\bibitem[\protect\citeauthoryear{Imamura et~al.}{2001}]{imamura2001cell}
\begin{barticle}
\bauthor{\bsnm{Imamura}, \binits{K.}},
\bauthor{\bsnm{Ogura}, \binits{T.}},
\bauthor{\bsnm{Kishimoto}, \binits{A.}},
\bauthor{\bsnm{Kaminishi}, \binits{M.}},
\bauthor{\bsnm{Esumi}, \binits{H.}}:
\batitle{Cell cycle regulation via p53 phosphorylation by a 5'-amp activated protein kinase activator, 5-aminoimidazole-4-carboxamide-1-$\beta$-d-ribofuranoside, in a human hepatocellular carcinoma cell line}.
\bjtitle{Biochemical and biophysical research communications}
\bvolume{287}(\bissue{2}),
\bfpage{562}--\blpage{567}
(\byear{2001})
\end{barticle}
\endbibitem

\bibitem[\protect\citeauthoryear{Inoki et~al.}{2003}]{inoki2003tsc2}
\begin{barticle}
\bauthor{\bsnm{Inoki}, \binits{K.}},
\bauthor{\bsnm{Zhu}, \binits{T.}},
\bauthor{\bsnm{Guan}, \binits{K.-L.}}:
\batitle{Tsc2 mediates cellular energy response to control cell growth and survival}.
\bjtitle{Cell}
\bvolume{115}(\bissue{5}),
\bfpage{577}--\blpage{590}
(\byear{2003})
\end{barticle}
\endbibitem

\bibitem[\protect\citeauthoryear{Javed et~al.}{1997}]{javed1997purification}
\begin{barticle}
\bauthor{\bsnm{Javed}, \binits{M.H.}},
\bauthor{\bsnm{Azimuddin}, \binits{S.M.}},
\bauthor{\bsnm{Hussain}, \binits{A.N.}},
\bauthor{\bsnm{Ahmed}, \binits{A.}},
\bauthor{\bsnm{Ishaq}, \binits{M.}}:
\batitle{Purification and characterization of lactate dehydrogenase from varanus liver}.
\bjtitle{Experimental \& Molecular Medicine}
\bvolume{29}(\bissue{1}),
\bfpage{25}--\blpage{30}
(\byear{1997})
\end{barticle}
\endbibitem

\bibitem[\protect\citeauthoryear{Jiang et~al.}{2014}]{jiang2014regulation}
\begin{barticle}
\bauthor{\bsnm{Jiang}, \binits{P.}},
\bauthor{\bsnm{Du}, \binits{W.}},
\bauthor{\bsnm{Wu}, \binits{M.}}:
\batitle{Regulation of the pentose phosphate pathway in cancer}.
\bjtitle{Protein \& cell}
\bvolume{5}(\bissue{8}),
\bfpage{592}--\blpage{602}
(\byear{2014})
\end{barticle}
\endbibitem

\bibitem[\protect\citeauthoryear{Jin et~al.}{2020}]{jin2020hemistepsin}
\begin{barticle}
\bauthor{\bsnm{Jin}, \binits{L.}},
\bauthor{\bsnm{Kim}, \binits{E.-Y.}},
\bauthor{\bsnm{Chung}, \binits{T.-W.}},
\bauthor{\bsnm{Han}, \binits{C.W.}},
\bauthor{\bsnm{Park}, \binits{S.Y.}},
\bauthor{\bsnm{Han}, \binits{J.H.}},
\bauthor{\bsnm{Bae}, \binits{S.-J.}},
\bauthor{\bsnm{Lee}, \binits{J.R.}},
\bauthor{\bsnm{Kim}, \binits{Y.W.}},
\bauthor{\bsnm{Jang}, \binits{S.B.}}, \betal:
\batitle{Hemistepsin a suppresses colorectal cancer growth through inhibiting pyruvate dehydrogenase kinase activity}.
\bjtitle{Scientific Reports}
\bvolume{10}(\bissue{1}),
\bfpage{21940}
(\byear{2020})
\end{barticle}
\endbibitem

\bibitem[\protect\citeauthoryear{Jones et~al.}{2005}]{jones2005amp}
\begin{barticle}
\bauthor{\bsnm{Jones}, \binits{R.G.}},
\bauthor{\bsnm{Plas}, \binits{D.R.}},
\bauthor{\bsnm{Kubek}, \binits{S.}},
\bauthor{\bsnm{Buzzai}, \binits{M.}},
\bauthor{\bsnm{Mu}, \binits{J.}},
\bauthor{\bsnm{Xu}, \binits{Y.}},
\bauthor{\bsnm{Birnbaum}, \binits{M.J.}},
\bauthor{\bsnm{Thompson}, \binits{C.B.}}:
\batitle{Amp-activated protein kinase induces a p53-dependent metabolic checkpoint}.
\bjtitle{Molecular cell}
\bvolume{18}(\bissue{3}),
\bfpage{283}--\blpage{293}
(\byear{2005})
\end{barticle}
\endbibitem

\bibitem[\protect\citeauthoryear{Kawauchi et~al.}{2008}]{kawauchi2008p53}
\begin{barticle}
\bauthor{\bsnm{Kawauchi}, \binits{K.}},
\bauthor{\bsnm{Araki}, \binits{K.}},
\bauthor{\bsnm{Tobiume}, \binits{K.}},
\bauthor{\bsnm{Tanaka}, \binits{N.}}:
\batitle{p53 regulates glucose metabolism through an ikk-nf-$\kappa$b pathway and inhibits cell transformation}.
\bjtitle{Nature cell biology}
\bvolume{10}(\bissue{5}),
\bfpage{611}--\blpage{618}
(\byear{2008})
\end{barticle}
\endbibitem

\bibitem[\protect\citeauthoryear{Kight and Fleming}{1995}]{kight1995oxidation}
\begin{barticle}
\bauthor{\bsnm{Kight}, \binits{C.E.}},
\bauthor{\bsnm{Fleming}, \binits{S.E.}}:
\batitle{Oxidation of glucose carbon entering the tca cycle is reduced by glutamine in small intestine epithelial cells}.
\bjtitle{American Journal of Physiology-Gastrointestinal and Liver Physiology}
\bvolume{268}(\bissue{6}),
\bfpage{879}--\blpage{888}
(\byear{1995})
\end{barticle}
\endbibitem

\bibitem[\protect\citeauthoryear{Kubaichuk and Kietzmann}{2023}]{kubaichuk2023usp10}
\begin{barticle}
\bauthor{\bsnm{Kubaichuk}, \binits{K.}},
\bauthor{\bsnm{Kietzmann}, \binits{T.}}:
\batitle{Usp10 contributes to colon carcinogenesis via mtor/s6k mediated hif-1$\alpha$ but not hif-2$\alpha$ protein synthesis}.
\bjtitle{Cells}
\bvolume{12}(\bissue{12}),
\bfpage{1585}
(\byear{2023})
\end{barticle}
\endbibitem

\bibitem[\protect\citeauthoryear{Khayat et~al.}{1998}]{khayat1998unique}
\begin{barticle}
\bauthor{\bsnm{Khayat}, \binits{Z.A.}},
\bauthor{\bsnm{McCALL}, \binits{A.L.}},
\bauthor{\bsnm{KLIP}, \binits{A.}}:
\batitle{Unique mechanism of glut3 glucose transporter regulation by prolonged energy demand: increased protein half-life}.
\bjtitle{Biochemical Journal}
\bvolume{333}(\bissue{3}),
\bfpage{713}--\blpage{718}
(\byear{1998})
\end{barticle}
\endbibitem

\bibitem[\protect\citeauthoryear{Kallio et~al.}{1998}]{kallio1998signal}
\begin{barticle}
\bauthor{\bsnm{Kallio}, \binits{P.J.}},
\bauthor{\bsnm{Okamoto}, \binits{K.}},
\bauthor{\bsnm{O'Brien}, \binits{S.}},
\bauthor{\bsnm{Carrero}, \binits{P.}},
\bauthor{\bsnm{Makino}, \binits{Y.}},
\bauthor{\bsnm{Tanaka}, \binits{H.}},
\bauthor{\bsnm{Poellinger}, \binits{L.}}:
\batitle{Signal transduction in hypoxic cells: inducible nuclear translocation and recruitment of thecbp/p300 coactivator by the hypoxia-induciblefactor-1$\alpha$}.
\bjtitle{The EMBO journal}
\bvolume{17}(\bissue{22}),
\bfpage{6573}--\blpage{6586}
(\byear{1998})
\end{barticle}
\endbibitem

\bibitem[\protect\citeauthoryear{Koho et~al.}{2008}]{koho2008lactate}
\begin{barticle}
\bauthor{\bsnm{Koho}, \binits{N.M.}},
\bauthor{\bsnm{Raekallio}, \binits{M.}},
\bauthor{\bsnm{Kuusela}, \binits{E.}},
\bauthor{\bsnm{Vuolle}, \binits{J.}},
\bauthor{\bsnm{P{\"o}s{\"o}}, \binits{A.R.}}:
\batitle{Lactate transport in canine red blood cells}.
\bjtitle{American journal of veterinary research}
\bvolume{69}(\bissue{8}),
\bfpage{1091}--\blpage{1096}
(\byear{2008})
\end{barticle}
\endbibitem

\bibitem[\protect\citeauthoryear{Kim et~al.}{2006}]{kim2006hif}
\begin{barticle}
\bauthor{\bsnm{Kim}, \binits{J.-w.}},
\bauthor{\bsnm{Tchernyshyov}, \binits{I.}},
\bauthor{\bsnm{Semenza}, \binits{G.L.}},
\bauthor{\bsnm{Dang}, \binits{C.V.}}:
\batitle{Hif-1-mediated expression of pyruvate dehydrogenase kinase: a metabolic switch required for cellular adaptation to hypoxia}.
\bjtitle{Cell metabolism}
\bvolume{3}(\bissue{3}),
\bfpage{177}--\blpage{185}
(\byear{2006})
\end{barticle}
\endbibitem

\bibitem[\protect\citeauthoryear{Kuby}{2019}]{kuby2019study}
\begin{botherref}
\oauthor{\bsnm{Kuby}, \binits{S.A.}}:
A study of enzymes: Enzyme catalysts, kinetics, and substrate binding.
CRC Press
(2019)
\end{botherref}
\endbibitem

\bibitem[\protect\citeauthoryear{Liang and Clarke}{2001}]{liang2001regulation}
\begin{barticle}
\bauthor{\bsnm{Liang}, \binits{S.-H.}},
\bauthor{\bsnm{Clarke}, \binits{M.F.}}:
\batitle{Regulation of p53 localization}.
\bjtitle{European journal of biochemistry}
\bvolume{268}(\bissue{10}),
\bfpage{2779}--\blpage{2783}
(\byear{2001})
\end{barticle}
\endbibitem

\bibitem[\protect\citeauthoryear{Li et~al.}{2017}]{li2017mitochondrial}
\begin{barticle}
\bauthor{\bsnm{Li}, \binits{X.}},
\bauthor{\bsnm{Han}, \binits{G.}},
\bauthor{\bsnm{Li}, \binits{X.}},
\bauthor{\bsnm{Kan}, \binits{Q.}},
\bauthor{\bsnm{Fan}, \binits{Z.}},
\bauthor{\bsnm{Li}, \binits{Y.}},
\bauthor{\bsnm{Ji}, \binits{Y.}},
\bauthor{\bsnm{Zhao}, \binits{J.}},
\bauthor{\bsnm{Zhang}, \binits{M.}},
\bauthor{\bsnm{Grigalavicius}, \binits{M.}}, \betal:
\batitle{Mitochondrial pyruvate carrier function determines cell stemness and metabolic reprogramming in cancer cells}.
\bjtitle{Oncotarget}
\bvolume{8}(\bissue{28}),
\bfpage{46363}
(\byear{2017})
\end{barticle}
\endbibitem

\bibitem[\protect\citeauthoryear{Liang et~al.}{2020}]{liang2020dichloroacetate}
\begin{barticle}
\bauthor{\bsnm{Liang}, \binits{Y.}},
\bauthor{\bsnm{Hou}, \binits{L.}},
\bauthor{\bsnm{Li}, \binits{L.}},
\bauthor{\bsnm{Li}, \binits{L.}},
\bauthor{\bsnm{Zhu}, \binits{L.}},
\bauthor{\bsnm{Wang}, \binits{Y.}},
\bauthor{\bsnm{Huang}, \binits{X.}},
\bauthor{\bsnm{Hou}, \binits{Y.}},
\bauthor{\bsnm{Zhu}, \binits{D.}},
\bauthor{\bsnm{Zou}, \binits{H.}}, \betal:
\batitle{Dichloroacetate restores colorectal cancer chemosensitivity through the p53/mir-149-3p/pdk2-mediated glucose metabolic pathway}.
\bjtitle{Oncogene}
\bvolume{39}(\bissue{2}),
\bfpage{469}--\blpage{485}
(\byear{2020})
\end{barticle}
\endbibitem

\bibitem[\protect\citeauthoryear{Lee et~al.}{2015}]{lee2015p53}
\begin{barticle}
\bauthor{\bsnm{Lee}, \binits{P.}},
\bauthor{\bsnm{Hock}, \binits{A.}},
\bauthor{\bsnm{Vousden}, \binits{K.}},
\bauthor{\bsnm{Cheung}, \binits{E.}}:
\batitle{p53-and p73-independent activation of tigar expression in vivo}.
\bjtitle{Cell death \& disease}
\bvolume{6}(\bissue{8}),
\bfpage{1842}--\blpage{1842}
(\byear{2015})
\end{barticle}
\endbibitem

\bibitem[\protect\citeauthoryear{Lu et~al.}{2011}]{lu2011overexpression}
\begin{barticle}
\bauthor{\bsnm{Lu}, \binits{C.-W.}},
\bauthor{\bsnm{Lin}, \binits{S.-C.}},
\bauthor{\bsnm{Chien}, \binits{C.-W.}},
\bauthor{\bsnm{Lin}, \binits{S.-C.}},
\bauthor{\bsnm{Lee}, \binits{C.-T.}},
\bauthor{\bsnm{Lin}, \binits{B.-W.}},
\bauthor{\bsnm{Lee}, \binits{J.-C.}},
\bauthor{\bsnm{Tsai}, \binits{S.-J.}}:
\batitle{Overexpression of pyruvate dehydrogenase kinase 3 increases drug resistance and early recurrence in colon cancer}.
\bjtitle{The American journal of pathology}
\bvolume{179}(\bissue{3}),
\bfpage{1405}--\blpage{1414}
(\byear{2011})
\end{barticle}
\endbibitem

\bibitem[\protect\citeauthoryear{Lien et~al.}{2016}]{lien2016metabolic}
\begin{botherref}
\oauthor{\bsnm{Lien}, \binits{E.C.}},
\oauthor{\bsnm{Lyssiotis}, \binits{C.A.}},
\oauthor{\bsnm{Cantley}, \binits{L.C.}}:
Metabolic reprogramming by the pi3k-akt-mtor pathway in cancer.
Metabolism in Cancer,
39--72
(2016)
\end{botherref}
\endbibitem

\bibitem[\protect\citeauthoryear{Liang et~al.}{2013}]{liang2013regulation}
\begin{barticle}
\bauthor{\bsnm{Liang}, \binits{Y.}},
\bauthor{\bsnm{Liu}, \binits{J.}},
\bauthor{\bsnm{Feng}, \binits{Z.}}:
\batitle{The regulation of cellular metabolism by tumor suppressor p53}.
\bjtitle{Cell \& bioscience}
\bvolume{3},
\bfpage{1}--\blpage{10}
(\byear{2013})
\end{barticle}
\endbibitem

\bibitem[\protect\citeauthoryear{Lannutti et~al.}{2011}]{lannutti2011cal}
\begin{barticle}
\bauthor{\bsnm{Lannutti}, \binits{B.J.}},
\bauthor{\bsnm{Meadows}, \binits{S.A.}},
\bauthor{\bsnm{Herman}, \binits{S.E.}},
\bauthor{\bsnm{Kashishian}, \binits{A.}},
\bauthor{\bsnm{Steiner}, \binits{B.}},
\bauthor{\bsnm{Johnson}, \binits{A.J.}},
\bauthor{\bsnm{Byrd}, \binits{J.C.}},
\bauthor{\bsnm{Tyner}, \binits{J.W.}},
\bauthor{\bsnm{Loriaux}, \binits{M.M.}},
\bauthor{\bsnm{Deininger}, \binits{M.}}, \betal:
\batitle{Cal-101, a p110$\delta$ selective phosphatidylinositol-3-kinase inhibitor for the treatment of b-cell malignancies, inhibits pi3k signaling and cellular viability}.
\bjtitle{Blood, The Journal of the American Society of Hematology}
\bvolume{117}(\bissue{2}),
\bfpage{591}--\blpage{594}
(\byear{2011})
\end{barticle}
\endbibitem

\bibitem[\protect\citeauthoryear{Liu et~al.}{2013}]{liu2013bay}
\begin{barticle}
\bauthor{\bsnm{Liu}, \binits{N.}},
\bauthor{\bsnm{Rowley}, \binits{B.R.}},
\bauthor{\bsnm{Bull}, \binits{C.O.}},
\bauthor{\bsnm{Schneider}, \binits{C.}},
\bauthor{\bsnm{Haegebarth}, \binits{A.}},
\bauthor{\bsnm{Schatz}, \binits{C.A.}},
\bauthor{\bsnm{Fracasso}, \binits{P.R.}},
\bauthor{\bsnm{Wilkie}, \binits{D.P.}},
\bauthor{\bsnm{Hentemann}, \binits{M.}},
\bauthor{\bsnm{Wilhelm}, \binits{S.M.}}, \betal:
\batitle{Bay 80-6946 is a highly selective intravenous pi3k inhibitor with potent p110$\alpha$ and p110$\delta$ activities in tumor cell lines and xenograft models}.
\bjtitle{Molecular cancer therapeutics}
\bvolume{12}(\bissue{11}),
\bfpage{2319}--\blpage{2330}
(\byear{2013})
\end{barticle}
\endbibitem

\bibitem[\protect\citeauthoryear{Lago et~al.}{2011}]{lago2011p53}
\begin{barticle}
\bauthor{\bsnm{Lago}, \binits{C.U.}},
\bauthor{\bsnm{Sung}, \binits{H.J.}},
\bauthor{\bsnm{Ma}, \binits{W.}},
\bauthor{\bsnm{Wang}, \binits{P.-y.}},
\bauthor{\bsnm{Hwang}, \binits{P.M.}}:
\batitle{p53, aerobic metabolism, and cancer}.
\bjtitle{Antioxidants \& Redox Signaling}
\bvolume{15}(\bissue{6}),
\bfpage{1739}--\blpage{1748}
(\byear{2011})
\end{barticle}
\endbibitem

\bibitem[\protect\citeauthoryear{Li et~al.}{2015}]{li2015targeting}
\begin{barticle}
\bauthor{\bsnm{Li}, \binits{W.}},
\bauthor{\bsnm{Saud}, \binits{S.M.}},
\bauthor{\bsnm{Young}, \binits{M.R.}},
\bauthor{\bsnm{Chen}, \binits{G.}},
\bauthor{\bsnm{Hua}, \binits{B.}}:
\batitle{Targeting ampk for cancer prevention and treatment}.
\bjtitle{Oncotarget}
\bvolume{6}(\bissue{10}),
\bfpage{7365}
(\byear{2015})
\end{barticle}
\endbibitem

\bibitem[\protect\citeauthoryear{Laughner et~al.}{2001}]{laughner2001her2}
\begin{barticle}
\bauthor{\bsnm{Laughner}, \binits{E.}},
\bauthor{\bsnm{Taghavi}, \binits{P.}},
\bauthor{\bsnm{Chiles}, \binits{K.}},
\bauthor{\bsnm{Mahon}, \binits{P.C.}},
\bauthor{\bsnm{Semenza}, \binits{G.L.}}:
\batitle{Her2 (neu) signaling increases the rate of hypoxia-inducible factor 1$\alpha$ (hif-1$\alpha$) synthesis: novel mechanism for hif-1-mediated vascular endothelial growth factor expression}.
\bjtitle{Molecular and cellular biology}
\bvolume{21}(\bissue{12}),
\bfpage{3995}--\blpage{4004}
(\byear{2001})
\end{barticle}
\endbibitem

\bibitem[\protect\citeauthoryear{Lee et~al.}{2008}]{lee2008nuclear}
\begin{barticle}
\bauthor{\bsnm{Lee}, \binits{S.B.}},
\bauthor{\bsnm{Xuan~Nguyen}, \binits{T.L.}},
\bauthor{\bsnm{Choi}, \binits{J.W.}},
\bauthor{\bsnm{Lee}, \binits{K.-H.}},
\bauthor{\bsnm{Cho}, \binits{S.-W.}},
\bauthor{\bsnm{Liu}, \binits{Z.}},
\bauthor{\bsnm{Ye}, \binits{K.}},
\bauthor{\bsnm{Bae}, \binits{S.S.}},
\bauthor{\bsnm{Ahn}, \binits{J.-Y.}}:
\batitle{Nuclear akt interacts with b23/npm and protects it from proteolytic cleavage, enhancing cell survival}.
\bjtitle{Proceedings of the National Academy of Sciences}
\bvolume{105}(\bissue{43}),
\bfpage{16584}--\blpage{16589}
(\byear{2008})
\end{barticle}
\endbibitem

\bibitem[\protect\citeauthoryear{Liu et~al.}{2019}]{liu2019tumor}
\begin{barticle}
\bauthor{\bsnm{Liu}, \binits{J.}},
\bauthor{\bsnm{Zhang}, \binits{C.}},
\bauthor{\bsnm{Hu}, \binits{W.}},
\bauthor{\bsnm{Feng}, \binits{Z.}}:
\batitle{Tumor suppressor p53 and metabolism}.
\bjtitle{Journal of molecular cell biology}
\bvolume{11}(\bissue{4}),
\bfpage{284}--\blpage{292}
(\byear{2019})
\end{barticle}
\endbibitem

\bibitem[\protect\citeauthoryear{Metzen et~al.}{2003}]{metzen2003intracellular}
\begin{barticle}
\bauthor{\bsnm{Metzen}, \binits{E.}},
\bauthor{\bsnm{Berchner-Pfannschmidt}, \binits{U.}},
\bauthor{\bsnm{Stengel}, \binits{P.}},
\bauthor{\bsnm{Marxsen}, \binits{J.H.}},
\bauthor{\bsnm{Stolze}, \binits{I.}},
\bauthor{\bsnm{Klinger}, \binits{M.}},
\bauthor{\bsnm{Huang}, \binits{W.Q.}},
\bauthor{\bsnm{Wotzlaw}, \binits{C.}},
\bauthor{\bsnm{Hellwig-Burgel}, \binits{T.}},
\bauthor{\bsnm{Jelkmann}, \binits{W.}}, \betal:
\batitle{Intracellular localisation of human hif-1$\alpha$ hydroxylases: implications for oxygen sensing}.
\bjtitle{Journal of cell science}
\bvolume{116}(\bissue{7}),
\bfpage{1319}--\blpage{1326}
(\byear{2003})
\end{barticle}
\endbibitem

\bibitem[\protect\citeauthoryear{Mayo and Donner}{2001}]{mayo2001phosphatidylinositol}
\begin{barticle}
\bauthor{\bsnm{Mayo}, \binits{L.D.}},
\bauthor{\bsnm{Donner}, \binits{D.B.}}:
\batitle{A phosphatidylinositol 3-kinase/akt pathway promotes translocation of mdm2 from the cytoplasm to the nucleus}.
\bjtitle{Proceedings of the National Academy of Sciences}
\bvolume{98}(\bissue{20}),
\bfpage{11598}--\blpage{11603}
(\byear{2001})
\end{barticle}
\endbibitem

\bibitem[\protect\citeauthoryear{Mayo et~al.}{2002}]{mayo2002pten}
\begin{barticle}
\bauthor{\bsnm{Mayo}, \binits{L.D.}},
\bauthor{\bsnm{Dixon}, \binits{J.E.}},
\bauthor{\bsnm{Durden}, \binits{D.L.}},
\bauthor{\bsnm{Tonks}, \binits{N.K.}},
\bauthor{\bsnm{Donner}, \binits{D.B.}}:
\batitle{Pten protects p53 from mdm2 and sensitizes cancer cells to chemotherapy}.
\bjtitle{Journal of Biological Chemistry}
\bvolume{277}(\bissue{7}),
\bfpage{5484}--\blpage{5489}
(\byear{2002})
\end{barticle}
\endbibitem

\bibitem[\protect\citeauthoryear{Marchenko et~al.}{2010}]{marchenko2010stress}
\begin{barticle}
\bauthor{\bsnm{Marchenko}, \binits{N.}},
\bauthor{\bsnm{Hanel}, \binits{W.}},
\bauthor{\bsnm{Li}, \binits{D.}},
\bauthor{\bsnm{Becker}, \binits{K.}},
\bauthor{\bsnm{Reich}, \binits{N.}},
\bauthor{\bsnm{Moll}, \binits{U.M.}}:
\batitle{Stress-mediated nuclear stabilization of p53 is regulated by ubiquitination and importin-$\alpha$3 binding}.
\bjtitle{Cell Death \& Differentiation}
\bvolume{17}(\bissue{2}),
\bfpage{255}--\blpage{267}
(\byear{2010})
\end{barticle}
\endbibitem

\bibitem[\protect\citeauthoryear{Maxfield et~al.}{2004}]{maxfield2004cox17}
\begin{barticle}
\bauthor{\bsnm{Maxfield}, \binits{A.B.}},
\bauthor{\bsnm{Heaton}, \binits{D.N.}},
\bauthor{\bsnm{Winge}, \binits{D.R.}}:
\batitle{Cox17 is functional when tethered to the mitochondrial inner membrane}.
\bjtitle{Journal of Biological Chemistry}
\bvolume{279}(\bissue{7}),
\bfpage{5072}--\blpage{5080}
(\byear{2004})
\end{barticle}
\endbibitem

\bibitem[\protect\citeauthoryear{Mamun et~al.}{2020}]{mamun2020hypoxia}
\begin{barticle}
\bauthor{\bsnm{Mamun}, \binits{A.A.}},
\bauthor{\bsnm{Hayashi}, \binits{H.}},
\bauthor{\bsnm{Yamamura}, \binits{A.}},
\bauthor{\bsnm{Nayeem}, \binits{M.J.}},
\bauthor{\bsnm{Sato}, \binits{M.}}:
\batitle{Hypoxia induces the translocation of glucose transporter 1 to the plasma membrane in vascular endothelial cells}.
\bjtitle{The Journal of Physiological Sciences}
\bvolume{70}(\bissue{1}),
\bfpage{1}--\blpage{15}
(\byear{2020})
\end{barticle}
\endbibitem

\bibitem[\protect\citeauthoryear{Matoba et~al.}{2006}]{matoba2006p53}
\begin{barticle}
\bauthor{\bsnm{Matoba}, \binits{S.}},
\bauthor{\bsnm{Kang}, \binits{J.-G.}},
\bauthor{\bsnm{Patino}, \binits{W.D.}},
\bauthor{\bsnm{Wragg}, \binits{A.}},
\bauthor{\bsnm{Boehm}, \binits{M.}},
\bauthor{\bsnm{Gavrilova}, \binits{O.}},
\bauthor{\bsnm{Hurley}, \binits{P.J.}},
\bauthor{\bsnm{Bunz}, \binits{F.}},
\bauthor{\bsnm{Hwang}, \binits{P.M.}}:
\batitle{p53 regulates mitochondrial respiration}.
\bjtitle{Science}
\bvolume{312}(\bissue{5780}),
\bfpage{1650}--\blpage{1653}
(\byear{2006})
\end{barticle}
\endbibitem

\bibitem[\protect\citeauthoryear{Maddalena et~al.}{2015}]{maddalena2015evaluation}
\begin{barticle}
\bauthor{\bsnm{Maddalena}, \binits{F.}},
\bauthor{\bsnm{Lettini}, \binits{G.}},
\bauthor{\bsnm{Gallicchio}, \binits{R.}},
\bauthor{\bsnm{Sisinni}, \binits{L.}},
\bauthor{\bsnm{Simeon}, \binits{V.}},
\bauthor{\bsnm{Nardelli}, \binits{A.}},
\bauthor{\bsnm{Venetucci}, \binits{A.A.}},
\bauthor{\bsnm{Storto}, \binits{G.}},
\bauthor{\bsnm{Landriscina}, \binits{M.}}:
\batitle{Evaluation of glucose uptake in normal and cancer cell lines by positron emission tomography}.
\bjtitle{Molecular imaging}
\bvolume{14}(\bissue{9}),
\bfpage{7290}--\blpage{2015}
(\byear{2015})
\end{barticle}
\endbibitem

\bibitem[\protect\citeauthoryear{Malinowsky et~al.}{2014}]{malinowsky2014activation}
\begin{barticle}
\bauthor{\bsnm{Malinowsky}, \binits{K.}},
\bauthor{\bsnm{Nitsche}, \binits{U.}},
\bauthor{\bsnm{Janssen}, \binits{K.}},
\bauthor{\bsnm{Bader}, \binits{F.}},
\bauthor{\bsnm{Sp{\"a}th}, \binits{C.}},
\bauthor{\bsnm{Drecoll}, \binits{E.}},
\bauthor{\bsnm{Keller}, \binits{G.}},
\bauthor{\bsnm{H{\"o}fler}, \binits{H.}},
\bauthor{\bsnm{Slotta-Huspenina}, \binits{J.}},
\bauthor{\bsnm{Becker}, \binits{K.}}:
\batitle{Activation of the pi3k/akt pathway correlates with prognosis in stage ii colon cancer}.
\bjtitle{British journal of cancer}
\bvolume{110}(\bissue{8}),
\bfpage{2081}--\blpage{2089}
(\byear{2014})
\end{barticle}
\endbibitem

\bibitem[\protect\citeauthoryear{Morris}{1991}]{morris1991factorial}
\begin{barticle}
\bauthor{\bsnm{Morris}, \binits{M.D.}}:
\batitle{Factorial sampling plans for preliminary computational experiments}.
\bjtitle{Technometrics}
\bvolume{33}(\bissue{2}),
\bfpage{161}--\blpage{174}
(\byear{1991})
\end{barticle}
\endbibitem

\bibitem[\protect\citeauthoryear{Mart{\'\i}nez-Reyes and Chandel}{2020}]{martinez2020mitochondrial}
\begin{barticle}
\bauthor{\bsnm{Mart{\'\i}nez-Reyes}, \binits{I.}},
\bauthor{\bsnm{Chandel}, \binits{N.S.}}:
\batitle{Mitochondrial tca cycle metabolites control physiology and disease}.
\bjtitle{Nature communications}
\bvolume{11}(\bissue{1}),
\bfpage{102}
(\byear{2020})
\end{barticle}
\endbibitem

\bibitem[\protect\citeauthoryear{Mayo et~al.}{2005}]{mayo2005phosphorylation}
\begin{barticle}
\bauthor{\bsnm{Mayo}, \binits{L.D.}},
\bauthor{\bsnm{Seo}, \binits{Y.R.}},
\bauthor{\bsnm{Jackson}, \binits{M.W.}},
\bauthor{\bsnm{Smith}, \binits{M.L.}},
\bauthor{\bsnm{Guzman}, \binits{J.R.}},
\bauthor{\bsnm{Korgaonkar}, \binits{C.K.}},
\bauthor{\bsnm{Donner}, \binits{D.B.}}:
\batitle{Phosphorylation of human p53 at serine 46 determines promoter selection and whether apoptosis is attenuated or amplified}.
\bjtitle{Journal of Biological Chemistry}
\bvolume{280}(\bissue{28}),
\bfpage{25953}--\blpage{25959}
(\byear{2005})
\end{barticle}
\endbibitem

\bibitem[\protect\citeauthoryear{Ma et~al.}{2005}]{ma2005plausible}
\begin{barticle}
\bauthor{\bsnm{Ma}, \binits{L.}},
\bauthor{\bsnm{Wagner}, \binits{J.}},
\bauthor{\bsnm{Rice}, \binits{J.J.}},
\bauthor{\bsnm{Hu}, \binits{W.}},
\bauthor{\bsnm{Levine}, \binits{A.J.}},
\bauthor{\bsnm{Stolovitzky}, \binits{G.A.}}:
\batitle{A plausible model for the digital response of p53 to dna damage}.
\bjtitle{Proceedings of the National Academy of Sciences}
\bvolume{102}(\bissue{40}),
\bfpage{14266}--\blpage{14271}
(\byear{2005})
\end{barticle}
\endbibitem

\bibitem[\protect\citeauthoryear{Nag et~al.}{2013}]{nag2013mdm2}
\begin{barticle}
\bauthor{\bsnm{Nag}, \binits{S.}},
\bauthor{\bsnm{Qin}, \binits{J.}},
\bauthor{\bsnm{Srivenugopal}, \binits{K.S.}},
\bauthor{\bsnm{Wang}, \binits{M.}},
\bauthor{\bsnm{Zhang}, \binits{R.}}:
\batitle{The mdm2-p53 pathway revisited}.
\bjtitle{Journal of biomedical research}
\bvolume{27}(\bissue{4}),
\bfpage{254}
(\byear{2013})
\end{barticle}
\endbibitem

\bibitem[\protect\citeauthoryear{Oakhill et~al.}{2011}]{oakhill2011ampk}
\begin{barticle}
\bauthor{\bsnm{Oakhill}, \binits{J.S.}},
\bauthor{\bsnm{Steel}, \binits{R.}},
\bauthor{\bsnm{Chen}, \binits{Z.-P.}},
\bauthor{\bsnm{Scott}, \binits{J.W.}},
\bauthor{\bsnm{Ling}, \binits{N.}},
\bauthor{\bsnm{Tam}, \binits{S.}},
\bauthor{\bsnm{Kemp}, \binits{B.E.}}:
\batitle{Ampk is a direct adenylate charge-regulated protein kinase}.
\bjtitle{Science}
\bvolume{332}(\bissue{6036}),
\bfpage{1433}--\blpage{1435}
(\byear{2011})
\end{barticle}
\endbibitem

\bibitem[\protect\citeauthoryear{Qutub and Popel}{2008}]{qutub2008reactive}
\begin{botherref}
\oauthor{\bsnm{Qutub}, \binits{A.A.}},
\oauthor{\bsnm{Popel}, \binits{A.S.}}:
Reactive oxygen species regulate hypoxia-inducible factor 1$\alpha$ differentially in cancer and ischemia.
Molecular and cellular biology
(2008)
\end{botherref}
\endbibitem

\bibitem[\protect\citeauthoryear{Rodrigues et~al.}{2015}]{rodrigues2015dichloroacetate}
\begin{barticle}
\bauthor{\bsnm{Rodrigues}, \binits{A.S.}},
\bauthor{\bsnm{Correia}, \binits{M.}},
\bauthor{\bsnm{Gomes}, \binits{A.}},
\bauthor{\bsnm{Pereira}, \binits{S.L.}},
\bauthor{\bsnm{Perestrelo}, \binits{T.}},
\bauthor{\bsnm{Sousa}, \binits{M.I.}},
\bauthor{\bsnm{Ramalho-Santos}, \binits{J.}}:
\batitle{Dichloroacetate, the pyruvate dehydrogenase complex and the modulation of mesc pluripotency}.
\bjtitle{PLoS One}
\bvolume{10}(\bissue{7}),
\bfpage{0131663}
(\byear{2015})
\end{barticle}
\endbibitem

\bibitem[\protect\citeauthoryear{Rosner and Hengstschl{\"a}ger}{2008}]{rosner2008cytoplasmic}
\begin{barticle}
\bauthor{\bsnm{Rosner}, \binits{M.}},
\bauthor{\bsnm{Hengstschl{\"a}ger}, \binits{M.}}:
\batitle{Cytoplasmic and nuclear distribution of the protein complexes mtorc1 and mtorc2: rapamycin triggers dephosphorylation and delocalization of the mtorc2 components rictor and sin1}.
\bjtitle{Human molecular genetics}
\bvolume{17}(\bissue{19}),
\bfpage{2934}--\blpage{2948}
(\byear{2008})
\end{barticle}
\endbibitem

\bibitem[\protect\citeauthoryear{Rahman and Hasan}{2015}]{rahman2015cancer}
\begin{barticle}
\bauthor{\bsnm{Rahman}, \binits{M.}},
\bauthor{\bsnm{Hasan}, \binits{M.R.}}:
\batitle{Cancer metabolism and drug resistance}.
\bjtitle{Metabolites}
\bvolume{5}(\bissue{4}),
\bfpage{571}--\blpage{600}
(\byear{2015})
\end{barticle}
\endbibitem

\bibitem[\protect\citeauthoryear{Ruiz-Iglesias and Ma{\~n}es}{2021}]{ruiz2021importance}
\begin{barticle}
\bauthor{\bsnm{Ruiz-Iglesias}, \binits{A.}},
\bauthor{\bsnm{Ma{\~n}es}, \binits{S.}}:
\batitle{The importance of mitochondrial pyruvate carrier in cancer cell metabolism and tumorigenesis}.
\bjtitle{Cancers}
\bvolume{13}(\bissue{7}),
\bfpage{1488}
(\byear{2021})
\end{barticle}
\endbibitem

\bibitem[\protect\citeauthoryear{Rosenstein et~al.}{2018}]{rosenstein2018clinical}
\begin{barticle}
\bauthor{\bsnm{Rosenstein}, \binits{P.G.}},
\bauthor{\bsnm{Tennent-Brown}, \binits{B.S.}},
\bauthor{\bsnm{Hughes}, \binits{D.}}:
\batitle{Clinical use of plasma lactate concentration. part 1: Physiology, pathophysiology, and measurement}.
\bjtitle{Journal of Veterinary Emergency and Critical Care}
\bvolume{28}(\bissue{2}),
\bfpage{85}--\blpage{105}
(\byear{2018})
\end{barticle}
\endbibitem

\bibitem[\protect\citeauthoryear{Reid et~al.}{1977}]{reid1977pyruvate}
\begin{barticle}
\bauthor{\bsnm{Reid}, \binits{E.E.}},
\bauthor{\bsnm{Thompson}, \binits{P.}},
\bauthor{\bsnm{Lyttle}, \binits{C.R.}},
\bauthor{\bsnm{Dennis}, \binits{D.T.}}:
\batitle{Pyruvate dehydrogenase complex from higher plant mitochondria and proplastids}.
\bjtitle{Plant Physiology}
\bvolume{59}(\bissue{5}),
\bfpage{842}--\blpage{848}
(\byear{1977})
\end{barticle}
\endbibitem

\bibitem[\protect\citeauthoryear{Reckzeh and Waldmann}{2020}]{reckzeh2020small}
\begin{barticle}
\bauthor{\bsnm{Reckzeh}, \binits{E.S.}},
\bauthor{\bsnm{Waldmann}, \binits{H.}}:
\batitle{Small-molecule inhibition of glucose transporters glut-1--4}.
\bjtitle{Chembiochem}
\bvolume{21}(\bissue{1-2}),
\bfpage{45}--\blpage{52}
(\byear{2020})
\end{barticle}
\endbibitem

\bibitem[\protect\citeauthoryear{Schwartzenberg-Bar-Yoseph et~al.}{2004}]{schwartzenberg2004tumor}
\begin{barticle}
\bauthor{\bsnm{Schwartzenberg-Bar-Yoseph}, \binits{F.}},
\bauthor{\bsnm{Armoni}, \binits{M.}},
\bauthor{\bsnm{Karnieli}, \binits{E.}}:
\batitle{The tumor suppressor p53 down-regulates glucose transporters glut1 and glut4 gene expression}.
\bjtitle{Cancer research}
\bvolume{64}(\bissue{7}),
\bfpage{2627}--\blpage{2633}
(\byear{2004})
\end{barticle}
\endbibitem

\bibitem[\protect\citeauthoryear{Sun et~al.}{2012}]{sun2012amino}
\begin{barticle}
\bauthor{\bsnm{Sun}, \binits{Z.}},
\bauthor{\bsnm{Do}, \binits{P.M.}},
\bauthor{\bsnm{Rhee}, \binits{M.S.}},
\bauthor{\bsnm{Govindasamy}, \binits{L.}},
\bauthor{\bsnm{Wang}, \binits{Q.}},
\bauthor{\bsnm{Ingram}, \binits{L.O.}},
\bauthor{\bsnm{Shanmugam}, \binits{K.}}:
\batitle{Amino acid substitutions at glutamate-354 in dihydrolipoamide dehydrogenase of escherichia coli lower the sensitivity of pyruvate dehydrogenase to nadh}.
\bjtitle{Microbiology}
\bvolume{158}(\bissue{5}),
\bfpage{1350}--\blpage{1358}
(\byear{2012})
\end{barticle}
\endbibitem

\bibitem[\protect\citeauthoryear{Sanli et~al.}{2012}]{sanli2012sestrin2}
\begin{barticle}
\bauthor{\bsnm{Sanli}, \binits{T.}},
\bauthor{\bsnm{Linher-Melville}, \binits{K.}},
\bauthor{\bsnm{Tsakiridis}, \binits{T.}},
\bauthor{\bsnm{Singh}, \binits{G.}}:
\batitle{Sestrin2 modulates ampk subunit expression and its response to ionizing radiation in breast cancer cells}.
\bjtitle{PloS one}
\bvolume{7}(\bissue{2}),
\bfpage{32035}
(\byear{2012})
\end{barticle}
\endbibitem

\bibitem[\protect\citeauthoryear{Simabuco et~al.}{2018}]{simabuco2018p53}
\begin{barticle}
\bauthor{\bsnm{Simabuco}, \binits{F.M.}},
\bauthor{\bsnm{Morale}, \binits{M.G.}},
\bauthor{\bsnm{Pavan}, \binits{I.C.}},
\bauthor{\bsnm{Morelli}, \binits{A.P.}},
\bauthor{\bsnm{Silva}, \binits{F.R.}},
\bauthor{\bsnm{Tamura}, \binits{R.E.}}:
\batitle{p53 and metabolism: from mechanism to therapeutics}.
\bjtitle{Oncotarget}
\bvolume{9}(\bissue{34}),
\bfpage{23780}
(\byear{2018})
\end{barticle}
\endbibitem

\bibitem[\protect\citeauthoryear{Stambolic et~al.}{2001}]{stambolic2001regulation}
\begin{barticle}
\bauthor{\bsnm{Stambolic}, \binits{V.}},
\bauthor{\bsnm{MacPherson}, \binits{D.}},
\bauthor{\bsnm{Sas}, \binits{D.}},
\bauthor{\bsnm{Lin}, \binits{Y.}},
\bauthor{\bsnm{Snow}, \binits{B.}},
\bauthor{\bsnm{Jang}, \binits{Y.}},
\bauthor{\bsnm{Benchimol}, \binits{S.}},
\bauthor{\bsnm{Mak}, \binits{T.}}:
\batitle{Regulation of pten transcription by p53}.
\bjtitle{Molecular cell}
\bvolume{8}(\bissue{2}),
\bfpage{317}--\blpage{325}
(\byear{2001})
\end{barticle}
\endbibitem

\bibitem[\protect\citeauthoryear{Sargeant and P{\^a}quet}{1993}]{sargeant1993effect}
\begin{barticle}
\bauthor{\bsnm{Sargeant}, \binits{R.J.}},
\bauthor{\bsnm{P{\^a}quet}, \binits{M.R.}}:
\batitle{Effect of insulin on the rates of synthesis and degradation of glut1 and glut4 glucose transporters in 3t3-l1 adipocytes}.
\bjtitle{Biochemical Journal}
\bvolume{290}(\bissue{3}),
\bfpage{913}--\blpage{919}
(\byear{1993})
\end{barticle}
\endbibitem

\bibitem[\protect\citeauthoryear{Sun}{2015}]{sun2015general}
\begin{barticle}
\bauthor{\bsnm{Sun}, \binits{Z.}}:
\batitle{The general information of the tumor suppressor gene p53 and the protein p53}.
\bjtitle{Journal of Cancer Prevention \& Current Research}
\bvolume{3}(\bissue{1}),
\bfpage{1}--\blpage{13}
(\byear{2015})
\end{barticle}
\endbibitem

\bibitem[\protect\citeauthoryear{Szablewski}{2013}]{szablewski2013expression}
\begin{barticle}
\bauthor{\bsnm{Szablewski}, \binits{L.}}:
\batitle{Expression of glucose transporters in cancers}.
\bjtitle{Biochimica et Biophysica Acta (BBA)-Reviews on Cancer}
\bvolume{1835}(\bissue{2}),
\bfpage{164}--\blpage{169}
(\byear{2013})
\end{barticle}
\endbibitem

\bibitem[\protect\citeauthoryear{Schmid et~al.}{2004}]{schmid2004p300}
\begin{barticle}
\bauthor{\bsnm{Schmid}, \binits{T.}},
\bauthor{\bsnm{Zhou}, \binits{J.}},
\bauthor{\bsnm{K{\"o}hl}, \binits{R.}},
\bauthor{\bsnm{Br{\"u}ne}, \binits{B.}}:
\batitle{p300 relieves p53-evoked transcriptional repression of hypoxia-inducible factor-1 (hif-1)}.
\bjtitle{Biochemical Journal}
\bvolume{380}(\bissue{1}),
\bfpage{289}--\blpage{295}
(\byear{2004})
\end{barticle}
\endbibitem

\bibitem[\protect\citeauthoryear{Tang et~al.}{2021}]{tang2021structure}
\begin{barticle}
\bauthor{\bsnm{Tang}, \binits{J.}},
\bauthor{\bsnm{Chen}, \binits{L.}},
\bauthor{\bsnm{Qin}, \binits{Z.-h.}},
\bauthor{\bsnm{Sheng}, \binits{R.}}:
\batitle{Structure, regulation, and biological functions of tigar and its role in diseases}.
\bjtitle{Acta Pharmacologica Sinica}
\bvolume{42}(\bissue{10}),
\bfpage{1547}--\blpage{1555}
(\byear{2021})
\end{barticle}
\endbibitem

\bibitem[\protect\citeauthoryear{Treins et~al.}{2005}]{treins2005regulation}
\begin{barticle}
\bauthor{\bsnm{Treins}, \binits{C.}},
\bauthor{\bsnm{Giorgetti-Peraldi}, \binits{S.}},
\bauthor{\bsnm{Murdaca}, \binits{J.}},
\bauthor{\bsnm{Monthouel-Kartmann}, \binits{M.-N.}},
\bauthor{\bsnm{Van~Obberghen}, \binits{E.}}:
\batitle{Regulation of hypoxia-inducible factor (hif)-1 activity and expression of hif hydroxylases in response to insulin-like growth factor i}.
\bjtitle{Molecular endocrinology}
\bvolume{19}(\bissue{5}),
\bfpage{1304}--\blpage{1317}
(\byear{2005})
\end{barticle}
\endbibitem

\bibitem[\protect\citeauthoryear{Tian et~al.}{2017}]{tian2017modeling}
\begin{barticle}
\bauthor{\bsnm{Tian}, \binits{X.}},
\bauthor{\bsnm{Huang}, \binits{B.}},
\bauthor{\bsnm{Zhang}, \binits{X.-P.}},
\bauthor{\bsnm{Lu}, \binits{M.}},
\bauthor{\bsnm{Liu}, \binits{F.}},
\bauthor{\bsnm{Onuchic}, \binits{J.N.}},
\bauthor{\bsnm{Wang}, \binits{W.}}:
\batitle{Modeling the response of a tumor-suppressive network to mitogenic and oncogenic signals}.
\bjtitle{Proceedings of the National Academy of Sciences}
\bvolume{114}(\bissue{21}),
\bfpage{5337}--\blpage{5342}
(\byear{2017})
\end{barticle}
\endbibitem

\bibitem[\protect\citeauthoryear{Toews}{1966}]{toews1966kinetic}
\begin{barticle}
\bauthor{\bsnm{Toews}, \binits{C.}}:
\batitle{Kinetic studies with skeletal-muscle hexokinase}.
\bjtitle{Biochemical Journal}
\bvolume{100}(\bissue{3}),
\bfpage{739}--\blpage{744}
(\byear{1966})
\end{barticle}
\endbibitem

\bibitem[\protect\citeauthoryear{Talaiezadeh et~al.}{2015}]{talaiezadeh2015kinetic}
\begin{barticle}
\bauthor{\bsnm{Talaiezadeh}, \binits{A.}},
\bauthor{\bsnm{Shahriari}, \binits{A.}},
\bauthor{\bsnm{Tabandeh}, \binits{M.R.}},
\bauthor{\bsnm{Fathizadeh}, \binits{P.}},
\bauthor{\bsnm{Mansouri}, \binits{S.}}:
\batitle{Kinetic characterization of lactate dehydrogenase in normal and malignant human breast tissues}.
\bjtitle{Cancer cell international}
\bvolume{15}(\bissue{1}),
\bfpage{1}--\blpage{9}
(\byear{2015})
\end{barticle}
\endbibitem

\bibitem[\protect\citeauthoryear{Usvalampi et~al.}{2021}]{usvalampi2021production}
\begin{barticle}
\bauthor{\bsnm{Usvalampi}, \binits{A.}},
\bauthor{\bsnm{Li}, \binits{H.}},
\bauthor{\bsnm{Frey}, \binits{A.D.}}:
\batitle{Production of glucose 6-phosphate from a cellulosic feedstock in a one pot multi-enzyme synthesis}.
\bjtitle{Frontiers in Bioengineering and Biotechnology}
\bvolume{9},
\bfpage{678038}
(\byear{2021})
\end{barticle}
\endbibitem

\bibitem[\protect\citeauthoryear{Vara et~al.}{2004}]{vara2004pi3k}
\begin{barticle}
\bauthor{\bsnm{Vara}, \binits{J.{\'A}.F.}},
\bauthor{\bsnm{Casado}, \binits{E.}},
\bauthor{\bsnm{Castro}, \binits{J.}},
\bauthor{\bsnm{Cejas}, \binits{P.}},
\bauthor{\bsnm{Belda-Iniesta}, \binits{C.}},
\bauthor{\bsnm{Gonz{\'a}lez-Bar{\'o}n}, \binits{M.}}:
\batitle{Pi3k/akt signalling pathway and cancer}.
\bjtitle{Cancer treatment reviews}
\bvolume{30}(\bissue{2}),
\bfpage{193}--\blpage{204}
(\byear{2004})
\end{barticle}
\endbibitem

\bibitem[\protect\citeauthoryear{Valvona et~al.}{2016}]{valvona2016regulation}
\begin{barticle}
\bauthor{\bsnm{Valvona}, \binits{C.J.}},
\bauthor{\bsnm{Fillmore}, \binits{H.L.}},
\bauthor{\bsnm{Nunn}, \binits{P.B.}},
\bauthor{\bsnm{Pilkington}, \binits{G.J.}}:
\batitle{The regulation and function of lactate dehydrogenase a: therapeutic potential in brain tumor}.
\bjtitle{Brain pathology}
\bvolume{26}(\bissue{1}),
\bfpage{3}--\blpage{17}
(\byear{2016})
\end{barticle}
\endbibitem

\bibitem[\protect\citeauthoryear{Vousden and Ryan}{2009}]{vousden2009p53}
\begin{barticle}
\bauthor{\bsnm{Vousden}, \binits{K.H.}},
\bauthor{\bsnm{Ryan}, \binits{K.M.}}:
\batitle{p53 and metabolism}.
\bjtitle{Nature Reviews Cancer}
\bvolume{9}(\bissue{10}),
\bfpage{691}--\blpage{700}
(\byear{2009})
\end{barticle}
\endbibitem

\bibitem[\protect\citeauthoryear{Wee and Aguda}{2006}]{wee2006akt}
\begin{barticle}
\bauthor{\bsnm{Wee}, \binits{K.B.}},
\bauthor{\bsnm{Aguda}, \binits{B.D.}}:
\batitle{Akt versus p53 in a network of oncogenes and tumor suppressor genes regulating cell survival and death}.
\bjtitle{Biophysical journal}
\bvolume{91}(\bissue{3}),
\bfpage{857}--\blpage{865}
(\byear{2006})
\end{barticle}
\endbibitem

\bibitem[\protect\citeauthoryear{Wanka et~al.}{2012}]{wanka2012synthesis}
\begin{barticle}
\bauthor{\bsnm{Wanka}, \binits{C.}},
\bauthor{\bsnm{Brucker}, \binits{D.P.}},
\bauthor{\bsnm{B{\"a}hr}, \binits{O.}},
\bauthor{\bsnm{Ronellenfitsch}, \binits{M.}},
\bauthor{\bsnm{Weller}, \binits{M.}},
\bauthor{\bsnm{Steinbach}, \binits{J.P.}},
\bauthor{\bsnm{Rieger}, \binits{J.}}:
\batitle{Synthesis of cytochrome c oxidase 2: a p53-dependent metabolic regulator that promotes respiratory function and protects glioma and colon cancer cells from hypoxia-induced cell death}.
\bjtitle{Oncogene}
\bvolume{31}(\bissue{33}),
\bfpage{3764}--\blpage{3776}
(\byear{2012})
\end{barticle}
\endbibitem

\bibitem[\protect\citeauthoryear{Woolbright et~al.}{2019}]{woolbright2019metabolic}
\begin{barticle}
\bauthor{\bsnm{Woolbright}, \binits{B.L.}},
\bauthor{\bsnm{Rajendran}, \binits{G.}},
\bauthor{\bsnm{Harris}, \binits{R.A.}},
\bauthor{\bsnm{Taylor~III}, \binits{J.A.}}:
\batitle{Metabolic flexibility in cancer: targeting the pyruvate dehydrogenase kinase: pyruvate dehydrogenase axis}.
\bjtitle{Molecular Cancer Therapeutics}
\bvolume{18}(\bissue{10}),
\bfpage{1673}--\blpage{1681}
(\byear{2019})
\end{barticle}
\endbibitem

\bibitem[\protect\citeauthoryear{Wang et~al.}{2021}]{wang2021pyruvate}
\begin{botherref}
\oauthor{\bsnm{Wang}, \binits{X.}},
\oauthor{\bsnm{Shen}, \binits{X.}},
\oauthor{\bsnm{Yan}, \binits{Y.}},
\oauthor{\bsnm{Li}, \binits{H.}}:
Pyruvate dehydrogenase kinases (pdks): An overview toward clinical applications.
Bioscience Reports
\textbf{41}(4)
(2021)
\end{botherref}
\endbibitem

\bibitem[\protect\citeauthoryear{Wang et~al.}{2013}]{wang2013april}
\begin{barticle}
\bauthor{\bsnm{Wang}, \binits{G.}},
\bauthor{\bsnm{Wang}, \binits{F.}},
\bauthor{\bsnm{Ding}, \binits{W.}},
\bauthor{\bsnm{Wang}, \binits{J.}},
\bauthor{\bsnm{Jing}, \binits{R.}},
\bauthor{\bsnm{Li}, \binits{H.}},
\bauthor{\bsnm{Wang}, \binits{X.}},
\bauthor{\bsnm{Wang}, \binits{Y.}},
\bauthor{\bsnm{Ju}, \binits{S.}},
\bauthor{\bsnm{Wang}, \binits{H.}}:
\batitle{April induces tumorigenesis and metastasis of colorectal cancer cells via activation of the pi3k/akt pathway}.
\bjtitle{PloS one}
\bvolume{8}(\bissue{1}),
\bfpage{55298}
(\byear{2013})
\end{barticle}
\endbibitem

\bibitem[\protect\citeauthoryear{Wood et~al.}{2007}]{wood2007hypoxia}
\begin{barticle}
\bauthor{\bsnm{Wood}, \binits{I.S.}},
\bauthor{\bsnm{Wang}, \binits{B.}},
\bauthor{\bsnm{Lorente-Cebri{\'a}n}, \binits{S.}},
\bauthor{\bsnm{Trayhurn}, \binits{P.}}:
\batitle{Hypoxia increases expression of selective facilitative glucose transporters (glut) and 2-deoxy-d-glucose uptake in human adipocytes}.
\bjtitle{Biochemical and biophysical research communications}
\bvolume{361}(\bissue{2}),
\bfpage{468}--\blpage{473}
(\byear{2007})
\end{barticle}
\endbibitem

\bibitem[\protect\citeauthoryear{Wu et~al.}{2016}]{wu2016lactic}
\begin{barticle}
\bauthor{\bsnm{Wu}, \binits{H.}},
\bauthor{\bsnm{Ying}, \binits{M.}},
\bauthor{\bsnm{Hu}, \binits{X.}}:
\batitle{Lactic acidosis switches cancer cells from aerobic glycolysis back to dominant oxidative phosphorylation}.
\bjtitle{Oncotarget}
\bvolume{7}(\bissue{26}),
\bfpage{40621}
(\byear{2016})
\end{barticle}
\endbibitem

\bibitem[\protect\citeauthoryear{Xu et~al.}{2012}]{xu2012akt}
\begin{botherref}
\oauthor{\bsnm{Xu}, \binits{N.}},
\oauthor{\bsnm{Lao}, \binits{Y.}},
\oauthor{\bsnm{Zhang}, \binits{Y.}},
\oauthor{\bsnm{Gillespie}, \binits{D.A.}}:
Akt: a double-edged sword in cell proliferation and genome stability.
Journal of oncology
\textbf{2012}
(2012)
\end{botherref}
\endbibitem

\bibitem[\protect\citeauthoryear{Xu et~al.}{2019}]{xu2019hypoxia}
\begin{barticle}
\bauthor{\bsnm{Xu}, \binits{K.}},
\bauthor{\bsnm{Zhan}, \binits{Y.}},
\bauthor{\bsnm{Yuan}, \binits{Z.}},
\bauthor{\bsnm{Qiu}, \binits{Y.}},
\bauthor{\bsnm{Wang}, \binits{H.}},
\bauthor{\bsnm{Fan}, \binits{G.}},
\bauthor{\bsnm{Wang}, \binits{J.}},
\bauthor{\bsnm{Li}, \binits{W.}},
\bauthor{\bsnm{Cao}, \binits{Y.}},
\bauthor{\bsnm{Shen}, \binits{X.}}, \betal:
\batitle{Hypoxia induces drug resistance in colorectal cancer through the hif-1$\alpha$/mir-338-5p/il-6 feedback loop}.
\bjtitle{Molecular Therapy}
\bvolume{27}(\bissue{10}),
\bfpage{1810}--\blpage{1824}
(\byear{2019})
\end{barticle}
\endbibitem

\bibitem[\protect\citeauthoryear{Yizhak et~al.}{2014}]{yizhak2014computational}
\begin{barticle}
\bauthor{\bsnm{Yizhak}, \binits{K.}},
\bauthor{\bsnm{Le~D{\'e}v{\'e}dec}, \binits{S.E.}},
\bauthor{\bsnm{Rogkoti}, \binits{V.M.}},
\bauthor{\bsnm{Baenke}, \binits{F.}},
\bauthor{\bsnm{De~Boer}, \binits{V.C.}},
\bauthor{\bsnm{Frezza}, \binits{C.}},
\bauthor{\bsnm{Schulze}, \binits{A.}},
\bauthor{\bsnm{Van De~Water}, \binits{B.}},
\bauthor{\bsnm{Ruppin}, \binits{E.}}:
\batitle{A computational study of the warburg effect identifies metabolic targets inhibiting cancer migration}.
\bjtitle{Molecular systems biology}
\bvolume{10}(\bissue{8}),
\bfpage{744}
(\byear{2014})
\end{barticle}
\endbibitem

\bibitem[\protect\citeauthoryear{Yu et~al.}{2017}]{yu2017modeling}
\begin{barticle}
\bauthor{\bsnm{Yu}, \binits{L.}},
\bauthor{\bsnm{Lu}, \binits{M.}},
\bauthor{\bsnm{Jia}, \binits{D.}},
\bauthor{\bsnm{Ma}, \binits{J.}},
\bauthor{\bsnm{Ben-Jacob}, \binits{E.}},
\bauthor{\bsnm{Levine}, \binits{H.}},
\bauthor{\bsnm{Kaipparettu}, \binits{B.A.}},
\bauthor{\bsnm{Onuchic}, \binits{J.N.}}:
\batitle{Modeling the genetic regulation of cancer metabolism: interplay between glycolysis and oxidative phosphorylation}.
\bjtitle{Cancer research}
\bvolume{77}(\bissue{7}),
\bfpage{1564}--\blpage{1574}
(\byear{2017})
\end{barticle}
\endbibitem

\bibitem[\protect\citeauthoryear{Yang and Sauve}{2021}]{yang2021assays}
\begin{botherref}
\oauthor{\bsnm{Yang}, \binits{Y.}},
\oauthor{\bsnm{Sauve}, \binits{A.A.}}:
Assays for determination of cellular and mitochondrial nad+ and nadh content.
Mitochondrial Regulation: Methods and Protocols,
271--285
(2021)
\end{botherref}
\endbibitem

\bibitem[\protect\citeauthoryear{Yang et~al.}{2009}]{yang2009e3}
\begin{barticle}
\bauthor{\bsnm{Yang}, \binits{W.-L.}},
\bauthor{\bsnm{Wang}, \binits{J.}},
\bauthor{\bsnm{Chan}, \binits{C.-H.}},
\bauthor{\bsnm{Lee}, \binits{S.-W.}},
\bauthor{\bsnm{Campos}, \binits{A.D.}},
\bauthor{\bsnm{Lamothe}, \binits{B.}},
\bauthor{\bsnm{Hur}, \binits{L.}},
\bauthor{\bsnm{Grabiner}, \binits{B.C.}},
\bauthor{\bsnm{Lin}, \binits{X.}},
\bauthor{\bsnm{Darnay}, \binits{B.G.}}, \betal:
\batitle{The e3 ligase traf6 regulates akt ubiquitination and activation}.
\bjtitle{Science}
\bvolume{325}(\bissue{5944}),
\bfpage{1134}--\blpage{1138}
(\byear{2009})
\end{barticle}
\endbibitem

\bibitem[\protect\citeauthoryear{Zhang et~al.}{2007}]{zhang2007exploring}
\begin{barticle}
\bauthor{\bsnm{Zhang}, \binits{T.}},
\bauthor{\bsnm{Brazhnik}, \binits{P.}},
\bauthor{\bsnm{Tyson}, \binits{J.J.}}:
\batitle{Exploring mechanisms of the dna-damage response: p53 pulses and their possible relevance to apoptosis}.
\bjtitle{Cell cycle}
\bvolume{6}(\bissue{1}),
\bfpage{85}--\blpage{94}
(\byear{2007})
\end{barticle}
\endbibitem

\bibitem[\protect\citeauthoryear{Zhang et~al.}{2009}]{zhang2009computational}
\begin{barticle}
\bauthor{\bsnm{Zhang}, \binits{T.}},
\bauthor{\bsnm{Brazhnik}, \binits{P.}},
\bauthor{\bsnm{Tyson}, \binits{J.J.}}:
\batitle{Computational analysis of dynamical responses to the intrinsic pathway of programmed cell death}.
\bjtitle{Biophysical journal}
\bvolume{97}(\bissue{2}),
\bfpage{415}--\blpage{434}
(\byear{2009})
\end{barticle}
\endbibitem

\bibitem[\protect\citeauthoryear{Zhong et~al.}{2000}]{zhong2000modulation}
\begin{barticle}
\bauthor{\bsnm{Zhong}, \binits{H.}},
\bauthor{\bsnm{Chiles}, \binits{K.}},
\bauthor{\bsnm{Feldser}, \binits{D.}},
\bauthor{\bsnm{Laughner}, \binits{E.}},
\bauthor{\bsnm{Hanrahan}, \binits{C.}},
\bauthor{\bsnm{Georgescu}, \binits{M.-M.}},
\bauthor{\bsnm{Simons}, \binits{J.W.}},
\bauthor{\bsnm{Semenza}, \binits{G.L.}}:
\batitle{Modulation of hypoxia-inducible factor 1$\alpha$ expression by the epidermal growth factor/phosphatidylinositol 3-kinase/pten/akt/frap pathway in human prostate cancer cells: implications for tumor angiogenesis and therapeutics}.
\bjtitle{Cancer research}
\bvolume{60}(\bissue{6}),
\bfpage{1541}--\blpage{1545}
(\byear{2000})
\end{barticle}
\endbibitem

\bibitem[\protect\citeauthoryear{Zhong et~al.}{1999}]{zhong1999overexpression}
\begin{barticle}
\bauthor{\bsnm{Zhong}, \binits{H.}},
\bauthor{\bsnm{De~Marzo}, \binits{A.M.}},
\bauthor{\bsnm{Laughner}, \binits{E.}},
\bauthor{\bsnm{Lim}, \binits{M.}},
\bauthor{\bsnm{Hilton}, \binits{D.A.}},
\bauthor{\bsnm{Zagzag}, \binits{D.}},
\bauthor{\bsnm{Buechler}, \binits{P.}},
\bauthor{\bsnm{Isaacs}, \binits{W.B.}},
\bauthor{\bsnm{Semenza}, \binits{G.L.}},
\bauthor{\bsnm{Simons}, \binits{J.W.}}:
\batitle{Overexpression of hypoxia-inducible factor 1$\alpha$ in common human cancers and their metastases}.
\bjtitle{Cancer research}
\bvolume{59}(\bissue{22}),
\bfpage{5830}--\blpage{5835}
(\byear{1999})
\end{barticle}
\endbibitem

\bibitem[\protect\citeauthoryear{Zerfaoui et~al.}{2021}]{zerfaoui2021new}
\begin{barticle}
\bauthor{\bsnm{Zerfaoui}, \binits{M.}},
\bauthor{\bsnm{Dokunmu}, \binits{T.M.}},
\bauthor{\bsnm{Toraih}, \binits{E.A.}},
\bauthor{\bsnm{Rezk}, \binits{B.M.}},
\bauthor{\bsnm{Abd~Elmageed}, \binits{Z.Y.}},
\bauthor{\bsnm{Kandil}, \binits{E.}}:
\batitle{New insights into the link between melanoma and thyroid cancer: role of nucleocytoplasmic trafficking}.
\bjtitle{Cells}
\bvolume{10}(\bissue{2}),
\bfpage{367}
(\byear{2021})
\end{barticle}
\endbibitem

\bibitem[\protect\citeauthoryear{Zhang et~al.}{2011}]{zhang2011two}
\begin{barticle}
\bauthor{\bsnm{Zhang}, \binits{X.-P.}},
\bauthor{\bsnm{Liu}, \binits{F.}},
\bauthor{\bsnm{Wang}, \binits{W.}}:
\batitle{Two-phase dynamics of p53 in the dna damage response}.
\bjtitle{Proceedings of the National Academy of Sciences}
\bvolume{108}(\bissue{22}),
\bfpage{8990}--\blpage{8995}
(\byear{2011})
\end{barticle}
\endbibitem

\bibitem[\protect\citeauthoryear{Zhang et~al.}{2010}]{zhang2010role}
\begin{barticle}
\bauthor{\bsnm{Zhang}, \binits{X.-d.}},
\bauthor{\bsnm{Qin}, \binits{Z.-h.}},
\bauthor{\bsnm{Wang}, \binits{J.}}:
\batitle{The role of p53 in cell metabolism}.
\bjtitle{Acta Pharmacologica Sinica}
\bvolume{31}(\bissue{9}),
\bfpage{1208}--\blpage{1212}
(\byear{2010})
\end{barticle}
\endbibitem

\bibitem[\protect\citeauthoryear{Zimmerman et~al.}{2011}]{zimmerman2011cellular}
\begin{botherref}
\oauthor{\bsnm{Zimmerman}, \binits{J.J.}},
\oauthor{\bsnm{Saint Andr{\'e}-von~Arnim}, \binits{A.}},
\oauthor{\bsnm{McLaughlin}, \binits{J.}}:
Cellular respiration. in: Pediatric critical care.
Elsevier,
1058--1072
(2011)
\end{botherref}
\endbibitem

\bibitem[\protect\citeauthoryear{Zeng et~al.}{2021}]{zeng2021e3}
\begin{barticle}
\bauthor{\bsnm{Zeng}, \binits{S.}},
\bauthor{\bsnm{Zhao}, \binits{Z.}},
\bauthor{\bsnm{Zheng}, \binits{S.}},
\bauthor{\bsnm{Wu}, \binits{M.}},
\bauthor{\bsnm{Song}, \binits{X.}},
\bauthor{\bsnm{Li}, \binits{Y.}},
\bauthor{\bsnm{Zheng}, \binits{Y.}},
\bauthor{\bsnm{Liu}, \binits{B.}},
\bauthor{\bsnm{Chen}, \binits{L.}},
\bauthor{\bsnm{Gao}, \binits{C.}}, \betal:
\batitle{The e3 ubiquitin ligase trim31 is involved in cerebral ischemic injury by promoting degradation of tigar}.
\bjtitle{Redox Biology}
\bvolume{45},
\bfpage{102058}
(\byear{2021})
\end{barticle}
\endbibitem

\end{thebibliography}
\newpage
\begin{appendices}
\section{Detailed Model}\label{secA1}
\subsection{Model assumptions}
This section provides a detailed description of the cellular events and the molecule interactions considered while constructing our model.
\subsubsection{Growth factors activate PI3K/AKT pathway}
Under normal physiological conditions, the PI3K/AKT pathway activation is tightly regulated, mainly dependent on external growth signals and nutrient availability \citep{danielsen2015portrait}. This activation process is initiated when extracellular growth molecules bind to specific receptors in the cell membrane. This binding event triggers receptor activation, which subsequently activates intracellular phosphoinositide 3-kinase (PI3K) to catalyse the conversion of phosphatidylinositol 4,5-bisphosphate (PIP2) lipids into phosphatidylinositol 3,4,5-trisphosphate (PIP3). Following this, protein kinase B (AKT) undergoes phosphorylation at threonine-308 upon binding to PIP3, which results in its activation. Once activated, AKT regulates various cellular processes controlling cell survival, metabolism, and growth \citep{danielsen2015portrait,vara2004pi3k,carnero2014pten}.\\

In recent years, extensive research has demonstrated that components of the PI3K/AKT signalling pathway are frequently disrupted in human cancers, resulting in its sustained activation \citep{danielsen2015portrait,vara2004pi3k,malinowsky2014activation,wang2013april}. Colorectal cancer, in particular, has exhibited a high occurrence of PI3K/AKT pathway activation, with reports indicating its involvement in approximately 70\% of colorectal cancer cases \citep{malinowsky2014activation}. Thus, our system assumes persistent activation of the PI3K/AKT pathway in cancer cells.\\

In terms of cellular compartmentalisation, it is established that PIP2 and PIP3 are localised to the plasma membrane \citep{vara2004pi3k,carnero2014pten}, while the protein AKT predominantly resides in the cytoplasm, with some presence in the nucleus \citep{yang2009e3,lee2008nuclear}. However, the activation of AKT by PI3K primarily occurs in the cytoplasm, leading to the accumulation of activated AKT in this particular compartment. Therefore, our model assumes that all components of this pathway, including PIP2, PIP3, and AKT, function as cytoplasmic proteins. 
\subsubsection{AKT mediates mTOR activating}
Upon activation of AKT, various downstream substrates are phosphorylated. One critical effector of AKT is the mechanistic target of rapamycin (mTOR) \citep{dan2014akt,inoki2002tsc2}. The activation of mTORC1 is vital for the control of cellular processes such as cell growth and metabolism, primarily through its ability to regulate the mRNA translation \citep{dan2014akt,inoki2002tsc2,duvel2010activation}.\\

In normal conditions, the TSC1-TSC2 complex acts as an inhibitor of mTORC1 activation; however, the function of this complex is negatively regulated by AKT phosphorylation. Activated AKT phosphorylates TSC2, preventing the formation of the TSC1-TSC2 complex, which, in turn, leads to the activation of mTORC1. Once mTORC1 is activated, it directly phosphorylates the ribosomal protein S6 kinases (S6K1 and S6K2) and the eukaryotic initiation factor 4E (eIF4E)-binding proteins (4E-BP1 and 4E-BP2) controlling the initiation of cap-dependent translation \citep{dan2014akt,inoki2002tsc2,duvel2010activation}.\\

For simplicity, in our model, we assume that the mTORC1 activation is modulated by the phosphorylation of AKT in response to growth factors without explicitly incorporating the intermediate molecules involved in this process. Further, since mTORC1 is predominantly cytoplasmic \citep{rosner2008cytoplasmic}, and the upstream and downstream mTORC1 activation occurs in this compartment, our model incorporates only the cytoplasmic mTORC1.   
\subsubsection{mTOR induces the expression of HIF1$\alpha$}
Hypoxia-inducible factor 1 (HIF1) is a transcription factor consisting of two subunits, HIF1$\alpha$ and HIF1$\beta$. The $\beta$-subunit is constitutively present in the nucleus, while the $\alpha$-subunit is rarely detectable in normoxia but strikingly induced in hypoxic conditions \citep{valvona2016regulation,laughner2001her2,golias2019microenvironmental,metzen2003intracellular}. The HIF1$\alpha$ gene is continuously transcribed and translated, but its expression stays low due to rapid destruction via the ubiquitin-proteasome pathway in an oxygen-dependent manner \citep{valvona2016regulation,laughner2001her2,golias2019microenvironmental}.\\

The oxygen-dependent turnover of HIF1$\alpha$ is controlled by a family of prolyl 4-hydroxylases (PHDs) that uses oxygen as a substrate to hydroxylate HIF1$\alpha$ in its oxygen-dependent degradation domain. Prolyl hydroxylation triggers the recognition of HIF1$\alpha$ by the product of the Von Hippel-Lindau tumour suppressor gene (pVHL), which acts as an E3 ubiquitin ligase mediating the ubiquitination and subsequent degradation of HIF1$\alpha$ \citep{valvona2016regulation,golias2019microenvironmental}.\\

Upon hypoxia, when oxygen availability is low, the activity of PHDs decreases, thereby diminishing HIF1$\alpha$ recognition by pVHL and subsequent degradation. Consequently, HIF1$\alpha$ accumulates and translocates into the nucleus to form a heterodimeric complex with HIF1$\beta$. This complex binds to specific DNA sequences and activates the transcription of target genes involved in cellular responses to hypoxia \citep{valvona2016regulation,laughner2001her2}.\\

However, HIF1$\alpha$ is often stabilised in cancer cells even in nonhypoxic conditions \citep{laughner2001her2,valvona2016regulation,zhong2000modulation,zhong1999overexpression}, suggesting the involvement of other factors in its activation. Experimental investigations have revealed that the expression of the HIF1$\alpha$ subunit and its nuclear translocation can be induced under nonhypoxic conditions through the activation of growth factors-dependent signalling pathway \citep{valvona2016regulation,laughner2001her2,zhong2000modulation,treins2005regulation}. More specifically, these studies provided evidence that HIF1$\alpha$ induction in cancer is directly regulated by mTORC1 activation, which alone is sufficient to stimulate an increase in HIF1$\alpha$ protein levels through activation of cap-dependent translation \citep{duvel2010activation,laughner2001her2,hudson2002regulation,treins2005regulation}.\\

In light of these findings, our model assumes that mTORC1 activation promotes HIF1$\alpha$ protein synthesis and nuclear translocation, while the HIF1$\alpha$ oxygen-dependent degradation is unaffected in nonhypoxic conditions. Considering the complexity of the HIF1$\alpha$ oxygen-dependent degradation mechanism, we set the HIF1$\alpha$ degradation rate based on the half-life of hydroxylated HIF1$\alpha$ under normal conditions \citep{golias2019microenvironmental}, without incorporating the intricate molecular process. 
\subsubsection{HIF1 promotes glycolysis pathway}
HIF1 is a critical mediator of cellular responses, activating the transcription of target genes that regulate various processes, including angiogenesis, glycolysis, and cell survival. In our system, we focus on some of the HIF1 downstream targets that play a critical role in shifting the cell towards glycolytic metabolism. These involve glucose transporter-1 and -3 (GLUT1) and (GLUT3), respectively \citep{ancey2018glucose}, lactate dehydrogenase (LDH) \citep{valvona2016regulation}, and pyruvate dehydrogenase kinase-1 and -3 (PDK1) and (PDK3), respectively \citep{anwar2021targeting,lu2011overexpression,wang2021pyruvate,kim2006hif}.\\
\bmhead{GLUT1/3}
Glucose transporters are a group of membrane proteins that catalyse the glucose transport across the plasma membrane \citep{schwartzenberg2004tumor,szablewski2013expression,ancey2018glucose,mamun2020hypoxia}. Once inside the cell, glucose undergoes glycolysis, a process that converts a molecule of glucose into two molecules of pyruvate, generating two net ATP and two reduced nicotinamide adenine dinucleotide (NADH) molecules \cite{valvona2016regulation,golias2019microenvironmental}. Subsequently, pyruvate can either be converted into lactate in the cytoplasm, oxidising NADH back to NAD$^{+}$, or transported into the mitochondria, in the presence of oxygen, for further energy production through OXPHOS, resulting in approximately 36 ATP molecules \citep{valvona2016regulation,golias2019microenvironmental}.\\

Tumour cells are known for exhibiting accelerated metabolic rates and high glucose demand \citep{schwartzenberg2004tumor,ancey2018glucose}. Consequently, increasing GLUT expression is essential to provide heightened glucose uptake, meeting the elevated metabolic requirements in cancer \citep{schwartzenberg2004tumor,ancey2018glucose}. GLUT1 and -3, a downstream target of HIF1 \citep{ancey2018glucose}, are particularly significant in this context as their upregulation is consistently observed in many cancer types \citep{schwartzenberg2004tumor,szablewski2013expression,ancey2018glucose,mamun2020hypoxia}, and have been selected as targets to completely block glucose uptake in cancer cells \citep{reckzeh2020small}.\\

GLUT1 is the predominant isoform of glucose transporter found in nearly all types of cells \citep{szablewski2013expression,mamun2020hypoxia}, and GLUT3 exhibits the highest affinity for glucose \citep{day2013factors}. Accordingly, our study incorporates these two types, considering their central role in controlling the glucose uptake of both normal and cancer cells.\\
\bmhead{LDH}
Another important downstream target of HIF1 is LDH, a tetrameric enzyme predominantly located in the cytoplasm facilitating the conversion of pyruvate to lactate \citep{valvona2016regulation,anadon2014biomarkers}.\\

Pyruvate derived from glycolysis typically enters the mitochondria to generate ATP more efficiently. However, when HIF1 is activated, the LDH protein level increases, directing pyruvate away from the mitochondria by catalysing its conversion to lactate \citep{valvona2016regulation}. This metabolic shift allows cells to produce ATP through glycolysis, albeit in a less efficient manner, while consuming more glucose \citep{valvona2016regulation}. Therefore, the activation of HIF1 in our system is assumed to support glycolysis by inducing LDH enzyme.\\
\bmhead{PDK1/3}
Pyruvate is metabolised within the mitochondria via pyruvate dehydrogenase (PDH) activity. The PDH complex, located in the mitochondrial matrix, catalyses the oxidative decarboxylation of pyruvate to acetyl-CoA, NADH, and CO2 \citep{wang2021pyruvate,woolbright2019metabolic,rodrigues2015dichloroacetate}. Acetyl-CoA then enters the tricarboxylic acid (TCA) cycle, where it undergoes further metabolism, resulting in the eventual formation of ATP by the electron transport chain (ETC) \citep{woolbright2019metabolic}.\\

PDK family exerts significant regulatory control over PDH function through their capacity to phosphorylate its E1-$\alpha$ subunit at three different sites: Ser293, Ser300, and Ser232 \citep{wang2021pyruvate,woolbright2019metabolic,rodrigues2015dichloroacetate}. Phosphorylation of PDH at any of these sites inhibits its decarboxylation activity, disrupting pyruvate oxidation \citep{wang2021pyruvate,woolbright2019metabolic,rodrigues2015dichloroacetate}.\\

HIF1 enhances the promoter activities of two PDK family members, namely PDK1 and PDK3 \citep{anwar2021targeting,lu2011overexpression,wang2021pyruvate,kim2006hif}, commonly overexpressed in tumours \citep{jin2020hemistepsin,lu2011overexpression,anwar2021targeting}. Elevated PDK expression and subsequent PDH phosphorylation contribute to diverting pyruvate to lactate to dispose of excess pyruvate when the mitochondrial oxidative capacity is limited. Consequently, the activation of HIF1 in our model is assumed to trigger the inhibitory regulation of the PDH activity by increasing PDK1 and PDK3 expression levels, both situated within the mitochondrial matrix  \citep{wang2021pyruvate,woolbright2019metabolic,rodrigues2015dichloroacetate}.
\subsubsection{Metabolic stress activates AMPK}
AMP-activated protein kinase (AMPK) acts as a crucial energy sensor, maintaining cellular energy homeostasis \citep{hardie2011sensing,hardie2012ampk,faubert2015amp,li2015targeting}. It is activated when cellular ATP levels drop under different forms of metabolic stress, including glucose deprivation, ischemia, hypoxia, or oxidative stress \citep{hardie2011sensing,hardie2012ampk,faubert2015amp,li2015targeting}. Upon these stresses, AMPK undergoes phosphorylation at Thr-172 to restore cellular energy balance by suppressing ATP-consuming processes and promoting ATP production \citep{hardie2011sensing,hardie2012ampk,faubert2015amp,li2015targeting}.\\

Considering deregulated cellular energetics a hallmark of cancer \citep{hanahan2011hallmarks}, our model incorporates AMPK activation in cancer cells. This activation restrains cancer growth and promotes ATP production efficiency mainly by impeding mTORC1 activation via TSC2 phosphorylation and triggering p53 activation \citep{faubert2015amp,li2015targeting,inoki2003tsc2}.
\subsubsection{AMPK activates tumour suppressor gene p53}
p53 is a major tumour suppressor gene that plays a critical role in preventing the propagation of abnormal cells. It functions as a transcription factor, controlling the expression of various genes involved in cell cycle regulation, repair DNA, apoptosis, and cellular metabolism \citep{sun2015general,nag2013mdm2,matoba2006p53}. The p53 protein is continuously produced from the TP53 tumour suppressor gene \citep{sun2015general}. However, its levels are kept low by the action of murine double minute 2 (MDM2), an E3 ubiquitin ligase that promotes p53 degradation via the ubiquitin-proteasome pathway \citep{nag2013mdm2,haupt1997mdm2}.\\

Under conditions of metabolic stress, AMPK  phosphorylates p53 on Ser15, promoting its stabilisation by disrupting its binding to MDM2 \citep{jones2005amp,imamura2001cell}. Thus, p53 accumulates and translocates to the nucleus, activating the transcription of its target genes.\\

In the basal state, p53 primarily resides in the cytoplasm, where it interacts with MDM2, preventing its nuclear translocation \citep{liang2001regulation,zerfaoui2021new}. Therefore, in our model, p53 is considered a cytoplasmic protein under unstressed conditions. However, upon AMPK activation, it is stabilised and translocated into the nucleus, where it binds DNA and transcriptionally acts as a tetramer.
\subsubsection{p53 induces genes negatively regulate its activation}
p53 triggers the production of proteins such as MDM2 and wild-type p53-induced phosphatase 1 (WIP1) \citep{barak1993mdm2,batchelor2011stimulus}, which form a negative feedback loop reducing its stability \citep{nag2013mdm2,haupt1997mdm2,barak1993mdm2,batchelor2011stimulus}.\\
\bmhead{MDM2}
MDM2 is a p53-negative regulator that can be further induced by p53 activation \citep{barak1993mdm2}. It is predominantly located in the cytoplasm of unstressed cells \citep{marchenko2010stress}. However, in response to growth factors signalling, AKT phosphorylates MDM2 on Ser166 and -186, promoting its translocation from the cytoplasm into the nucleus, thereby inhibiting the transcriptional function of p53 \citep{mayo2001phosphatidylinositol,xu2012akt}.\\

In our system, we assume that p53 activation triggers increased production of the MDM2 protein, which is primarily cytoplasmic but migrates to the nucleus upon AKT phosphorylation.\\
\bmhead{WIP1}
Another target of p53 that can negatively affect its stability is WIP1. WIP1  resides exclusively in the nucleus \citep{fiscella1997wip1}, where it dephosphorylates nuclear p53, rendering p53 more susceptible to MDM2-mediated degradation \citep{batchelor2011stimulus}.\\

On the other hand, p53 undergoes nuclear export regulated by its nuclear export signals (NES). The first NES is within the tetramerization domain, masked when p53 forms a tetramer. The second NES is in the MDM2-binding domain, which can be attenuated by phosphorylating Ser15 via AMPK \citep{liang2001regulation,marchenko2010stress}. Thus, phosphorylation and tetramerization can inhibit p53 nuclear export by masking the NESs.\\

Accordingly, in the model, we assume that phosphorylated p53 forms a tetramer in the nucleus, yet WIP1-mediated dephosphorylation triggers p53 nuclear export by unmasking the NES, leading to its relocation to the cytoplasm.
\subsubsection{p53 induces genes to promote its activation, simultaneously inhibiting HIF1}
As previously discussed, the mTOR signalling pathway significantly influences HIF1 activation. Therefore, the ability of p53 to inhibit mTORC1 activation aligns with its function as a negative regulator of HIF1. This inhibition is mainly mediated by three p53 target genes: sestrin1 (SESN1), sestrin2 (SESN2), and phosphatase and tensin homolog (PTEN) \citep{budanov2008p53,sanli2012sestrin2,feng2010regulation}.\\
\bmhead{SESN1/2}
p53 induces the expression of SESN1 and SESN2 proteins \citep{budanov2008p53,sanli2012sestrin2}, which interact with the $\alpha$-subunits of AMPK, resulting in AMPK phosphorylation on Thr172. This phosphorylation activates AMPK, which subsequently inhibits mTORC1 activity \citep{budanov2008p53,sanli2012sestrin2,feng2010regulation}. Additionally, this activation forms a positive feedback loop through the phosphorylation of p53 by active AMPK, further supporting the p53 function \citep{feng2010regulation}. For simplicity, in our system, we assume that the activation of p53 directly induces AMPK activation without explicitly incorporating SESN1/2.\\
\bmhead{PTEN}
Another crucial target gene of p53 involved in regulating mTORC1 activity is PTEN \citep{stambolic2001regulation}. Upon induction by p53, PTEN promotes the degradation of PIP3 to PIP2, effectively suppressing the PI3K/AKT/mTOR pathway \citep{mayo2002pten,carnero2014pten,feng2010regulation}. The regulation of PTEN by p53 also serves as a positive feedback loop. PTEN prevents the AKT-dependent MDM2 translocation to the nucleus, boosting the nuclear p53 transcriptional activity \citep{mayo2002pten,mayo2005phosphorylation}. In the model, PTEN is considered to be in the cytoplasm, negatively regulating the PI3K/AKT/mTOR pathway \citep{iijima2004novel}.
\subsubsection{p53 suppresses glycolysis and enhances mitochondrial respiration}
p53 is a key target in cancer cells, exerting its influence on multiple levels. It inhibits the glycolytic pathway not only by disrupting HIF1 activation but also by regulating genes directly involved in glucose metabolic pathways, including GLUT1 and -3 \citep{ancey2018glucose,schwartzenberg2004tumor,kawauchi2008p53}, PDK2 \citep{anwar2021targeting,liang2020dichloroacetate}, TP53-inducible glycolysis and apoptosis regulator (TIGAR) \citep{bensaad2006tigar,lee2015p53}, and synthesis of cytochrome c oxidase 2 (SCO2) \citep{matoba2006p53}.\\ 
\bmhead{GLUT1/3}
Previously, we emphasised the significance of GLUT1 and -3 as preeminent actors in the enhanced cellular glucose uptake and the accelerated metabolism in cancer cells \citep{ancey2018glucose,reckzeh2020small}.\\

p53, as a transcription factor, can either activate or suppress the expression of specific genes. In the case of GLUT1 and -3, p53 acts as a suppressor, diminishing their protein levels \citep{ancey2018glucose,schwartzenberg2004tumor,kawauchi2008p53}. This repression exerted by p53 is assumed to reduce the baseline levels of GLUT1 and -3 in our model, inhibiting the glycolytic pathway induced by HIF1.\\
\bmhead{TIGAR}
Within the cell, glucose undergoes an irreversible conversion to glucose-6-phosphate (G6P), which can then follow either the glycolysis pathway, forming fructose-6-phosphate (F6P), or the pentose phosphate pathway (PPP) \citep{jiang2014regulation}. The decision between these pathways is influenced by TIGAR, a p53 target.\\

TIGAR curtails the activity of the enzyme guiding F6P towards the next glycolytic step. This inhibition leads to the buildup of F6P, allowing its isomerisation back to G6P and consequently diminishing glycolysis flux \citep{bensaad2006tigar,lee2015p53}. Thus, TIGAR functions as a cytoplasmic protein in our model \citep{tang2021structure}, triggered by p53 to slow down the glycolysis rate.\\
\bmhead{PDK2}
The irreversible pyruvate decarboxylation, catalysed by the PDH complex in the mitochondria, is a critical step in determining the metabolic fate of pyruvate towards OXPHOS \citep{anwar2021targeting,wang2021pyruvate,woolbright2019metabolic,rodrigues2015dichloroacetate}.\\

As mentioned earlier, HIF1 activation in cancer cells upregulates PDK, particularly PDK1 and -3, inhibiting PDH complex activity \citep{anwar2021targeting,wang2021pyruvate,woolbright2019metabolic,rodrigues2015dichloroacetate,lu2011overexpression}. However, p53 counteracts HIF1's inhibitory effect on PDH by suppressing the expression of another PDK member called PDK2 \citep{anwar2021targeting,liang2020dichloroacetate}. Consequently, the activation of p53 in our model is assumed to repress the PDK2 protein synthesis, promoting mitochondrial respiration over glycolysis.\\
\bmhead{SCO2}
Within the mitochondria, pyruvate is converted into acetyl-CoA, entering the TCA cycle to undergo a series of chemical reactions. Each round of the TCA cycle yields one energy molecule, three NADH molecules, and one reduced flavin adenine dinucleotide (FADH2) molecule \citep{martinez2020mitochondrial}. The NADH and FADH2 produced are then oxidised back into NAD$^{+}$ and FAD via ETC protein complexes (I, II, III, and IV) embedded in the inner mitochondrial membrane \citep{ahmad2018biochemistry}. Complex I and II facilitate NADH and FADH2 oxidation, respectively, and transfer the received electrons to Complex III and then to Complex IV \citep{ahmad2018biochemistry}. As electrons traverse these complexes, protons pump from the mitochondrial matrix to the intermembrane space. Each pair of electrons from NADH pumps ten protons (4 from Complex I, 4 from Complex III, and 2 from Complex IV), while FADH2 pumps only six (4 from Complex III, and 2 from Complex IV) \citep{ahmad2018biochemistry}.\\

Complex IV acts as the ultimate electron acceptor in this chain. It transfers these electrons to molecular oxygen, promoting the reduction of oxygen to water \citep{ahmad2018biochemistry}. However, electron transfer efficiency within Complex IV is highly regulated by the p53 target, SCO2. SCO2 is a mitochondrial protein \citep{maxfield2004cox17}, that is essential for the proper assembly and maturation of Complex IV, ensuring its optimal functionality. Any deficiency or malfunction of SCO2 can impair Complex IV function and its ability to efficiently consume oxygen in the final step, disrupting the smooth flow of electrons through the ETC \citep{matoba2006p53,wanka2012synthesis}.\\

The proton movement across the membrane while transferring electrons establishes an electrochemical gradient, creating a higher concentration of protons in the intermembrane space compared to the matrix. Consequently, protons flow back into the mitochondrial matrix through Complex V, ATP synthase, driving the ATP synthesis (for every four protons, one ATP is produced) \citep{ahmad2018biochemistry}. \\

Based on the above, each acetyl-CoA entering the TCA cycle in our model produces one ATP, one FADH2, and three NADH molecules \citep{martinez2020mitochondrial}. Considering that each NADH and FADH2 contributes nearly ten and six protons, respectively, and every four protons result in one ATP molecule, in our model, NADH and FADH2 are assumed to yield 2.5 and 1.5 ATP molecules, respectively \citep{ahmad2018biochemistry}. Regarding oxygen consumption, each NADH or FADH2 transfers a pair of electrons to Complex IV, which is then converted to water (H$_{2}$O), consuming 0.5 oxygen (O$_{2}$) molecules \citep{ahmad2018biochemistry}. Therefore, we assume that each NADH or FADH2 entering the ETC consumes 0.5 oxygen molecules. Finally, p53 activation in our model is assumed to maintain mitochondrial respiration by inducing the synthesis of SCO2, supporting Complex IV function and enhancing ETC activity.
\subsection{Model reactions}
The phosphorylation and dephosphorylation processes occurring at the molecular level for p53, MDM2, AMPK, PIP2/3, AKT, mTOR, and PDH—impacting their locations, functions, or activations— are represented in our model as enzyme-catalysed reactions employing monosubstrate Michaelis-Menten kinetics. Likewise, the ubiquitination process, facilitating species degradation, involving the p53 ubiquitination by MDM2, is modelled using the same kinetic framework. This modelling choice is based on the analogous catalytic roles of a protein kinase, phosphatase, or ubiquitin ligase and an enzyme in converting a substrate into a product. The general form of this reaction is expressed as:
\begin{align*}
v_{1(S)}=V_{max}\bigg(\frac{[S]}{{[S]}+K_{m}}\bigg),
\end{align*}
where $V_{max}$ denotes the maximum speed of the reaction, often regulated by the total concentration of the protein enacting the phosphorylation, dephosphorylation, or ubiquitination processes. $K_{m}$ is a Michaelis-Menten constant that represents the substrate level $[S]$ (the species undergoing phosphorylation, dephosphorylation, or ubiquitination) at which half of the maximum reaction velocity is achieved. The speed of this reaction increases linearly with the substrate $[S]$ when the substrate concentrations are low, while saturating and achieving the maximum speed for large concentrations when $[S] \gg K_{m}$.\\

p53 is a tetramer transcription factor that activates the production of proteins, including MDM2, WIP1, PTEN, SCO2, and TIGAR, while suppressing the synthesis of others, such as GLUT1/3 and PDK2. Accordingly, the production rates controlled by the tetrameric p53 in the nucleus are modelled by a Hill function with coefficient four, represented as:
\begin{align*}
v_{2(S)}^{+}=V_{max}\bigg(\frac{[S]^h}{{[S]^h}+K_{m}^h}\bigg),\quad for \; activation\\
v_{2(S)}^{-}=V_{max}\bigg(\frac{K_{m}^h}{{[S]^h}+K_{m}^h}\bigg),\quad for\; inhibition
\end{align*}
here $V_{max}$ denotes the maximum velocity, with $[S]$ representing the concentration of nuclear p53. The parameter $h$, set to four, is the Hill coefficient that determines the steepness of the Hill function, while $K_{m}$ signifies the activation/inhibition threshold constant, where the p53 influence kicks in by exceeding this threshold constant.\\

All species in our model undergoing modifications exclusively post-translation without affecting their concentration levels, encompassing processes like phosphorylation, reduction, and oxidation, are assumed to be at a steady-state point, where the total protein concentration across all forms remains constant over time. This applies to various species, such as AMPK/AMPK*, PIP2/PIP3, AKT/AKT*, mTOR/mTOR*, ADP/ATP, NAD$^+$/NADH, and FAD/FADH2.\\

The diffusivity of glucose and lactate molecules across the cell membrane, facilitated by transporters within the cellular membrane, is represented by a net flux. This flux is determined by subtracting the amount of substrate moving out of the cell from the amount moving into, depending on the substrate concentration inside and outside the cell. Both inward and outward fluxes are modelled using the Michaelis-Menten equation, accounting for the saturation process when all transporters become saturated by the substrate (glucose/lactate). The general function can be expressed as follows:
\begin{align*}
v_{3(S)}=V_{max}\bigg(\frac{[S_{out}]}{{[S_{out}]}+K_{m}}-\frac{[S_{in}]}{{[S_{in}]}+K_{m}}\bigg),
\end{align*}
in this reaction, the first and second terms describe the influx and efflux, respectively, where $[S_{in}]$, and $[S_{out}]$ denote the concentration of the substrate (glucose/lactate) in and out the cell. $K_{m}$ is the Michaelis-Menten constant, reflecting the affinity of each substrate to its transporter.\\

In the metabolic pathways, some reactions involve more than one substrate and product, such as the first and last step of the glycolysis pathway: (Glucose + ATP $\rightarrow$ G6P + ADP) and (Pyruvate + NADH $\rightarrow$ Lactate + NAD$^+$). These two reactions were classified to follow an order Bi-Bi sequential mechanism, where Hexokinase binds the ATP molecule first then Glucose, releasing G6P first and then ATP \citep{toews1966kinetic}. Similarly, the LDH enzyme binds the NADH first followed by the Pyruvate molecule to produce Lactate first and then NAD$^+$ \citep{chang1991kinetic}. Accordingly, these two reactions are represented in our model by the order Bi-Bi Michaelis–Menten equation \citep{toews1966kinetic,chang1991kinetic,kuby2019study}, which can be given by: 
\begin{align*}
v_{4(S_{1},S_{2})}=V_{max}\bigg(\frac{[S_{1}][S_{2}]}{[S_{1}][S_{2}]+K_{s1}[S_{2}]+K_{s2}[S_{1}]+K_{s1}K_{s2}}\bigg),
\end{align*}
in which $[S_{1}]$ and $[S_{2}]$ represent the first and second substrate concentrations (ATP/NADH and
Glucose/Pyruvate, respectively), whereas $K_{s1}$ and $K_{s2}$ represent the enzyme's Km values for their respective ligands.\\

On the other hand, the reaction catalysed by the PDH enzyme, involving the conversion of Pyruvate and NAD$^+$ into Acetyl-CoA and NADH, was assumed to follow a multisite ping-pong mechanism \citep{reid1977pyruvate}. This mechanism is characterised by the alternating binding of substrates and the release of products in a stepwise manner. Thus, the kinetics of this reaction is modelled in our system using ping-pong Michaelis–Menten equation \citep{reid1977pyruvate,kuby2019study}, expressed as:
\begin{align*}
v_{5(S_{1},S_{2})}=V_{max}\bigg(\frac{[S_{1}][S_{2}]}{[S_{1}][S_{2}]+K_{s1}[S_{2}]+K_{s2}[S_{1}]}\bigg),
\end{align*}
where $[S_{1}]$ and $[S_{2}]$ denote the concentrations of Pyruvate and NAD$^+$, respectively. And the parameters $K_{s1}$ and $K_{s2}$ correspond to the enzyme's affinity constants (Km values) for Pyruvate and NAD$^+$, respectively.\\

Pyruvate, a product of cytoplasmic glycolysis, is transported into the mitochondria via the mitochondrial pyruvate carrier (MPC) located in the outer mitochondrial membrane \citep{ruiz2021importance}. This pyruvate mitochondrial import is integrated into our model as a constant rate influenced negatively by mitochondrial pyruvate concentrations to prevent excessive accumulation within the mitochondrial compartment. Furthermore, cytoplasmic NADH typically travels into the mitochondria through the Malate-Aspartate Shuttle (MAS) for oxidation in the respiratory chain \citep{bhagavan2002medical}. The NADH mitochondrial import by MAS is significant for more efficient ATP production and maintaining a high NAD$^+$/NADH ratio in the cytosol \citep{bhagavan2002medical}. To streamline our model, we simplified the representation of NADH movement, considering a direct influx into the mitochondria at a constant rate subject to inhibition regulation by the mitochondrial NADH level. These mitochondrial import reactions can be generally expressed as:
\begin{align*}
v_{6(S)}=k[S_{c}]\bigg(\frac{K_{m}}{{[S_{m}]}+K_{m}}\bigg),
\end{align*}
where $k$ represents the mitochondrial import rate, and $[S_{c}]$ and $[S_{m}]$ designate the concentration of Pyruvate/NADH in the cytoplasm and mitochondria, respectively. $K_{m}$ here serves as the inhibition threshold coefficient for mitochondrial import.\\

In our system, glycolysis, the TCA cycle, and the ETC are individually modelled as one-step processes, considering the maximum velocity of each. These processes follow kinetics reactions, saturating at high levels of its respective initial substrates, G6P for glycolysis, Acetyl-CoA for TCA cycle, and NADH/FADH2 for ETC. However, as these processes involve ADP phosphorylation and NAD$^+$/FAD reduction, deficiency of these molecules can impact the overall reaction speed. Therefore, the maximal velocity is assumed to be governed by the total levels of ADP, NAD$^+$, and FAD if they are participants in the reaction. Comprehensively, the reaction can be described as:
\begin{align*}
v_{7(S)}=V_{max}\bigg(\frac{[S]}{{[S]}+K_{m}}\bigg)\bigg(\frac{[A]}{{[A]}+K_{a}}\bigg)\bigg(\frac{[N]}{{[N]}+K_{n}}\bigg)\bigg(\frac{[F]}{{[F]}+K_{f}}\bigg),
\end{align*}
where $[S]$ and $K_{m}$ represent the initial substrate concentration (G6P, Acetyl-CoA, or NADH/FADH2) and their Michaelis-Mentent constant. $[A]$, $[N]$, and $[F]$ denotes the concentration of ADP, NAD$^+$, and FAD, respectively, with $K_{a}$, $K_{n}$, and $K_{f}$ accounting for their threshold constants. Not all terms have to be involved in each reaction; only the term where its corresponding molecule participates. Moreover, owing to TIGAR's noncompetitive inhibition on the glycolysis pathway, we adjust the first term in glycolysis as follows:
\begin{align*}
\left(\frac{[S]}{{[S]\bigg(1+\frac{[I]}{K_{i}}\bigg)}+K_{m}\bigg(1+\frac{[I]}{K_{i}}\bigg)}\right)
\end{align*}
Mathematically, noncompetitive inhibition influences the maximum reaction speed ($V_{max}$) \citep{kuby2019study}, where $[I]$ represents the inhibitor concentration, TIGAR in our case, along with the inhibition threshold constant $[K_{i}]$.\\

Finally, in all metabolic reactions when the catalysing protein (whether transporter or enzyme) is part of our system, the maximum velocity is regulated by the catalysing protein concentration ($V_{max}=k_{cat}[E]_{tot}$). Except in the case of the LDH enzyme, where its influence on the reaction speed is characterised by a Hill function with a hill coefficient of four (similar to $v_{2(S)}^{+}$), considering the required tetramer formation before catalysing the reaction \citep{fan2011tyrosine,valvona2016regulation}.
\subsection{Model equations}
The constructed differential equations, alongside their corresponding chemical reactions and parameter values, are explicitly presented below.
\subsubsection{Cytoplasmic and nuclear p53 equations}
Let the concentration of cytoplasmic p53, nuclear p53, active cytoplasmic AMPK, cytoplasmic MDM2, nuclear MDM2, and nuclear WIP1 be denoted by $P53_{c}$, $P53_{n}$, $Ampk_{c}^*$, $Mdm2_{c}$, $Mdm2_{n}$, and $Wip1_{n}$, respectively.
\bigskip
\begin{flushleft}
\textbf{\underline{Chemical reactions:}}
\end{flushleft}
\begin{align*}
\bullet\quad                          &\xrightarrow{k_{1}} \quad  P53_{c}\\
P53_{c}\quad + \quad Ampk_{c}^*\quad  &\xrightarrow{k_{2}} \quad  P53_{n}\quad + \quad Ampk_{c}^*\\
P53_{c}\quad + \quad Mdm2_{c}  \quad  &\xrightarrow{k_{3}} \quad  \bullet\quad + \quad Mdm2_{c}\\
P53_{n}\quad + \quad Wip1_{n}  \quad  &\xrightarrow{k_{4}} \quad  P53_{c}\quad + \quad Wip1_{n}\\
P53_{c}\quad                          &\xrightarrow{k_{5}} \quad  \bullet\\
P53_{n}\quad + \quad Mdm2_{n} \quad   &\xrightarrow{k_{6}} \quad  \bullet\quad + \quad Mdm2_{n} 
\end{align*}
\begin{flushleft}
\textbf{\underline{Equations:}}
\end{flushleft}
\begin{align}
\frac{dP53_{c}}{dt}&= k_{1}-k_{2}Ampk_{c}^*\bigg(\frac{P53_{c}}{P53_{c}+K_{p1}}\bigg)-k_{3}Mdm2_{c}\bigg(\frac{P53_{c}}{P53_{c}+K_{p2}}\bigg)\notag\\
&+k_{4}Wip1_{n}\bigg(\frac{P53_{n}}{P53_{n}+K_{p3}}\bigg)-k_{5}P53_{c},\label{1A}\\
\frac{dP53_{n}}{dt}&=k_{2}Ampk_{c}^*\bigg(\frac{P53_{c}}{P53_{c}+K_{p1}}\bigg)-k_{6}Mdm2_{n}\bigg(\frac{P53_{n}}{P53_{n}+K_{p4}}\bigg)\notag\\
&-k_{4}Wip1_{n}\bigg(\frac{P53_{n}}{P53_{n}+K_{p3}}\bigg).\label{2A}
\end{align}
\bigskip
\begin{flushleft}
\textbf{\underline{Parameter values:}}
\end{flushleft}
\begin{table}[h]
\centering
\resizebox{\textwidth}{!}
{
\begin{tabular}{@{}llccc@{}}
\toprule
\textbf{Parameter} & \textbf{Description}& \textbf{Value} &\textbf{Unit} & \textbf{Reference} \\
\midrule
$k_{1}$  & p53$_c$ basal production rate                           & 0.2                  & $\mu$M/min & \citep{abukwaik2023interplay}\\
$k_{2}$  & AMPK$_{c}^*$-dependent p53$_c$ phosphorylation rate     & 0.35\footnotemark[1] & /min       & Assumed\\
$K_{p1}$ & M.C. of AMPK$_{c}^*$-dependent p53$_c$ phosphorylation  & 1                    & $\mu$M     & Assumed\\
$k_{3}$  & MDM2$_c$-dependent p53$_c$ degradation rate             & 0.75                 & /min       & \citep{abukwaik2023interplay}\\
$K_{p2}$ & M.C. of MDM2$_c$-dependent p53$_c$ degradation          & 0.03                 & $\mu$M     & \citep{abukwaik2023interplay}\\
$k_{4}$  & WIP1$_n$-dependent p53$_n$ dephosphorylation rate       & 0.2                  & /min       & \citep{abukwaik2023interplay}\\
$K_{p3}$ & M.C. of WIP1$_n$-dependent p53$_n$ dephosphorylation    & 0.05                 & $\mu$M     & \citep{abukwaik2023interplay}\\
$k_{5}$  & p53$_c$ basal degradation rate                          & 0.02                 & /min       & \citep{abukwaik2023interplay}\\
$k_{6}$  & MDM2$_n$-dependent p53$_n$ degradation rate             & 0.02                 & /min       & \citep{abukwaik2023interplay}\\
$K_{p4}$ & M.C. of MDM2$_n$-dependent p53$_n$ degradation          & 0.3                  & $\mu$M     & \citep{abukwaik2023interplay}\\
\botrule
\end{tabular}
}
\caption{Parameter values of Eqs. (\ref{1A}) and (\ref{2A}), where M.C. denotes the Michaelis Constant \footnotetext[1]{This parameter varies by cell type. It is set to be 0.35 in normal and wild-type cancer cells but reduced to 0 in mutated cancer cells to ensure no response of p53 in this cell type.}}
\label{tab1A}
\end{table}
\subsubsection{Cytoplasmic and nuclear MDM2 equations}
Let $Akt_{c}^*$ denote the active cytoplasmic AKT concentration.
\bigskip
\begin{flushleft}
\textbf{\underline{Chemical reactions:}}
\end{flushleft}
\begin{align*}
\bullet\quad                         &\xrightarrow{k_{7}}   \quad Mdm2_{c}\\
P53_{n}\quad                         &\xrightarrow{k_{8}}   \quad P53_{n}\quad +\quad Mdm2_{c}\\
Mdm2_{c}\quad +\quad Akt_{c}^*\quad  &\xrightarrow{k_{9}}   \quad Mdm2_{n}\quad +\quad Akt_{c}^*\\
Mdm2_{n}\quad                        &\xrightarrow{k_{10}}  \quad Mdm2_{c}\\
Mdm2_{c}\quad                        &\xrightarrow{k_{11}}  \quad \bullet\\
Mdm2_{n}\quad                        &\xrightarrow{k_{11}}  \quad \bullet
\end{align*}
\newpage
\begin{flushleft}
\textbf{\underline{Equations:}}
\end{flushleft}
\begin{align}
\frac{dMdm2_{c}}{dt}&= k_{7}+k_{8}\bigg(\frac{{P53_{n}}^h}{{P53_{n}}^h+{K_{p53}}^h}\bigg)-k_{9}Akt_{c}^*\bigg(\frac{Mdm2_{c}}{Mdm2_{c}+K_{m1}}\bigg)\notag\\
&+k_{10}\bigg(\frac{Mdm2_{n}}{Mdm2_{n}+K_{m2}}\bigg)-k_{11}Mdm2_{c},\label{3A}\\
\frac{dMdm2_{n}}{dt}&=k_{9}Akt_{c}^*\bigg(\frac{Mdm2_{c}}{Mdm2_{c}+K_{m1}}\bigg)-k_{10}\bigg(\frac{Mdm2_{n}}{Mdm2_{n}+K_{m2}}\bigg)\notag\\
&-k_{11}Mdm2_{n}.\label{4A}
\end{align}
\bigskip
\begin{flushleft}
\textbf{\underline{Parameter values:}}
\end{flushleft}
\begin{table}[h]
\centering
\resizebox{\textwidth}{!}
{
\begin{tabular}{@{}llccc@{}}
\toprule
\textbf{Parameter} & \textbf{Description}& \textbf{Value} &\textbf{Unit} & \textbf{Reference} \\
\midrule
$k_{7}$   & MDM2$_c$ basal production rate                         & 0.002  & $\mu$M/min & \citep{abukwaik2023interplay}\\
$k_{8}$   & p53$_n$-dependent MDM2$_c$ production rate             & 0.024  & $\mu$M/min & \citep{abukwaik2023interplay}\\
$K_{p53}$ & T.C. of p53$_n$ transcription activation               & 0.25   & $\mu$M     & \citep{abukwaik2023interplay}\\
$h$       & Hill coefficient of p53$_n$ transcription activation   & 4      & -          & \citep{abukwaik2023interplay}\\
$k_{9}$   & AKT$_{c}^*$-dependent MDM2$_c$ phosphorylation rate    & 10     & /min       & \citep{wee2006akt}\\
$K_{m1}$  & M.C. of AKT$_{c}^*$-dependent MDM2$_c$ phosphorylation & 0.3    & $\mu$M     & \citep{wee2006akt}\\
$k_{10}$  & MDM2$_n$ dephosphorylation rate                        & 0.2    & $\mu$M/min & \citep{wee2006akt}\\
$K_{m2}$  & M.C. of MDM2$_n$ dephosphorylation                     & 0.1    & $\mu$M     & \citep{wee2006akt}\\
$k_{11}$  & MDM2 basal degradation rate                            & 0.0028 & /min       & \citep{abukwaik2023interplay}\\
\botrule
\end{tabular}
}
\caption{Parameter values of Eqs. (\ref{3A}) and (\ref{4A}). T.C. denotes Threshold Constant, while M.C. represents Michaelis Constant}
\label{tab2A}
\end{table}
\subsubsection{WIP1, PTEN, SCO2, and TIGAR equations}
Let the concentration of the cytoplasmic PTEN, mitochondrial SCO2, and cytoplasmic TIGAR be denoted by $Pten_{c}$, $Sco2_{m}$, and $Tigar_{c}$, respectively.
\bigskip
\begin{flushleft}
\textbf{\underline{Chemical reactions:}}
\end{flushleft}
\begin{align*}
\bullet\quad         &\xrightarrow{k_{12}} \quad Wip1_{n}\\
P53_{n}\quad         &\xrightarrow{k_{13}} \quad P53_{n}\quad + \quad Wip1_{n}\\
Wip1_{n}\quad        &\xrightarrow{k_{14}} \quad \bullet\\
\bullet\quad         &\xrightarrow{k_{15}} \quad Pten_{c}\\
P53_{n}\quad         &\xrightarrow{k_{16}} \quad P53_{n} \quad + \quad Pten_{c}\\
Pten_{c}\quad        &\xrightarrow{k_{17}} \quad \bullet\\
\bullet\quad         &\xrightarrow{k_{18}} \quad Sco2_{m}\\
P53_{n}\quad         &\xrightarrow{k_{19}} \quad P53_{n} \quad + \quad Sco2_{m}\\
Sco2_{m}\quad        &\xrightarrow{k_{20}} \quad \bullet\\
\bullet\quad         &\xrightarrow{k_{21}} \quad Tigar_{c}\\
P53_{n}\quad         &\xrightarrow{k_{22}} \quad P53_{n}\quad + \quad Tigar_{c}\\
Tigar_{c}\quad       &\xrightarrow{k_{23}} \quad \bullet  
\end{align*}
\begin{flushleft}
\textbf{\underline{Equations:}}
\end{flushleft}
\begin{align}
\frac{dWip1_{n}}{dt}&= k_{12}+k_{13}\bigg(\frac{{P53_{n}}^h}{{P53_{n}}^h+{K_{p53}}^h}\bigg)-k_{14}Wip1_{n},\label{5A}\\
\frac{dPten_{c}}{dt}&= k_{15}+k_{16}\bigg(\frac{{P53_{n}}^h}{{P53_{n}}^h+{K_{p53}}^h}\bigg)-k_{17}Pten_{c},\label{6A}\\
\frac{dSco2_{m}}{dt}&= k_{18}+k_{19}\bigg(\frac{{P53_{n}}^h}{{P53_{n}}^h+{K_{p53}}^h}\bigg)-k_{20}Sco2_{m},\label{7A}\\
\frac{dTigar_{c}}{dt}&= k_{21}+k_{22}\bigg(\frac{{P53_{n}}^h}{{P53_{n}}^h+{K_{p53}}^h}\bigg)-k_{23}Tigar_{c}.\label{8A}
\end{align}
\bigskip
\begin{flushleft}
\textbf{\underline{Parameter values:}}
\end{flushleft}
\begin{table}[h]
\centering
\resizebox{\textwidth}{!}
{
\begin{tabular}{@{}llccc@{}}
\toprule
\textbf{Parameter} & \textbf{Description}& \textbf{Value} &\textbf{Unit} & \textbf{Reference} \\
\midrule
$k_{12}$ & WIP1$_{n}$ basal production rate                   & 0.002   & $\mu$M/min& \citep{abukwaik2023interplay}\\
$k_{13}$ & p53$_n$-dependent WIP1$_{n}$ production rate       & 0.09    & $\mu$M/min& \citep{abukwaik2023interplay}\\
$k_{14}$ & WIP1$_{n}$ basal degradation rate                  & 0.02    & /min      & \citep{abukwaik2023interplay}\\
$k_{15}$ & PTEN$_{c}$ basal production rate                   & 0.001   & $\mu$M/min& \citep{wee2006akt}\\
$k_{16}$ & p53$_n$-dependent PTEN$_{c}$ production rate       & 0.006   & $\mu$M/min& \citep{wee2006akt}\\
$k_{17}$ & PTEN$_{c}$ basal degradation rate                  & 0.0063  & /min      & \citep{wee2006akt}\\
$k_{18}$ & SCO2$_{m}$ basal production rate                   & 0.002   & $\mu$M/min& Like $k_{12}$\\
$k_{19}$ & p53$_n$-dependent SCO2$_{m}$ production rate       & 0.007   & $\mu$M/min& Estimated \citep{wanka2012synthesis}\\
$k_{20}$ & SCO2$_{m}$ basal degradation rate                  & 0.02    & /min      & Like $k_{14}$\\
$k_{21}$ & TIGAR$_{c}$ basal production rate                  & 0.00004 & $\mu$M/min& Estimated \citep{al2016identification}\\
$k_{22}$ & p53$_n$-dependent TIGAR$_{c}$ production rate      & 0.009   & $\mu$M/min& Estimated \citep{lee2015p53}\\
$k_{23}$ & TIGAR$_{c}$ basal degradation rate                 & 0.0012  & /min      & Estimated \citep{zeng2021e3}\\
\botrule
\end{tabular}
}
\caption{Parameter values of Eqs. (\ref{5A})-(\ref{8A})}
\label{tab3A}
\end{table}
\subsubsection{Active AMPK, PIP3, AKT, and mTOR equations}
Let the concentration of cytoplasmic PIP3 and active cytoplasmic mTOR be denoted by $Pip3_{c}$ and $Mtor_{c}^*$, respectively, while $Ampk_{c}$, $Pip2_{c}$, $Akt_{c}$ and $Mtor_{c}$ represent the concentrations for the inactive form of the corresponding variable.
\newpage
\begin{flushleft}
\textbf{\underline{Chemical reactions:}}
\end{flushleft}
\begin{align*}
Ampk_{c}\quad                          &\xrightarrow{k_{24}} \quad Ampk_{c}^*\\
P53_{n}\quad +\quad Ampk_{c}\quad      &\xrightarrow{k_{25}} \quad P53_{n}\quad+\quad Ampk_{c}^*\\
Ampk_{c}^*\quad                        &\xrightarrow{k_{26}} \quad Ampk_{c}\\
Pip2_{c}\quad                          &\xrightarrow{k_{27}} \quad Pip3_{c}\\
Pip3_{c}\quad +\quad Pten_{c}\quad     &\xrightarrow{k_{28}} \quad Pip2_{c}\quad +\quad Pten_{c}\\
Akt_{c}\quad +\quad Pip3_{c}\quad      &\xrightarrow{k_{29}} \quad Akt_{c}^*\quad +\quad Pip3_{c}\\
Akt_{c}^*\quad                         &\xrightarrow{k_{30}} \quad Akt_{c}\\
Mtor_{c}\quad +\quad Akt_{c}^*\quad    &\xrightarrow{k_{31}} \quad Mtor_{c}^*\quad +\quad Akt_{c}^*\\
Mtor_{c}^*\quad +\quad Ampk_{c}^*\quad &\xrightarrow{k_{32}} \quad Mtor_{c}\quad +\quad Ampk_{c}^*\\
Mtor_{c}^*\quad                        &\xrightarrow{k_{33}} \quad Mtor_{c}
\end{align*}
\bigskip
\begin{flushleft}
\textbf{\underline{Equations:}}
\end{flushleft}
\begin{align}
\frac{dAmpk_{c}^*}{dt}&= k_{24}\bigg(\frac{Ampk_{tot}-Ampk_{c}^*}{Ampk_{tot}-Ampk_{c}^*+K_{a1}}\bigg)+k_{25}\bigg(\frac{{P53_{n}}^h}{{P53_{n}}^h+{K_{p53}}^h}\bigg)\notag\\
&\times\bigg(\frac{Ampk_{tot}-Ampk_{c}^*}{Ampk_{tot}-Ampk_{c}^*+K_{a1}}\bigg)-k_{26}\bigg(\frac{Ampk_{c}^*}{Ampk_{c}^*+K_{a2}}\bigg),\label{9A}\\
\frac{dPip3_{c}}{dt}&= k_{27}\bigg(\frac{Pip_{tot}-Pip3_{c}}{Pip_{tot}-Pip3_{c}+K_{pip1}}\bigg)-k_{28}Pten_{c}\bigg(\frac{Pip3_{c}}{Pip3_{c}+K_{pip2}}\bigg),\label{10A}\\
\frac{dAkt_{c}^*}{dt}&= k_{29}Pip3_{c}\bigg(\frac{Akt_{tot}-Akt_{c}^*}{Akt_{tot}-Akt_{c}^*+K_{akt1}}\bigg)-k_{30}\bigg(\frac{Akt_{c}^*}{Akt_{c}^*+K_{akt2}}\bigg),\label{11A}\\
\frac{dMtor_{c}^*}{dt}&= k_{31}Akt_{c}^*\bigg(\frac{Mtor_{tot}-Mtor_{c}^*}{Mtor_{tot}-Mtor_{c}^*+K_{mtor1}}\bigg)-[k_{32}Ampk_{c}^*+k_{33}]\notag\\
&\times\bigg(\frac{Mtor_{c}^*}{Mtor_{c}^*+K_{mtor2}}\bigg).\label{12A}
\end{align}
\newpage
\begin{flushleft}
\textbf{\underline{Parameter values:}}
\end{flushleft}
\begin{table}[h]
\centering
\resizebox{\textwidth}{!}
{
\begin{tabular}{@{}llccc@{}}
\toprule
\textbf{Parameter} & \textbf{Description}& \textbf{Value} &\textbf{Unit} & \textbf{Reference} \\
\midrule
$k_{24}$      & AMPK$_c$ phosphorylation rate                            & 0.001\footnotemark[1] & $\mu$M/min& Assumed\\
$Ampk_{tot}$  & The total concentration of all AMPK$_c$ forms            & 1                     & $\mu$M    & Like $Pip_{tot}$\\
$K_{a1}$      & M.C. of AMPK$_c$ phosphorylation                         & 0.2                   & $\mu$M    & Assumed\\
$k_{25}$      & p53$_n$-dependent AMPK$_c$ phosphorylation rate          & 0.001                 & $\mu$M/min& Like $k_{24}$\\
$k_{26}$      & AMPK$_{c}^*$ dephosphorylation rate                      & 0.0001                & $\mu$M/min& Assumed\\
$K_{a2}$      & M.C. of AMPKk$_{c}^*$ dephosphorylation                  & 0.5                   & $\mu$M    & Assumed\\
$k_{27}$      & PIP2$_c$ phosphorylation rate                            & 0.15\footnotemark[1]  & $\mu$M/min& \citep{wee2006akt}\\
$Pip_{tot}$   & The total concentration of all PIP$_c$ forms             & 1                     & $\mu$M    & \citep{wee2006akt}\\
$K_{pip1}$    & M.C. of PIP2$_c$ phosphorylation                         & 0.1                   & $\mu$M    & \citep{wee2006akt}\\
$k_{28}$      & PTEN$_c$-dependent PIP3$_c$ dephosphorylation rate       & 0.5                   & /min      & \citep{zhang2011two}\\
$K_{Pip2}$    & M.C. of PTEN$_c$-dependent PIP3$_c$ dephosphorylation    & 0.5                   & $\mu$M    & \citep{wee2006akt,zhang2011two}\\
$k_{29}$      & PIP3$_c$-dependent AKT$_c$ phosphorylation rate          & 0.25                  & /min      & \citep{zhang2011two}\\
$Akt_{tot}$   & The total concentration of all AKT$_c$ forms             & 1                     & $\mu$M    & \citep{zhang2011two}\\
$K_{akt1}$    & M.C. of PIP3$_c$-dependent AKT$_c$ phosphorylation       & 0.35                  & $\mu$M    & \citep{tian2017modeling}\\
$k_{30}$      & AKT$_{c}^*$ dephosphorylation rate                       & 0.1                   & $\mu$M/min& \citep{zhang2011two}\\
$K_{akt2}$    & M.C. of AKT$_{c}^*$ dephosphorylation                    & 0.2                   & $\mu$M    & \citep{zhang2011two}\\
$k_{31}$      & AKT$_{c}^*$-dependent mTOR$_c$ activation rate           & 0.25                  & /min      & Like $k_{29}$\\
$Mtor_{tot}$  & The total concentration of all mTOR$_c$ forms            & 1                     & $\mu$M    & Like $Pip_{tot}$\\
$K_{mtor1}$   & M.C. of AKT$_{c}^*$-dependent mTOR$_{c}$ activation      & 0.1                   & $\mu$M    & Like $K_{pip1}$\\
$k_{32}$      & AMPK$_{c}^*$-dependent mTOR$_{c}^*$ inactivation rate    & 0.2                   & /min      & Assumed\\
$K_{mtor2}$   & M.C. of mTOR$_{c}^*$ inactivation                        & 0.5                   & $\mu$M    & Like $K_{pip2}$\\
$k_{33}$      & AMPK$_{c}^*$-independent mTOR$_{c}^*$ inactivation rate  & 0.00001               & $\mu$M/min& Assumed\\
\botrule
\end{tabular}
}
\caption{Parameter values of Eqs. (\ref{9A})-(\ref{12A}). M.C. represents Michaelis Constant. \footnotetext[1]{In the model, metabolic stress and continuous activation of growth factor signals are considered exclusive to cancer cells, assuming that normal cells exist in a healthy and disorder-free environment. Therefore, these parameters are set to 0 in normal cells}}
\label{tab4A}
\end{table}
\subsubsection{Cytoplasmic and nuclear HIF1 equations}
Let $Hif1\alpha_{c}$ and $Hif1\alpha_{n}$ denote the concentrations of cytoplasmic and nuclear HIF1$\alpha$, respectively.
\bigskip
\begin{flushleft}
\textbf{\underline{Chemical reactions:}}
\end{flushleft}
\begin{align*}
\bullet\quad                               &\xrightarrow{k_{34}} \quad Hif1\alpha_{c}\\
Mtor_{c}^*\quad                            &\xrightarrow{k_{35}} \quad Mtor_{c}^*\quad + \quad Hif1\alpha_{c}\\
Mtor_{c}^*\quad +\quad Hif1\alpha_{c}\quad &\xrightarrow{k_{36}} \quad Mtor_{c}^*\quad + \quad Hif1\alpha_{n}\\
Hif1\alpha_{c}\quad                        &\xrightarrow{k_{37}} \quad \bullet\\
Hif1\alpha_{n}\quad                        &\xrightarrow{k_{37}} \quad \bullet\\
\end{align*}
\begin{flushleft}
\textbf{\underline{Equations:}}
\end{flushleft}
\begin{align}
\frac{dHif1\alpha_{c}}{dt} &=k_{34}+k_{35}Mtor_{c}^*-k_{36}Mtor_{c}^*Hif1\alpha_{c}-k_{37}Hif1\alpha_{c},\label{13A}\\
\frac{dHif1\alpha_{n}}{dt} &=k_{36}Mtor_{c}^*Hif1\alpha_{c}-k_{37}Hif1\alpha_{n}.\label{14A}
\end{align}
\bigskip
\begin{flushleft}
\textbf{\underline{Parameter values:}}
\end{flushleft}
\begin{table}[h]
\centering
\resizebox{\textwidth}{!}
{
\begin{tabular}{@{}llccc@{}}
\toprule
\textbf{Parameter} & \textbf{Description}& \textbf{Value} &\textbf{Unit} & \textbf{Reference} \\
\midrule
$k_{34}$  & HIF1$\alpha_c$ basal production rate                      & 0.00002  & $\mu$M/min & Assumed\\
$k_{35}$  & mTOR$_{c}^*$-dependent HIF1$\alpha_c$ induction rate      & 0.000045 & /min       & Estimated \citep{duvel2010activation,lu2011overexpression}\\
$k_{36}$  & mTOR$_{c}^*$-dependent HIF1$\alpha_c$ nuclear import rate & 1.45     & /$\mu$Mmin & Estimated \citep{treins2005regulation}\\
$k_{37}$  & HIF1$\alpha$ basal degradation rate                       & 0.1386   & /min       & Estimated \citep{golias2019microenvironmental}\\
\botrule
\end{tabular}
}
\caption{Parameter values of Eqs. (\ref{13A}) and (\ref{14A})}
\label{tab5A}
\end{table}
\subsubsection{GLUT1, GLUT3, PDK1/3, PDK2, and LDH equations}
Let the concentration of cytoplasmic GLUT1, cytoplasmic GLUT3, mitochondrial PDK1 and 3, mitochondrial PDK2, and cytoplasmic LDH be denoted by $Glut1_{c}$, $Glut3_{c}$, $Pdk13_{m}$, $Pdk2_{m}$, and $Ldh_{c}$, respectively.
\bigskip
\begin{flushleft}
\textbf{\underline{Chemical reactions:}}
\end{flushleft}
\begin{align*}
\bullet\quad                &\xrightarrow{k_{38}}  \quad Glut1_{c}\\
Hif1\alpha_{n}\quad         &\xrightarrow{k_{39}}  \quad Glut1_{c}\quad + \quad Hif1\alpha_{n}\\
Glut1_{c}\quad              &\xrightarrow{k_{40}}  \quad \bullet\\
\bullet\quad                &\xrightarrow{k_{41}}  \quad Glut3_{c}\\
Hif1\alpha_{n}\quad         &\xrightarrow{k_{42}}  \quad Glut3_{c}\quad + \quad Hif1\alpha_{n}\\
Glut3_{c}\quad              &\xrightarrow{k_{43}}  \quad \bullet\\
\bullet\quad                &\xrightarrow{k_{44}}  \quad Pdk13_{m}\\
Hif1\alpha_{n}\quad         &\xrightarrow{k_{45}}  \quad Pdk13_{m}\quad + \quad Hif1\alpha_{n}\\
Pdk13_{m}\quad              &\xrightarrow{k_{46}}  \quad \bullet\\
\bullet\quad                &\xrightarrow{k_{47}}  \quad Pdk2_{m}\\
Pdk2_{m}\quad               &\xrightarrow{k_{46}}  \quad \bullet\\
\bullet\quad                &\xrightarrow{k_{48}}  \quad Ldh_{c}\\
Hif1\alpha_{n}\quad         &\xrightarrow{k_{49}}  \quad Ldh_{c}\quad + \quad Hif1\alpha_{n}\\
Ldh_{c}\quad                &\xrightarrow{k_{50}}  \quad \bullet
\end{align*}
\newpage
\begin{flushleft}
\textbf{\underline{Equations:}}
\end{flushleft}
\begin{align}
\frac{dGlut1_{c}}{dt} &=k_{38}\bigg(\frac{{K_{p53}}^h}{{P53_{n}}^h+{K_{p53}}^h}\bigg)+k_{39}Hif1\alpha_{n}-k_{40}Glut1_{c},\label{15A}\\
\frac{dGlut3_{c}}{dt} &=k_{41}\bigg(\frac{{K_{p53}}^h}{{P53_{n}}^h+{K_{p53}}^h}\bigg)+k_{42}Hif1\alpha_{n}-k_{43}Glut3_{c},\label{16A}\\
\frac{dPdk13_{m}}{dt} &=k_{44}+k_{45}Hif1\alpha_{n}-k_{46}Pdk13_{m},\label{17A}\\
\frac{dPdk2_{m}}{dt} &=k_{47}\bigg(\frac{{K_{p53}}^h}{{P53_{n}}^h+{K_{p53}}^h}\bigg)-k_{46}Pdk2_{m},\label{18A}\\
\frac{dLdh_{c}}{dt} &=k_{48}+k_{49}Hif1\alpha_{n}-k_{50}Ldh_{c}.\label{19A}
\end{align}
\begin{flushleft}
\textbf{\underline{Parameter values:}}
\end{flushleft}
\begin{table}[h]
\centering
\resizebox{\textwidth}{!}
{
\begin{tabular}{@{}llccc@{}}
\toprule
\textbf{Parameter} & \textbf{Description}& \textbf{Value} &\textbf{Unit} & \textbf{Reference} \\
\midrule
$k_{38}$ & GLUT1$_{c}$ basal production rate                    & 0.00005 & $\mu$M/min   & Estimated \citep{schwartzenberg2004tumor}\\
$k_{39}$ & HIF1$\alpha_n$-dependent GLUT1$_{c}$ production rate & 0.48    & /min         & Estimated \citep{duvel2010activation}\\
$k_{40}$ & GLUT1$_{c}$ basal degradation rate                   & 0.0019  & /min         & Estimated \citep{khayat1998unique}\\
$k_{41}$ & GLUT3$_{c}$ basal production rate                    & 0.00001 & $\mu$M/min   & Assumed \\
$k_{42}$ & HIF1$\alpha_n$-dependent GLUT3$_{c}$ production rate & 0.095   & /min         & Estimated \citep{wood2007hypoxia}\\
$k_{43}$ & GLUT3$_{c}$ basal degradation rate                   & 0.00075 & /min         & Estimated \citep{khayat1998unique}\\
$k_{44}$ & PDK1,3$_{m}$ basal production rate                   & 0.0002  & $\mu$M/min   & Assumed\\
$k_{45}$ & HIF1$\alpha_n$-dependent PDK1,3$_{m}$ production rate& 0.7     & /min         & Estimated \citep{lu2011overexpression}\\
$k_{46}$ & PDK$_{m}$ basal degradation rate                     & 0.0019  & /min         & Estimated \citep{crewe2017regulation,huang2002regulation}\\
$k_{47}$ & PDK2$_{m}$ basal production rate                     & 0.0001  & $\mu$M/min   & Half $k_{44}$, \citep{liang2020dichloroacetate}\\
$k_{48}$ & LDH$_{c}$ basal production rate                      & 0.0001  & $\mu$M/min   & Like $k_{47}$\\
$k_{49}$ & HIF1$\alpha_n$-dependent LDH$_{c}$ production rate   & 0.85    & /min         & Estimated \citep{hu2006differential}\\
$k_{50}$ & LDH$_{c}$ basal degradation rate                     & 0.000825& /min         & Estimated \citep{garcia2019changes}\\
\botrule
\end{tabular}
}
\caption{Parameter values of Eqs. (\ref{15A})-(\ref{19A})}
\label{tab6A}
\end{table}
\subsubsection{Active and inactive PDH equations}
Let the concentration of active and inactive mitochondrial PDH be denoted by $Pdh_{m}^*$ and $Pdh_{m}$, respectively.
\bigskip
\begin{flushleft}
\textbf{\underline{Chemical reactions:}}
\end{flushleft}
\begin{align*}
\bullet\quad                           &\xrightarrow{k_{51}} \quad Pdh_{m}^*\\
Pdh_{m}^*\quad+\quad Pdk13_{m}\quad    &\xrightarrow{k_{52}} \quad Pdh_{m}\quad+\quad Pdk13_{m}\\
Pdh_{m}^*\quad+\quad Pdk2_{m}\quad     &\xrightarrow{k_{52}} \quad Pdh_{m}\quad+\quad Pdk2_{m}\\
Pdh_{m}\quad                           &\xrightarrow{k_{53}} \quad Pdh_{m}^*\\
Pdh_{m}^*\quad                         &\xrightarrow{k_{54}} \quad \bullet\\
Pdh_{m}\quad                           &\xrightarrow{k_{54}} \quad \bullet\\
\end{align*}
\begin{flushleft}
\textbf{\underline{Equations:}}
\end{flushleft}
\begin{align}
\frac{dPdh_{m}^*}{dt} &=k_{51}-k_{52}Pdk13_{m}\bigg(\frac{{Pdh_{m}^*}}{{Pdh_{m}^*}+{K_{pdh1}}}\bigg)-k_{52}Pdk2_{m}\notag\\
&\times\bigg(\frac{{Pdh_{m}^*}}{{Pdh_{m}^*}+{K_{pdh1}}}\bigg)+k_{53}\bigg(\frac{{Pdh_{m}}}{{Pdh_{m}}+{K_{pdh2}}}\bigg)-k_{54}Pdh_{m}^*,\label{20A}\\
\frac{dPdh_{m}}{dt} &=k_{52}Pdk13_{m}\bigg(\frac{{Pdh_{m}^*}}{{Pdh_{m}^*}+{K_{pdh1}}}\bigg)+k_{52}Pdk2_{m}\bigg(\frac{{Pdh_{m}^*}}{{Pdh_{m}^*}+{K_{pdh1}}}\bigg)\notag\\
&-k_{53}\bigg(\frac{{Pdh_{m}}}{{Pdh_{m}}+{K_{pdh2}}}\bigg)-k_{54}Pdh_{m}.\label{21A}
\end{align}
\bigskip
\begin{flushleft}
\textbf{\underline{Parameter values:}}
\end{flushleft}
\begin{table}[h]
\centering
\resizebox{\textwidth}{!}
{
\begin{tabular}{@{}llccc@{}}
\toprule
\textbf{Parameter} & \textbf{Description}& \textbf{Value} &\textbf{Unit} & \textbf{Reference} \\
\midrule
$k_{51}$   & PDH$_{m}^*$ basal production rate                          & 0.001      & $\mu$M/min   & Assumed\\
$k_{52}$   & PDK$_m$-dependent PDH$_{m}^*$ phosphorylation rate         & 0.017      & /min         & Estimated \citep{liang2020dichloroacetate}\\
$K_{pdh1}$ & M.C. of PDK$_m$-dependent PDH$_{m}^*$ phosphorylation      & 0.5        & $\mu$M       & Assumed\\
$k_{53}$   & PDH$_{m}$ dephosphorylation rate                           & 0.005      & $\mu$M/min   & Estimated \citep{liang2020dichloroacetate}\\
$K_{pdh2}$ & M.C. of PDH$_{m}$ dephosphorylation                        & 0.5        & $\mu$M       & Assumed\\
$k_{54}$   & PDH$_{m}$ basal degradation rate                           & 0.00028    & /min         & Estimated \citep{hu1983induction}\\
\botrule
\end{tabular}
}
\caption{Parameter values of Eqs. (\ref{20A}) and (\ref{21A}). M.C. denotes Michaelis Constant}
\label{tab7A}
\end{table}
\subsubsection{Metabolic equations}
Let the concentration of cytoplasmic glucose, G6P, pyruvate, NADH, NAD$^{+}$, and lactate be denoted by $Glucose_{c}$, $G6p_{c}$, $Pyruvate_{c}$, $Nadh_{c}$, $Nad_{c}$, and  $Lactate_{c}$, respectively. Meanwhile, the mitochondrial concentrations of pyruvate, acetyl-CoA, NADH, NAD$^{+}$, FADH2, and FAD are represented by $Pyruvate_{m}$, $Acetyl_{m}$, $Nadh_{m}$, $Nad_{m}$, $Fadh_{m}$, and $Fad_{m}$. Additionally, $Glucose_{out}$ and $Lactate_{out}$ respectively represent the concentration of glucose and lactate outside the cell, whereas $Atp$, $Adp$, and $O2_{con}$ indicate the total concentration of ATP, ADP, and oxygen consumption, respectively.
\bigskip 
\begin{flushleft}
\textbf{\underline{Chemical reactions:}}
\end{flushleft}
\begin{align*}
Glucose_{out}+Glut1_{c}        &\xrightarrow{\xleftarrow{k_{55}}}  Glucose_{c}+Glut1_{c}\\
Glucose_{out}+Glut3_{c}        &\xrightarrow{\xleftarrow{k_{55}}}  Glucose_{c}+Glut3_{c}\\
Glucose_{c}+Atp                &\xrightarrow{k_{56}}               G6p_{c}+Adp\\
G6p_{c}+2Nad_{c}+3Adp          &\xrightarrow{k_{57}}               2Pyruvate_{c}+2Nadh_{c}+3Atp\\
G6p_{c}                        &\xrightarrow{k_{58}}               PPP\\
Pyruvate_{c}+Nadh_{c}+Ldh_{c}  &\xrightarrow{k_{59}}               Lactate_{c}+Nad_{c}+Ldh_{c}\\
Pyruvate_{c}                   &\xrightarrow{k_{60}}               Pyruvate_{m}\\
Pyruvate_{m}+Nad_{m}+Pdh_{m}^* &\xrightarrow{k_{61}}               Acetyl_{m}+Nadh_{m}+Pdh_{m}^*\\
Acetyl_{m}+3Nad_{m}+Fad_{m}+Adp&\xrightarrow{k_{62}}               3Nadh_{m}+Fadh_{m}+Atp\\
Nadh_{c}                       &\xrightarrow{k_{63}}               Nadh_{m}\\
Nadh_{m}+2.5Adp+Sco2_{m}       &\xrightarrow{k_{64}}               Nad_{m}+2.5Atp+0.5O2_{con}+Sco2_{m}\\
Fadh_{m}+1.5Adp+Sco2_{m}       &\xrightarrow{k_{64}}               Fad_{m}+1.5Atp+0.5O2_{con}+Sco2_{m}\\
Lactate_{c}                    &\xrightarrow{\xleftarrow{k_{65}}}  Lactate_{out}\\
Lactate_{out}                  &\xrightarrow{k_{66}}               \bullet\\
Atp                            &\xrightarrow{k_{67}}               Adp
\end{align*}
\begin{flushleft}
\textbf{\underline{Equations:}}
\end{flushleft}
\begin{align}
\frac{dGlucose_{c}}{dt} &=k_{55}Glut1_{c}\bigg(\frac{{Glucose_{out}}}{{Glucose_{out}}+{K_{g1}}}-\frac{{Glucose_{c}}}{{Glucose_{c}}+{K_{g1}}}\bigg)\notag\\
&+k_{55}Glut3_{c}\bigg(\frac{{Glucose_{out}}}{{Glucose_{out}}+{K_{g2}}}-\frac{{Glucose_{c}}}{{Glucose_{c}}+{K_{g2}}}\bigg)\notag\\
&-k_{56}\bigg(\frac{{Glucose_{c}}{Atp}}{{Glucose_{c}}{Atp}+{K_{atp}}{Glucose_{c}}+{K_{g3}}{Atp}+{K_{atp}}{K_{g3}}}\bigg),\label{22A}\\
\frac{dG6p_{c}}{dt} &=k_{56}\bigg(\frac{{Glucose_{c}}{Atp}}{{Glucose_{c}}{Atp}+{K_{atp}}{Glucose_{c}}+{K_{g3}}{Atp}+{K_{atp}}{K_{g3}}}\bigg)\notag\\
&-k_{57}\left(\frac{G6p_{c}}{{G6p_{c}\bigg(1+\frac{Tigar_{c}}{K_{tig}}\bigg)}+K_{g4}\bigg(1+\frac{Tigar_{c}}{K_{tig}}\bigg)}\right)\notag\\
&\times\bigg(\frac{Nc_{tot}-Nadh_{c}}{Nc_{tot}-Nadh_{c}+K_{nadc}}\bigg)\bigg(\frac{A_{tot}-Atp}{A_{tot}-Atp+K_{adp}}\bigg)\notag\\
&-k_{58}G6p_{c},\label{23A}\\
\frac{dPyruvate_{c}}{dt} &=2k_{57}\left(\frac{G6p_{c}}{{G6p_{c}\bigg(1+\frac{Tigar_{c}}{K_{tig}}\bigg)}+K_{g4}\bigg(1+\frac{Tigar_{c}}{K_{tig}}\bigg)}\right)\notag\\
&\times\bigg(\frac{Nc_{tot}-Nadh_{c}}{Nc_{tot}-Nadh_{c}+K_{nadc}}\bigg)\bigg(\frac{A_{tot}-Atp}{A_{tot}-Atp+K_{adp}}\bigg)\notag\\
&-k_{59}\bigg(\frac{{Ldh_{c}}^m}{{Ldh_{c}}^m+{K_{l}}^m}\bigg)\notag\\
&\times\bigg(\frac{{Pyruvate_{c}}{Nadh_{c}}}{{Pyruvate_{c}}{Nadh_{c}}+{K_{nc}}{Pyruvate_{c}}+{K_{pyr1}}{Nadh_{c}}+{{K_{nc}}K_{pyr1}}}\bigg)\notag\\
&-k_{60}Pyruvate_{c}\bigg(\frac{{K_{pyr2}}}{{Pyruvate_{m}}+{K_{pyr2}}}\bigg),\label{24A}\\
\frac{dPyruvate_{m}}{dt} &=k_{60}Pyruvate_{c}\bigg(\frac{{K_{pyr2}}}{{Pyruvate_{m}}+{K_{pyr2}}}\bigg)-k_{61}Pdh_{m}^*\notag\\
&\times\bigg(\frac{{Pyruvate_{m}}({Nm_{tot}-Nadh_{m}})}{({Pyruvate_{m}}+{K_{pyr3}})({Nm_{tot}-Nadh_{m}})+{K_{nadm1}}{Pyruvate_{m}}}\bigg),\label{25A}\\
\frac{dAcetyl_{m}}{dt} &=k_{61}Pdh_{m}^*\notag\\
&\times\bigg(\frac{{Pyruvate_{m}}({Nm_{tot}-Nadh_{m}})}{({Pyruvate_{m}}+{K_{pyr3}})({Nm_{tot}-Nadh_{m}})+{K_{nadm1}}{Pyruvate_{m}}}\bigg)\notag\\
&-k_{62}\bigg(\frac{{Acetyl_{m}}}{{Acetyl_{m}}+{K_{ace}}}\bigg)\bigg(\frac{Nm_{tot}-Nadh_{m}}{Nm_{tot}-Nadh_{m}+K_{nadm2}}\bigg)\notag\\
&\times\bigg(\frac{Fm_{tot}-Fadh_{m}}{Fm_{tot}-Fadh_{m}+{K_{fadm}}}\bigg)\bigg(\frac{A_{tot}-Atp}{A_{tot}-Atp+{K_{adp}}}\bigg),\label{26A}\\
\frac{dNadh_{c}}{dt} &=2k_{57}\left(\frac{G6p_{c}}{{G6p_{c}\bigg(1+\frac{Tigar_{c}}{K_{tig}}\bigg)}+K_{g4}\bigg(1+\frac{Tigar_{c}}{K_{tig}}\bigg)}\right)\notag\\
&\times\bigg(\frac{Nc_{tot}-Nadh_{c}}{Nc_{tot}-Nadh_{c}+K_{nadc}}\bigg)\bigg(\frac{A_{tot}-Atp}{A_{tot}-Atp+K_{adp}}\bigg)\notag\\
&-k_{59}\bigg(\frac{{Ldh_{c}}^m}{{Ldh_{c}}^m+{K_{l}}^m}\bigg)\notag\\
&\times\bigg(\frac{{Pyruvate_{c}}{Nadh_{c}}}{{Pyruvate_{c}}{Nadh_{c}}+{K_{nc}}{Pyruvate_{c}}+{K_{pyr1}}{Nadh_{c}}+{{K_{nc}}K_{pyr1}}}\bigg)\notag\\
&-k_{63}Nadh_{c}\bigg(\frac{{K_{nm}}}{{Nadh_{m}}+{K_{nm}}}\bigg),\label{27A}\\
\frac{dNadh_{m}}{dt} &=k_{61}Pdh_{m}^*\notag\\
&\times\bigg(\frac{{Pyruvate_{m}}({Nm_{tot}-Nadh_{m}})}{({Pyruvate_{m}}+{K_{pyr3}})({Nm_{tot}-Nadh_{m}})+{K_{nadm1}}{Pyruvate_{m}}}\bigg)\notag\\
&+3k_{62}\bigg(\frac{{Acetyl_{m}}}{{Acetyl_{m}}+{K_{ace}}}\bigg)\bigg(\frac{Nm_{tot}-Nadh_{m}}{Nm_{tot}-Nadh_{m}+K_{nadm2}}\bigg)\notag\\
&\times\bigg(\frac{Fm_{tot}-Fadh_{m}}{Fm_{tot}-Fadh_{m}+{K_{fadm}}}\bigg)\bigg(\frac{A_{tot}-Atp}{A_{tot}-Atp+{K_{adp}}}\bigg)\notag\\
&+k_{63}Nadh_{c}\bigg(\frac{{K_{nm}}}{{Nadh_{m}}+{K_{nm}}}\bigg)-k_{64}Sco2_{m}\bigg(\frac{{Nadh_{m}}}{{Nadh_{m}}+{K_{e}}}\bigg)\notag\\
&\times\bigg(\frac{A_{tot}-Atp}{A_{tot}-Atp+K_{adp}}\bigg),\label{28A}\\
\frac{dFadh_{m}}{dt} &=k_{62}\bigg(\frac{{Acetyl_{m}}}{{Acetyl_{m}}+{K_{ace}}}\bigg)\bigg(\frac{Nm_{tot}-Nadh_{m}}{Nm_{tot}-Nadh_{m}+K_{nadm2}}\bigg)\notag\\
&\times\bigg(\frac{Fm_{tot}-Fadh_{m}}{Fm_{tot}-Fadh_{m}+{K_{fadm}}}\bigg)\bigg(\frac{A_{tot}-Atp}{A_{tot}-Atp+{K_{adp}}}\bigg)\notag\\
&-k_{64}Sco2_{m}\bigg(\frac{{Fadh_{m}}}{{Fadh_{m}}+{K_{e}}}\bigg)\bigg(\frac{A_{tot}-Atp}{A_{tot}-Atp+K_{adp}}\bigg),\label{29A}\\
\frac{dLactate_{c}}{dt} &=k_{59}\bigg(\frac{{Ldh_{c}}^m}{{Ldh_{c}}^m+{K_{l}}^m}\bigg)\notag\\
&\times\bigg(\frac{{Pyruvate_{c}}{Nadh_{c}}}{{Pyruvate_{c}}{Nadh_{c}}+{K_{nc}}{Pyruvate_{c}}+{K_{pyr1}}{Nadh_{c}}+{{K_{nc}}K_{pyr1}}}\bigg)\notag\\
&-k_{65}\bigg(\frac{{Lactate_{c}}}{{Lactate_{c}}+{K_{lac}}}-\frac{{Lactate_{out}}}{{Lactate_{out}}+{K_{lac}}}\bigg),\label{30A}\\
\frac{dLactate_{out}}{dt} &=k_{65}\bigg(\frac{{Lactate_{c}}}{{Lactate_{c}}+{K_{lac}}}-\frac{{Lactate_{out}}}{{Lactate_{out}}+{K_{lac}}}\bigg)\notag\\
&-k_{66}Lactate_{out},\label{31A}\\
\frac{dAtp}{dt} &=-k_{56}\bigg(\frac{{Glucose_{c}}{Atp}}{{Glucose_{c}}{Atp}+{K_{atp}}{Glucose_{c}}+{K_{g3}}{Atp}+{K_{atp}}{K_{g3}}}\bigg)\notag\\
&+3k_{57}\left(\frac{G6p_{c}}{{G6p_{c}\bigg(1+\frac{Tigar_{c}}{K_{tig}}\bigg)}+K_{g4}\bigg(1+\frac{Tigar_{c}}{K_{tig}}\bigg)}\right)\notag\\
&\times\bigg(\frac{Nc_{tot}-Nadh_{c}}{Nc_{tot}-Nadh_{c}+K_{nadc}}\bigg)\bigg(\frac{A_{tot}-Atp}{A_{tot}-Atp+K_{adp}}\bigg)\notag\\
&+k_{62}\bigg(\frac{{Acetyl_{m}}}{{Acetyl_{m}}+{K_{ace}}}\bigg)\bigg(\frac{Nm_{tot}-Nadh_{m}}{Nm_{tot}-Nadh_{m}+K_{nadm2}}\bigg)\notag\\
&\times\bigg(\frac{Fm_{tot}-Fadh_{m}}{Fm_{tot}-Fadh_{m}+{K_{fadm}}}\bigg)\bigg(\frac{A_{tot}-Atp}{A_{tot}-Atp+{K_{adp}}}\bigg)\notag\\
&+2.5k_{64}Sco2_{m}\bigg(\frac{{Nadh_{m}}}{{Nadh_{m}}+{K_{e}}}\bigg)\bigg(\frac{A_{tot}-Atp}{A_{tot}-Atp+K_{adp}}\bigg)\notag\\
&+1.5k_{64}Sco2_{m}\bigg(\frac{{Fadh_{m}}}{{Fadh_{m}}+{K_{e}}}\bigg)\bigg(\frac{A_{tot}-Atp}{A_{tot}-Atp+K_{adp}}\bigg)\notag\\
&-k_{67}Atp,\label{32A}\\
\frac{dO2_{con}}{dt} &=0.5k_{64}Sco2_{m}\bigg(\frac{{Nadh_{m}}}{{Nadh_{m}}+{K_{e}}}\bigg)\bigg(\frac{A_{tot}-Atp}{A_{tot}-Atp+K_{adp}}\bigg)\notag\\
&+0.5k_{64}Sco2_{m}\bigg(\frac{{Fadh_{m}}}{{Fadh_{m}}+{K_{e}}}\bigg)\bigg(\frac{A_{tot}-Atp}{A_{tot}-Atp+K_{adp}}\bigg)\label{33A}.
\end{align}
\begin{flushleft}
\textbf{\underline{Parameter values:}}
\end{flushleft}
\begin{table}[h]
\centering
\resizebox{\textwidth}{!}
{
\begin{tabular}{@{}llccc@{}}
\toprule
\textbf{Parameter} & \textbf{Description}& \textbf{Value} &\textbf{Unit} & \textbf{Reference}\\
\midrule
$k_{55}$       & GLUT$_c$-dependent Glucose transport rate                                           & 61     & /min      & \citep{wanka2012synthesis,maddalena2015evaluation}\\
$Glucose_{out}$& Glucose blood concentration                                                         & 5000   & $\mu$M    & \citep{grupe1995transgenic}\\
$K_{g1}$       & M.C. of GLUT1$_c$-dependent Glucose transport                                       & 3000   & $\mu$M    & \citep{day2013factors}\\
$K_{g2}$       & M.C. of GLUT3$_c$-dependent Glucose transport                                       & 1400   & $\mu$M    & \citep{day2013factors}\\
$k_{56}$       & Maximal HK-dependent G6P$_c$ and ADP formation rate                                 & 30     & $\mu$M/min& \citep{usvalampi2021production} \\
$K_{g3}$       & M.C. of HK-dependent G6P$_c$ and ADP formation for Glucose$_c$                      & 100    & $\mu$M    & \citep{usvalampi2021production,castillo2018effect,toews1966kinetic}\\
$K_{atp}$      & M.C. of HK-dependent G6P$_c$ and ADP formation for ATP                              & 1000   & $\mu$M    & \citep{usvalampi2021production,toews1966kinetic}\\
$k_{57}$       & Maximal Glycolysis rate                                                             & 40     & $\mu$M/min& \citep{ataullakhanov2002determines}\\
$K_{g4}$       & M.C. of Glycolysis for G6P$_c$                                                      & 1000   & $\mu$M    & Assumed\\
$K_{nadc}$     & T.C. for NAD$^+_{c}$ reduction into NADH$_c$                                        & 100    & $\mu$M    & Assumed\\
$K_{adp}$      & T.C. for ADP phosphorylation into ATP                                               & 100    & $\mu$M    & Assumed\\
$Nc_{tot}$     & The total concentration of NAD$^+_{c}$ and NADH$_{c}$                               & 3000   & $\mu$M    & \citep{yang2021assays}\\
$A_{tot}$      & The total concentration of ATP and ADP                                              & 5000   & $\mu$M    & \citep{zimmerman2011cellular}\\
$K_{tig}$      & T.C of Glycolysis inhibition by TIGAR$_c$                                           & 3.5    & $\mu$M    & Assumed\\
$k_{58}$       & G6P$_c$ undergoing the PPP rate                                                     & 0.0015 & /min      & Estimated \citep{kight1995oxidation}\\
$k_{59}$       & Maximal LDH$_c$-dependent Lactate$_c$ and NAD$^+_c$ formation rate                  & 300    & $\mu$M/min& \citep{javed1997purification}\\
$K_{l}$        & T.C. of LDH$_c$-dependent Lactate$_c$ and NAD$^+_c$ formation                       & 0.3    & $\mu$M    & Assumed\\
$m$            & Hill coefficient of LDH$_c$-dependent Lactate$_c$ and NAD$^+_c$ formation           & 4      & -         & LDH is a tetrameric enzyme \citep{fan2011tyrosine,valvona2016regulation}\\
$K_{pyr1}$     & M.C. of LDH$_c$-dependent Lactate$_c$ and NAD$^+_c$ formation for Pyruvate$_c$      & 630    & $\mu$M    & \citep{talaiezadeh2015kinetic}\\
$K_{nc}$       & M.C. of LDH$_c$-dependent Lactate$_c$ and NAD$^+_c$ formation for NADH$_c$          & 330    & $\mu$M    & \citep{talaiezadeh2015kinetic}\\
$k_{60}$       & Pyruvate$_c$ mitochondrial import rate                                              & 0.01   & /min      & Estimated \citep{li2017mitochondrial}\\
$K_{pyr2}$     & T.C. of Pyruvate$_c$ mitochondrial import                                           & 5      & $\mu$M    & Assumed\\
$k_{61}$       & PDH$_{m}^*$-dependent Acetyl-CoA$_m$ and NADH$_m$ formation rate                    & 95     & /min      & Estimated \citep{javed1997purification}\\
$K_{pyr3}$     & M.C. of PDH$_{m}^*$-dependent Acetyl-CoA$_m$ and NADH$_m$ formation for Pyruvate$_m$& 430    & $\mu$M    & \citep{sun2012amino}\\
$K_{nadm1}$    & M.C. of PDH$_{m}^*$-dependent Acetyl-CoA$_m$ and NADH$_m$ formation for NAD$^+_m$   & 280    & $\mu$M    & \citep{sun2012amino}\\
$Nm_{tot}$     & The total concentration of NAD$^+_{m}$ and NADH$_{m}$                               & 80     & $\mu$M    & \citep{yang2021assays}\\
$Fm_{tot}$     & The total concentration of FAD$_{m}$ and FADH2$_{m}$                                & 10     & $\mu$M    & \citep{yang2021assays}\\
$k_{62}$       & Maximal TCA cycle rate                                                              & 4      & $\mu$M/min& Estimated \citep{ataullakhanov2002determines,devic2016warburg}\\
$K_{ace}$      & M.C. of TCA cycle for Acetyl-CoA$_m$                                                & 500    & $\mu$M    & Assumed\\
$K_{nadm2}$    & T.C. for NAD$^+_{m}$ reduction to NADH$_{m}$                                        & 10     & $\mu$M    & Assumed\\
$K_{fadm}$     & T.C. for FAD$_{m}$ reduction to FADH2$_{m}$                                         & 1      & $\mu$M    & Assumed\\
$k_{63}$       & NADH$_c$ mitochondrial import rate                                                  & 0.1    & /min      & Assumed\\
$K_{nm}$       & T.C. of NADH$_c$ mitochondrial import                                 & 1      & $\mu$M    & Assumed\\
$k_{64}$       & SCO2$_m$-dependent ETC rate                                                         & 120    & /min      & Estimated \citep{ataullakhanov2002determines,devic2016warburg}\\
$K_{e}$        & M.C. of SCO2$_m$-dependent ETC                                                      & 10     & $\mu$M     & Assumed\\
$k_{65}$       & Maximal carrier-dependent Lactate$_{c}$ transport rate                              & 8300   & $\mu$M/min & \citep{koho2008lactate}\\
$K_{lac}$      & M.C. of carrier-dependent Lactate$_{c}$ transport                                   & 1000000& $\mu$M     & Assumed\\
$k_{66}$       & Lactate$_{out}$ degradation rate                                                    & 0.01155& /min       & Estimated \citep{rosenstein2018clinical}\\
$k_{67}$       & ATP basal consumption rate                                                          & 0.0066 & /min       & Assumed\\
\botrule
\end{tabular}
}
\caption{Parameter values of Eqs. (\ref{22A})-(\ref{33A}). T.C. denotes Threshold Constant, while M.C. represents Michaelis Constant}
\label{tab8A}
\end{table}
\subsection{Discussion of Parameter Values}
This section offers a detailed explanation of the methodologies and  mechanisms employed to derive several parameters within our model.\\
\begin{flushleft}
\textbf{\textit{k}}$_{\textbf{2}}$: AMPK$_{c}^*$-dependent p53$_c$ phosphorylation rate.
\end{flushleft}
For effective gene activation in response to stimuli, p53 must maintain its active (phosphorylated) state for a certain period. Consequently, we assumed a faster phosphorylation rate for p53 than its dephosphorylation rate ($k_{4}$), ensuring that p53 remains active long enough to complete the transcription of necessary genes. Furthermore, we discussed the impact of varying this parameter on the model outcomes through bifurcation diagrams, as shown in Fig. \ref{Fig7}.\\
\begin{flushleft}
\textbf{\textit{k}}$_{\textbf{18}}$ and \textbf{\textit{k}}$_{\textbf{20}}$: SCO2$_{m}$ basal production and degradation rates, respectively.
\end{flushleft}
Due to the lack of experimental data to estimate the production and degradation rates of SCO2, we have chosen to align them with those of WIP1. However, since SCO2 exclusively impacts the ETC activity in our model, we adjusted the process speed to reflect normal activity at the SCO2 steady-state level. Therefore, any variations in SCO2 concentrations—either increases or decreases—directly affect the ETC's baseline functioning, enhancing or diminishing it.\\
\begin{flushleft}
\textbf{\textit{k}}$_{\textbf{19}}$: p53$_n$-dependent SCO2$_{m}$ production rate. 
\end{flushleft}
Experimental findings by Wanka et al. reveal that in colon cancer cells possessing wild-type p53 (HCT116 p53$^{+/+}$), SCO2 levels are approximately 2.3 times higher than in cells with mutated p53 (HCT116 p53$^{-/-}$) \citep{wanka2012synthesis}. Based on this data, we estimated that the activation of p53 in wild-type cells elevates SCO2 levels by 2.3-fold.\\
\begin{flushleft}
\textbf{\textit{k}}$_{\textbf{21}}$: TIGAR$_{c}$ basal production rate.   
\end{flushleft}
TIGAR expression was undetectable in normal colon cells \citep{al2016identification}. Accordingly, we assumed a minimal basal production rate for TIGAR, resulting in negligible levels that do not exert TIGAR influence under normal conditions.\\
\begin{flushleft}
\textbf{\textit{k}}$_{\textbf{22}}$: p53$_n$-dependent TIGAR$_{c}$ production rate.   
\end{flushleft}
We used a least squares method to estimate the induction rate of TIGAR by p53, drawing on data from Lee et al., which demonstrated how various p53 levels affect TIGAR protein concentrations \citep{lee2015p53}.\\
\begin{flushleft}
\textbf{\textit{k}}$_{\textbf{23}}$: TIGAR$_{c}$ basal degradation rate.  
\end{flushleft}
We determined the TIGAR degradation rate based on its half-life, which is approximately 10 hours, as reported in \citep{zeng2021e3}.\\
\begin{flushleft}
\textbf{\textit{k}}$_{\textbf{24}}$: AMPK$_c$ phosphorylation rate.   
\end{flushleft}
p53 activation is a critical adaptive response to metabolic stress, triggered by the activation of AMPK \citep{jones2005amp,imamura2001cell}. Thus, for effective p53 response in stressed cells, AMPK activation levels must be sufficiently high. To determine this threshold, we analysed the impact of different AMPK activation rates on the nuclear accumulation of p53, aiming to identify the activation rate required for a robust p53 response, see Fig. \ref{Fig6}.\\
\begin{flushleft}
\textbf{\textit{k}}$_{\textbf{25}}$: p53$_n$-dependent AMPK$_c$ phosphorylation rate.  
\end{flushleft}
The kinetics of AMPK phosphorylation, whether initiated by metabolic stress or p53 activation, may differ based on the triggering event, cell type, and current physiological state. Despite these variations, since both mechanisms engage similar cofactors and protein-protein interactions and target the same phosphorylation site on AMPK, we assumed that the phosphorylation speed catalysed by p53 is consistent with that induced by metabolic stress ($k_{24}$).\\
\begin{flushleft}
\textbf{\textit{k}}$_{\textbf{26}}$: AMPK$_{c}^*$ dephosphorylation rate.    
\end{flushleft}
The activation process of AMPK also involves dephosphorylation inhibitory mechanisms that guarantee AMPK stays active sufficiently to re-establish cellular energy balance \citep{oakhill2011ampk}. Therefore, we assumed the AMPK dephosphorylation rate to be ten-fold slower than its phosphorylation rate ($k_{24}$/$k_{26}$=10), enabling cells to adapt swiftly to metabolic stress and gradually return to baseline once the stress is mitigated.\\
\begin{flushleft}
\textbf{\textit{k}}$_{\textbf{31}}$: AKT$_{c}^*$-dependent mTOR$_c$ activation rate.    
\end{flushleft}
Lacking experimental data to measure mTOR activation and inactivation rates, our study proceeds under the assumption that the speed of mTOR activation by AKT is akin to that of AKT activation by PIP3 ($k_{29}$). This assumption is grounded in the observation that both steps are integral components of the PI3K/AKT/mTOR signalling pathway, known for its rapid and tightly regulated response. While the precise kinetics of mTOR activation involve different mechanisms, the need for synchronised actions within the signalling cascade suggests these key activation events occur at comparable rates. This assumption simplifies our model, enabling us to explore the broader dynamics of the PI3K/AKT/mTOR pathway without being hindered by the lack of detailed kinetic data for each step.\\
\begin{flushleft}
\textbf{\textit{k}}$_{\textbf{32}}$ and \textbf{\textit{k}}$_{\textbf{33}}$: AMPK$_{c}^*$-dependent and independent mTOR$_c$ inactivation rates, respectively.    
\end{flushleft}
In line with our assumptions, we set the mTOR inactivation rate slightly lower than its activation rate ($k_{31}$), allowing cellular responses to persist adequately for desired physiological effects. Additionally, we introduced a minimal AMPK-independent inactivation rate for mTOR, ensuring its regression after stimuli removal, even in the absence of AMPK activity. Our assumption allows effective mTOR response to cellular signals and returns to a basal state when necessary.\\
\begin{flushleft}
\textbf{\textit{k}}$_{\textbf{34}}$: HIF1$\alpha_c$ basal production rate.  
\end{flushleft}
As we do not have direct laboratory measurements to estimate the HIF1$\alpha$ production rate confidently, we infer an appropriately low rate, guided by the rapid degradation mechanisms that maintain minimal HIF1$\alpha$ levels under normoxic conditions \citep{valvona2016regulation,laughner2001her2,golias2019microenvironmental}. Then, by establishing this low steady-state level as a baseline, we accurately model the influence of HIF1$\alpha$ on its target genes, ensuring that any deviation from this baseline—under conditions that inhibit its degradation or increase its synthesis—precisely reflects the increased activity of HIF1$\alpha$ on its target genes' expression.\\
\begin{flushleft}
\textbf{\textit{k}}$_{\textbf{35}}$: mTOR$_{c}^*$-dependent HIF1$\alpha_c$ induction rate.    
\end{flushleft}
mTOR signalling is recognized for its role in boosting HIF1$\alpha$ protein levels by promoting its mRNA translation \citep{laughner2001her2, hudson2002regulation, duvel2010activation}. Activation of mTOR has been observed to elevate HIF1$\alpha$ expression by about 2.3-fold, as seen in \citep{duvel2010activation}. The same increase in HIF1$\alpha$ levels was also evident in colon cancer cells compared to normal cells \citep{lu2011overexpression}. Based on these findings, we estimated the induction rate of HIF1$\alpha$ by mTOR to reflect a 2.3-fold increase in HIF1$\alpha$ levels.\\
\begin{flushleft}
\textbf{\textit{k}}$_{\textbf{36}}$: mTOR$_{c}^*$-dependent HIF1$\alpha_c$ nuclear import rate.   
\end{flushleft}
HIF1$\alpha$ is predominantly found in the cytoplasm under nonhypoxic conditions \citep{kallio1998signal}. However, the HIF1$\alpha$ protein induced by growth factors, specifically by mTOR, has been noted to localise exclusively within the nucleus \citep{treins2005regulation}. Hence, we set this parameter to ensure exclusive HIF1$\alpha$ nuclear localisation in response to growth factor signals.\\
\begin{flushleft}
\textbf{\textit{k}}$_{\textbf{37}}$: HIF1$\alpha$ basal degradation rate.    
\end{flushleft}
Hydroxylated HIF1$\alpha$ exhibits high instability in vitro, with a half-life of less than five minutes \citep{golias2019microenvironmental}. Accordingly, we considered that the half-life of HIF1$\alpha$ under nonhypoxic conditions is five minutes, estimating the HIF1$\alpha$ degradation rate at $k_{37}=0.1386$ / min.\\
\begin{flushleft}
\textbf{\textit{k}}$_{\textbf{38}}$: GLUT1$_{c}$ basal production rate.  
\end{flushleft}
The introduction of wild-type p53 expression vectors was found to dose-dependently decrease the GLUT1 promoter activity by up to 50\% of its basal levels \citep{schwartzenberg2004tumor}. Based on this evidence, and considering that p53 activation in our cancer cells induces an average level of p53, we hypothesise a mean reduction of approximately 35\% in the GLUT1 production rate (from 0.00005 $\mu$M/min to 0.000032 $\mu$M/min) due to p53 activation in cancerous environments.\\
\begin{flushleft}
\textbf{\textit{k}}$_{\textbf{39}}$: HIF1$\alpha_n$-dependent GLUT1$_{c}$ production rate.  
\end{flushleft}
mTOR activation elevated HIF1$\alpha$ expression by 2.3-fold, which subsequently induced a 3.4-fold increase in GLUT1 concentration \citep{duvel2010activation}. From these observations, we infer that a 2.3-fold rise in HIF1$\alpha$, triggered by mTOR activation, will lead to a 3.4-fold enhancement in GLUT1 levels.\\
\begin{flushleft}
\textbf{\textit{k}}$_{\textbf{40}}$ and \textbf{\textit{k}}$_{\textbf{43}}$: GLUT1$_{c}$ and GLUT3$_{c}$ basal degradation rate, respectively.   
\end{flushleft}
The half-life of the GLUT1 and GLUT3 proteins has been reported to be around 6 and 15 hours, respectively \citep{khayat1998unique}. Thus, we calculated their degradation rates to be 0.0019 and 0.00075 / min, respectively.\\
\begin{flushleft}
\textbf{\textit{k}}$_{\textbf{41}}$: GLUT3$_{c}$ basal production rate.    
\end{flushleft}
GLUT1 is known as the most abundantly expressed glucose transporter within the GLUT family, facilitating basal glucose uptake in nearly all cell types \citep{sargeant1993effect,schwartzenberg2004tumor,dai2020glut3}. In contrast, GLUT3 expression is more selective and less common under normal conditions. Accordingly, we made the assumption that the expression of GLUT3 protein is lower than that of GLUT1, estimating this parameter to yield approximately half of GLUT1 concentration in normal cells.\\
\begin{flushleft}
\textbf{\textit{k}}$_{\textbf{42}}$: HIF1$\alpha_n$-dependent GLUT3$_{c}$ production rate.   
\end{flushleft}
Given that both GLUT1 and GLUT3 respond similarly to HIF1$\alpha$ induction under hypoxia \citep{wood2007hypoxia}, we extend this pattern to predict a 3.4-fold increase in GLUT3 levels following a 2.3-fold rise in HIF1$\alpha$, mirroring the GLUT1 response.\\
\begin{flushleft}
\textbf{\textit{k}}$_{\textbf{44}}$, \textbf{\textit{k}}$_{\textbf{47}}$, \textbf{\textit{k}}$_{\textbf{48}}$, and \textbf{\textit{k}}$_{\textbf{51}}$: PDK1,3$_{m}$, PDK2$_{m}$, LDH$_{c}$, and PDH$_{m}^*$ basal production rate, respectively. 
\end{flushleft}
Under normal physiological conditions, LDH and PDK proteins are typically maintained at modest levels, consistent with their roles in metabolic regulation under non-stressed states. Consequently, we have set a low basal production rate for them at 0.0001 $\mu$M/min to reflect the minimal activity required for metabolic homeostasis. In contrast, for effective aerobic respiration, PDH levels must significantly exceed the levels of PDK to guarantee efficient conversion of pyruvate to acetyl-CoA within the mitochondria. Thus, we estimated the PDH production rate to be ten times that of PDK, 0.001 $\mu$M/min. \\

Despite this arbitrary production rate for these enzymes, we accurately reflect their impact on cellular metabolism. These enzymes catalyse key metabolic reactions in glucose metabolism, where their activities directly influence the corresponding reaction rates. Therefore, we derived the maximum velocity (V$_{max}$) of these reactions under standard conditions from literature and then correlated these with enzymes' basal steady-state levels in our model. This approach allows us to predict how variations in enzyme levels—increases or decreases under different physiological scenarios— affect reaction speeds and metabolic outcomes, even in the absence of exact production rate data.\\
\begin{flushleft}
\textbf{\textit{k}}$_{\textbf{45}}$: HIF1$\alpha_n$-dependent PDK1,3$_{m}$ production rate.  
\end{flushleft}
PDK3 levels were found to be roughly 2-fold higher in colon cancer cells than in normal cells, aligning with a 2.3-fold enhancement in HIF1$\alpha$ levels detected in these cancer cells \citep{lu2011overexpression}. Given this correlation, we estimate a 2-fold escalation in PDK3 expression in response to the 2.3-fold rise in HIF1$\alpha$. In a similar vein, PDK1 exhibited a comparable upsurge to PDK3 under hypoxic conditions in colorectal cancer cells \citep{lu2011overexpression}, leading us to anticipate an analogous increase in PDK1 in response to HIF1$\alpha$ elevation.\\
\begin{flushleft}
\textbf{\textit{k}}$_{\textbf{46}}$: PDK$_{m}$ basal degradation rate.    
\end{flushleft}
Measuring the PDK protein stability over time in cells treated with cycloheximide revealed that PDK1 and PDK2 levels remain stable for up to two hours \citep{crewe2017regulation}. Another study indicates the mRNA half-life of PDK2 extends beyond six hours \citep{huang2002regulation}. However, due to a lack of direct data specifying the half-life of each PDK protein, for simplicity, we assume that PDK1, PDK2, and PDK3 uniformly exhibit a half-life of six hours. This assumption leads to an estimated degradation rate of 0.0019 / min.\\
\begin{flushleft}
\textbf{\textit{k}}$_{\textbf{47}}$: PDK2$_{m}$ basal production rate.    
\end{flushleft}
The expression level of PDK2 is regulated by the p53 target gene miR-149-3p. Comparative analyses between HCT116 cells, which have a high miR-149-3p level, and HCT116/F cells, exhibiting reduced miR-149-3p levels due to loss of p53 function, showed that PDK2 expression is approximately 1.4-fold higher in HCT116/F cells \citep{liang2020dichloroacetate}. Consistent with this, the p53 activation in our cancer cells effectively reduces PDK2 basal expression to a comparable extent.\\
\begin{flushleft}
\textbf{\textit{k}}$_{\textbf{49}}$: HIF1$\alpha_n$-dependent LDH$_{c}$ production rate.    
\end{flushleft}
Under hypoxic conditions, LDH protein levels increased due to HIF1$\alpha$ induction at the same rate as GLUT1 \citep{hu2006differential}. Therefore, we estimated this rate to instigate a 3.4-fold boost in LDH protein level in response to a 2.3-fold rise in HIF1$\alpha$ expression.\\
\begin{flushleft}
\textbf{\textit{k}}$_{\textbf{50}}$: LDH$_{c}$ basal degradation rate.    
\end{flushleft}
The half-life of LDH in HCT116 cells is reported to be 14 hours \citep{garcia2019changes}, which corresponds to a degradation rate of 0.000825 / min.\\ 
\begin{flushleft}
\textbf{\textit{k}}$_{\textbf{52}}$ and \textbf{\textit{k}}$_{\textbf{53}}$: PDK$_m$-dependent PDH$_{m}^*$ phosphorylation and PDH$_{m}$ dephosphorylation rates, respectively.   
\end{flushleft}
According to a study investigating colon cancer cells with high miR-149-3p levels (HCT116) versus those with diminished miR-149-3p due to p53 loss (HCT116/F), the observed 1.4-fold increase in PDK2 levels in HCT116/F cells led to a higher PDH phosphorylation level by 1.8-fold than HCT116 \citep{liang2020dichloroacetate}. As a result, we estimated elevated PDK levels in our p53-mutated cancer cells to induce a 1.8-fold increase in PDH phosphorylation relative to cells with wild-type p53.\\
\begin{flushleft}
\textbf{\textit{k}}$_{\textbf{54}}$: PDH$_{m}$ basal degradation rate.   
\end{flushleft}
We calculated the PDH degradation rate by considering its half-life, which varies from 41 to 49 hours \citep{hu1983induction}, opting to use the shorter duration of 41 hours for our estimation.\\
\begin{flushleft}
\textbf{\textit{k}}$_{\textbf{58}}$: G6P$_c$ undergoing the PPP rate.   
\end{flushleft}
Studies indicate that under typical physiological conditions, glucose metabolism proceeds primarily via the glycolytic pathway, with a small fraction, around 5\%, being directed into the PPP \citep{kight1995oxidation}. Guided by this evidence, we have estimated this rate to drive a similar proportion of G6P to the PPP in normal conditions, with the remainder metabolised through glycolysis.\\
\begin{flushleft}
\textbf{\textit{k}}$_{\textbf{60}}$: Pyruvate$_c$ mitochondrial import rate.    
\end{flushleft}
Pyruvate concentrations were quantified within the mitochondria and cytoplasm in mouse prostate cancer cells \citep{li2017mitochondrial}. The data revealed that mitochondrial pyruvate concentration is significantly lower than that in the cytosol by approximately 80-fold. Drawing on this data, we estimated this rate to maintain pyruvate concentrations between the compartments relatively close to the experimental findings.\\
\begin{flushleft}
\textbf{\textit{k}}$_{\textbf{61}}$: PDH$_{m}^*$-dependent Acetyl-CoA$_m$ and NADH$_m$ formation rate.  
\end{flushleft}
We assumed that the rate of pyruvate conversion into acetyl-CoA matches the rate of pyruvate conversion into lactate, set at 300 $\mu$M/min. This rate aligns with a calculated constant of $k_{61}$ = [300 $\mu$M/min] / [the normal PDH$_{m}^*$ level (3.176 $\mu$M)], yielding a rate of 95 /min. Our assumption is based on the premise that, under certain conditions, the cell's metabolic machinery adjusts to utilize pyruvate efficiently for both anaerobic and aerobic pathways, allowing for comparable conversion rates.\\
\begin{flushleft}
\textbf{\textit{k}}$_{\textbf{62}}$: Maximal TCA cycle rate.    
\end{flushleft}
It has been reported that glycolysis operates at a rate approximately ten times faster than oxidative phosphorylation \citep{devic2016warburg}. Based on this insight, we inferred that the maximal rate of the TCA cycle is likely around ten times slower than glycolysis, leading to a rate of $k_{57}$/10 = 4 $\mu$M/min for the TCA cycle.\\
\begin{flushleft}
\textbf{\textit{k}}$_{\textbf{63}}$: NADH$_c$ mitochondrial import rate.   
\end{flushleft}
NADH enters the mitochondria through the Malate-Aspartate Shuttle (MAS), essential for preserving a high cytosolic NAD$^+$/NADH ratio \citep{bhagavan2002medical}. This shuttle is represented implicitly in our model without experimental data to define its rate. However, varying this parameter in our simulations showed the system's robustness, with no notable sensitivity to the model outcomes. Consequently, we have assigned an arbitrary rate of 0.1 /min.\\
\begin{flushleft}
\textbf{\textit{k}}$_{\textbf{64}}$: SCO2$_m$-dependent ETC rate.    
\end{flushleft}
The TCA cycle comprises a series of sequential chemical reactions, each depending on the completion of the previous step and catalysed by different enzymes. In contrast, the ETC primarily involves electron transfer and proton pumping, processes that can proceed rapidly once initiated. Given these characteristics, we assumed that the ETC operates relatively faster than the TCA cycle, estimating it to be three times quicker. Thus, we set $k_{64}$ = [12 $\mu$M/min] / [the standard SCO2 level (0.1 $\mu$M)] = 120 /min.\\
\begin{flushleft}
\textbf{\textit{k}}$_{\textbf{66}}$: Lactate$_{out}$ degradation rate.   
\end{flushleft}
This rate was determined based on the half-life of lactate in healthy cells, which is around 60 min \citep{rosenstein2018clinical}.\\
\begin{flushleft}
\textbf{\textit{k}}$_{\textbf{67}}$: ATP basal consumption rate.  
\end{flushleft}
This rate was selected arbitrarily to facilitate a comparative analysis of ATP steady-state levels across all three cell types: normal, cancer p53$^{+/+}$, and cancer p53$^{-/-}$. 
\subsection{Parameter Robustness Analysis}
\begin{figure}
\centering
\includegraphics[width=1\textwidth]{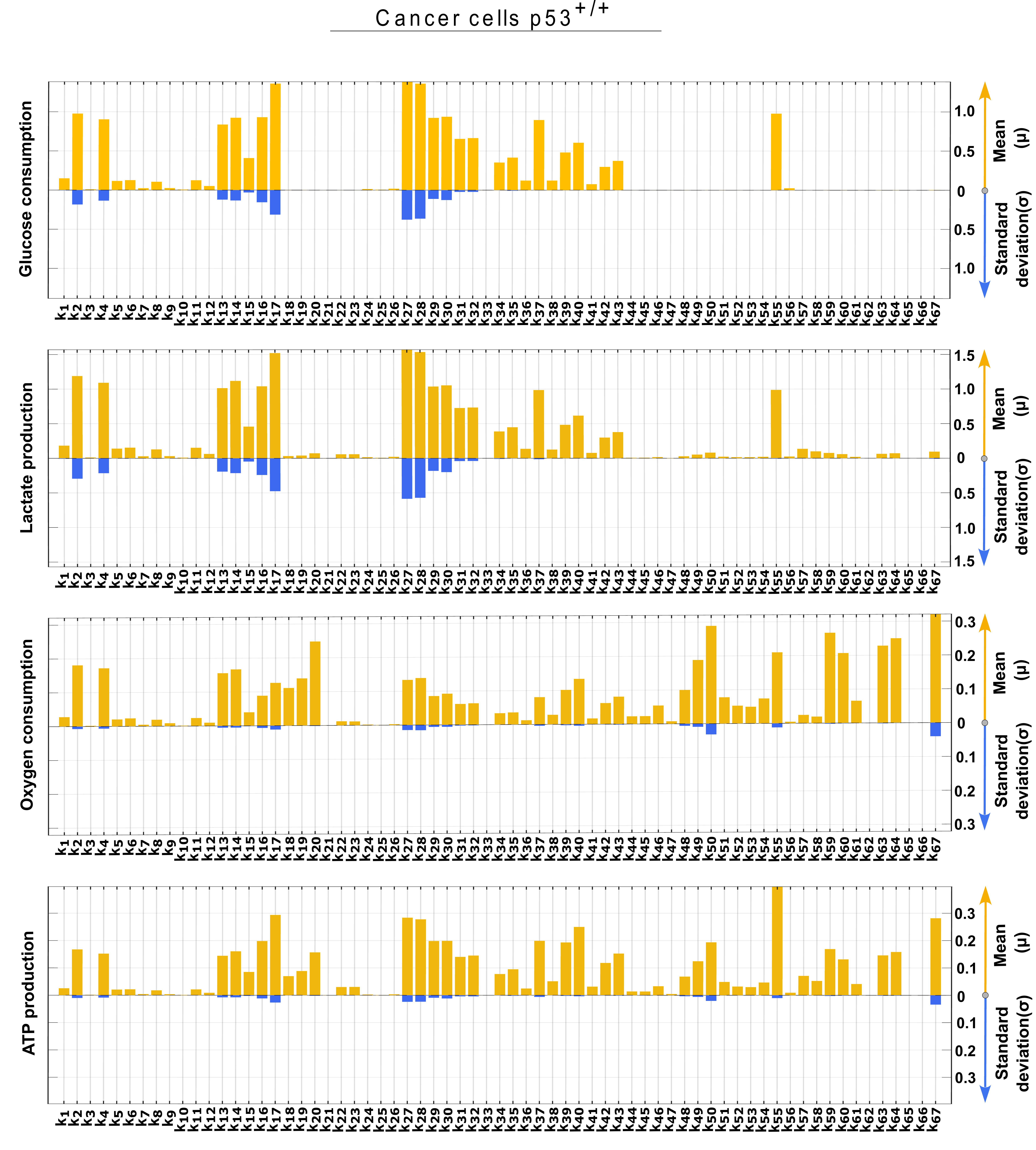}
\caption{Global sensitivity analysis for cancer cells p53$^{+/+}$ using Morris method. The mean (yellow) and standard deviation (blue) of the Elementary Effect are shown for each parameter for key model outcomes: glucose consumption, lactate production, oxygen consumption, and ATP production. A higher mean indicates greater influence of the parameter on the model outcome, while a higher standard deviation suggests greater interaction of the parameter with other parameters}
\label{Fig10}
\end{figure}

\begin{figure}
\centering
\includegraphics[width=1\textwidth]{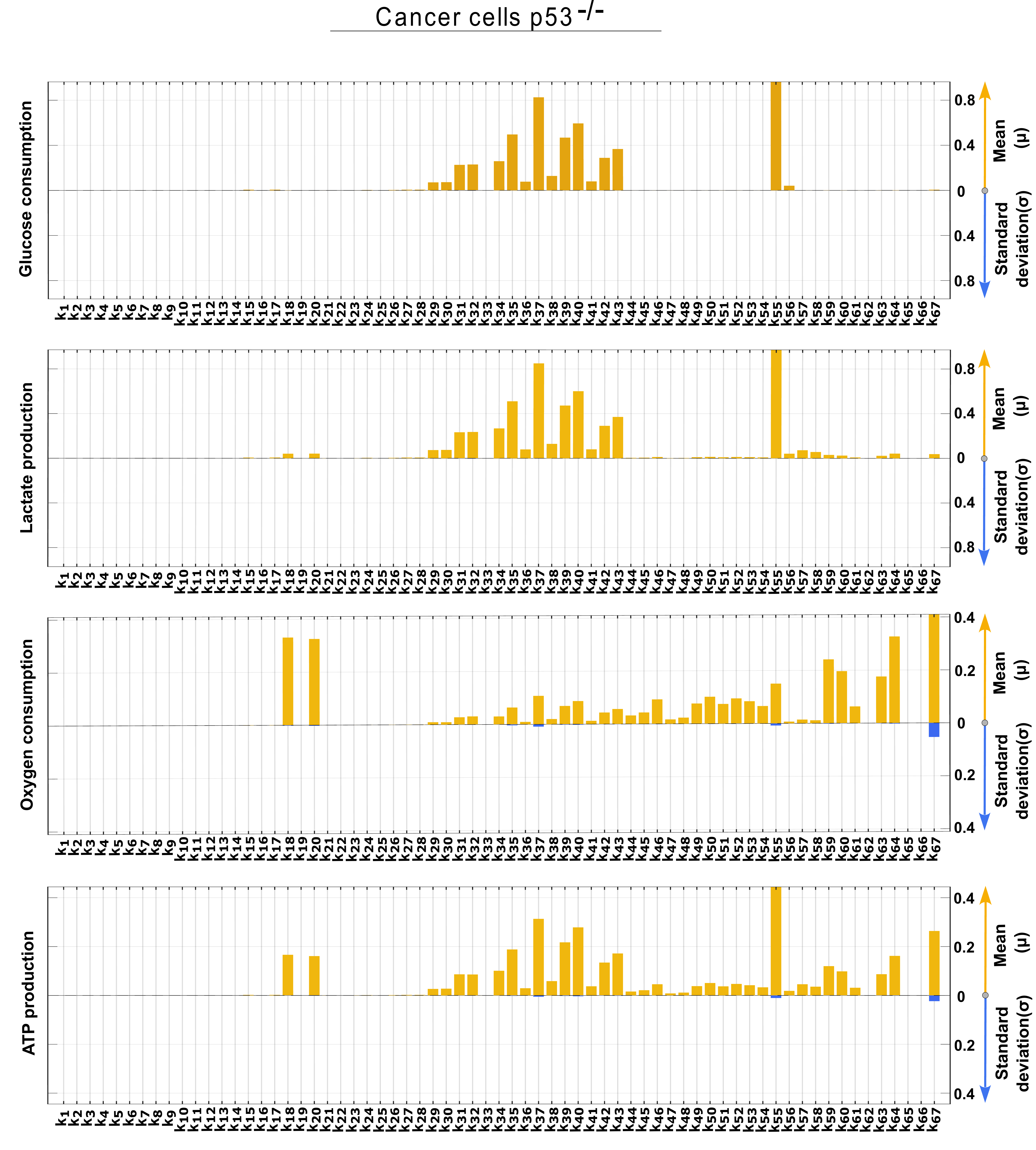}
\caption{Global sensitivity analysis for cancer cells p53$^{-/-}$ using Morris method. The mean (yellow) and standard deviation (blue) of the Elementary Effect are shown for each parameter for key model outcomes: glucose consumption, lactate production, oxygen consumption, and ATP production. A higher mean indicates greater influence of the parameter on the model outcome, while a higher standard deviation suggests greater interaction of the parameter with other parameters}
\label{Fig11}
\end{figure}

To investigate our model's robustness against parameter uncertainties, we performed a global sensitivity analysis across both cancer cell phenotypes using Morris method \citep{morris1991factorial}, see Fig. \ref{Fig10} for p53$^{+/+}$ and Fig. \ref{Fig11} for p53$^{-/-}$.\\

Morris method evaluates the Elementary Effects (EEs) of input parameters by randomly sampling the input space and conducting a series of one-at-a-time (OAT) experiments. In each experiment, one input parameter is varied while the others are kept fixed, and the resultant change in the output is observed. This process is repeated multiple times to explore different regions of the input space.\\

Following these experiments, the mean ($\mu$) and standard deviation ($\sigma$) of these EEs are determined, providing insights into the influence and interaction of each parameter on the model's outcomes. The higher $\mu_{i}$, the higher the overall impact of parameter $k_{i}$ on the output is, while a high standard deviation $\sigma_{i}$ indicates that $k_{i}$'s impact on the output is variable depending on the region of the parameter space, suggesting non-linearity or interactions with other parameters.\\

Applying this method to the main model's outcomes reveals that cancer cells with intact p53 are more sensitive to parameter variations, especially in terms of glucose consumption and lactate production, compared to p53-mutant cells. Conversely, processes like oxygen consumption and ATP production were more robust, with sensitivity indices below 0.5 in both cell types. This robustness suggests that these processes are maintained consistently to sustain cellular energy balance even under different conditions.\\

The parameters significantly affecting glucose consumption and lactate production outcomes in p53-wild type cells were primarily associated with the dynamics of p53 activation and deactivation by WIP1 ($k_{2}$, $k_{4}$, $k_{13}$, and $k_{14}$), as well as those regulating growth factor signalling pathways ($k_{16}$, $k_{17}$, $k_{27}$, $k_{28}$, $k_{29}$, and $k_{30}$). These pathways are intricately linked to the metabolic processes, and their sensitivity underscores the importance of p53 in maintaining metabolic homeostasis and responding to cellular stressors.\\

On the other hand, the oxygen-dependent degradation rate of HIF1 ($k_{37}$) and the rate of glucose transport ($k_{55}$) were critical for both cell types. These parameters, however, exhibited a low standard deviation, indicating that their impact is consistent and linear across different regions of the parameter space. This consistency implies that these parameters have a predictable and stable influence on cellular metabolism, regardless of the p53 status.
\end{appendices}
\end{document}